\documentclass[a4paper,11pt]{article}

\usepackage{jheppub} 
\usepackage[utf8]{inputenc}

\usepackage{amsmath, amssymb, graphicx, booktabs, bm, psfrag, color}
\usepackage{siunitx}
\usepackage{mathrsfs}  
\usepackage{tikz}
\usepackage[compat=1.1.0]{tikz-feynman}
\usepackage{array}
\usepackage[version=3]{mhchem}
\usepackage{subcaption}
\usepackage{multirow}
\usepackage{colortbl}
\usepackage{tablefootnote}
\usepackage{slashed}
\usepackage{enumitem}
\usepackage{float}
\usepackage{placeins}
\usepackage[T1]{fontenc} 

\makeatletter
\AtBeginDocument{%
  \expandafter\renewcommand\expandafter\subsection\expandafter
    {\expandafter\@fb@secFB\subsection}%
  \newcommand\@fb@secFB{\FloatBarrier
    \gdef\@fb@afterHHook{\@fb@topbarrier \gdef\@fb@afterHHook{}}}%
  \g@addto@macro\@afterheading{\@fb@afterHHook}%
  \gdef\@fb@afterHHook{}%
}
\makeatother

\graphicspath{{plots/}}

\title{\boldmath Reconsidering the \textit{One Leptoquark} solution: flavor
  anomalies and neutrino mass}

\author[a]{Yi Cai,}
\author[a]{John Gargalionis,}
\author[b]{Michael A. Schmidt}
\author[a]{and Raymond R. Volkas}

\affiliation[a]{ARC Centre of Excellence for Particle Physics at the
  Terascale,\\School of Physics, The University of Melbourne, Victoria 3010,
  Australia}
\affiliation[b]{ARC Centre of Excellence for Particle Physics at the
  Terascale,\\School of Physics, The University of Sydney, NSW 2006, Australia}

\emailAdd{yi.cai@unimelb.edu.au}
\emailAdd{garj@student.unimelb.edu.au}
\emailAdd{michael.schmidt@sydney.edu.au}
\emailAdd{raymondv@unimelb.edu.au}

\abstract{We reconsider a model introducing a scalar leptoquark $\phi \sim
  (\mathbf{3}, \mathbf{1}, -1/3)$ to explain recent deviations from the standard
  model in semileptonic $B$ decays. The leptoquark can accommodate the
  persistent tension in the decays $\bar{B}\rightarrow D^{(*)}\tau \bar{\nu}$ as
  long as its mass is lower than approximately $10 \text{ TeV}$, and we show
  that a sizeable Yukawa coupling to the right-chiral tau lepton is necessary
  for an acceptable explanation. A characteristic prediction of this scenario is
  a value of $R_{D^*}$ slightly smaller than the current world average.
  Agreement with the measured $\bar{B}\rightarrow D^{(*)}\tau \bar{\nu}$ rates
  is mildly compromised for parameter choices addressing the tensions in $b \to
  s \mu \mu$, where the model can significantly reduce the discrepancies in
  angular observables, branching ratios and the lepton-flavor-universality
  observables $R_K$ and $R_{K^*}$. The leptoquark can also reconcile the
  predicted and measured value of the anomalous magnetic moment of the muon and
  appears naturally in models of radiative neutrino mass derived from
  lepton-number violating effective operators. As a representative example, we
  incorporate the particle into an existing two-loop neutrino mass scenario
  derived from a dimension-nine operator. In this specific model, the structure
  of the neutrino mass matrix provides enough freedom to explain the small
  masses of the neutrinos in the region of parameter space dictated by agreement
  with the anomalies in $\bar{B}\rightarrow D^{(*)}\tau \bar{\nu}$, but not the
  $b \to s$ transition. This is achieved without excessive fine-tuning in the
  parameters important for neutrino mass.}

\begin{document} 
\maketitle
\flushbottom

\section{Introduction}

Recently, measurements in the decays of $B$ mesons have established a number of
significant and unresolved deviations from the predictions of the standard model
(SM). Many of these involve rare flavor changing neutral current (FCNC)
$b\rightarrow s$ transitions. An important example is the LHC\textit{b}
collaboration's measured suppression in the ratios
\begin{equation}\label{eq:RK}
  R_{K^{(*)}} = \frac{\Gamma(\bar B\to\bar K^{(*)}\mu^+\mu^-)}{\Gamma(\bar B\to\bar K^{(*)} e^+ e^-)},
\end{equation}
hinting towards a violation of lepton flavor universality (LFU). Although the
prediction of each individual decay rate is plagued by hadronic uncertainties,
these cancel out in the ratios $R_K$ and $R_{K^{*}}$ in the regime where
new-physics effects are small~\cite{Hiller:2003js, Capdevila:2017bsm,
  Capdevila:2016ivx}. In the SM the prediction of the observables outside of the
low-$q^2$ region is determined by physics which is wholly independent of the
flavor of the lepton pair in the final state, making $R_K$ and $R_{K^*}$ finely
sensitive to violations of LFU. LHC\textit{b} finds~\cite{Aaij:2014ora}
\begin{equation}\label{eq:RK2}
  R_K = 0.745\,_{-0.074}^{+0.090}\pm 0.036,
\end{equation}
for dilepton invariant mass squared range $1 \text{ GeV}^2 < q^2 < 6 \text{
  GeV}^2$, while the SM demands $R^{\text{SM}}_K = 1.0003 \pm
0.0001$~\cite{Bobeth:2007dw}. More recently, LHC\textit{b} have also measured
$R_{K^{*}}$~\cite{newRkstar}:
\begin{equation}\label{eq:RKstar}
    R_{K^*} =\begin{cases}
     0.660\,_{-0.070}^{+0.110}\pm 0.024 & \text{ for $0.045 \text{ GeV}^2 < q^2 < 1.1 \text{
  GeV}^2$} \\
    0.685\,_{-0.069}^{+0.113}\pm 0.047 & \text{ for $1.1 \text{ GeV}^2 < q^2 < 6 \text{
  GeV}^2$}
  \end{cases}, 
\end{equation}
a deviation from the SM prediction~\cite{Bordone:2016gaq} at the $10\%$ level
and a clear signpost to new physics. A number of analyses have argued that each
of these $\sim 2.5 \sigma$ discrepancies can be eliminated through a
four-fermion effective operator $(\bar{s}\gamma_\mu P_L b) (\bar{\mu}\gamma^\mu
P_L \mu)$, leading to new contributions to the muonic decay mode of the $B$
meson~\cite{Descotes-Genon:2013wba, Descotes-Genon:2015uva,
  PhysRevLett.113.241802, Hiller:2014yaa, Ghosh:2014awa, Hurth:2014vma,
  Altmannshofer:2014rta, Glashow:2014iga, Altmannshofer:2017yso,
  Capdevila:2017bsm, Ciuchini:2017mik, Geng:2017svp, DAmico:2017mtc}. Such an
effective operator can also ameliorate other tensions in the measurements of
angular observables and branching ratios involving the $b \rightarrow s$
transition, although these are subject to sizeable hadronic
uncertainties~\cite{Straub:2015ica, Lyon:2014hpa, Descotes-Genon:2014uoa,
  Jager:2014rwa}. Currently, global fits suggest new physics in
$(\bar{s}\gamma_\mu P_L b) (\bar{\mu}\gamma^\mu P_L \mu)$ is preferred at
between $4.2$ and $6.2\sigma$~\cite{Altmannshofer:2017yso, Capdevila:2017bsm,
  Ciuchini:2017mik, Geng:2017svp, DAmico:2017mtc} over the SM, and many
new-physics models attempting to explain this deviation exist.

Another intriguing anomaly is the long-standing deviation in the ratios
\begin{equation}
  R^{\tau/\ell}_{D^{(*)}} = \frac{\Gamma(\bar{B}\rightarrow
    D^{(*)}\tau \bar{\nu})}{\Gamma(\bar{B}\rightarrow
    D^{(*)}\ell \bar{\nu})},
\end{equation}
where $\ell \in \{e,\mu\}$, reported by the BaBar~\cite{Lees:2012xj,
  Lees:2013uzd}, Belle~\cite{Huschle:2015rga, Sato:2016svk, Hirose:2016wfn} and
LHC\textit{b}~\cite{Aaij:2015yra} collaborations. These measurements show a
remarkable degree of self-consistency and together amount to a deviation larger
than $4 \sigma$ from the SM expectation~\cite{Sakaki:2013bfa, Bardhan:2016uhr,
  Freytsis:2015qca, Choudhury:2016ulr, Bhattacharya:2016zcw,
  Bhattacharya:2015ida}. Measurements of the dilepton invariant mass
distribution disfavor many popular new physics scenarios (e.g. type-II two Higgs
doublet models~\cite{Lees:2012xj}) as candidate explanations. We present a
summary of the recent experimental results and SM predictions associated with $b
\to c \tau \nu$ in Table~\ref{tbl:rdrdstartable}.
\begin{table}[t]
  \centering
  \begin{tabular}{ccc}
    \toprule
    Experiment & $R_D$ & $R_{D^*}$ \\
    \midrule
    BaBar~\cite{Lees:2012xj} & $0.440 \pm 0.058 \pm 0.042$ & $0.332 \pm 0.024 \pm 0.018$\\
    \arrayrulecolor{black!30}\midrule
    \multirow{3}{*}{Belle~\cite{Huschle:2015rga, Sato:2016svk, Hirose:2016wfn}} & $0.375 \pm 0.064 \pm 0.026$ & $0.293 \pm 0.038 \pm 0.015$\\
               & --- & $0.302 \pm 0.030 \pm 0.011$\\
               & --- & $0.270 \pm 0.035^{+0.028}_{-0.025}$\\
    \midrule
    LHC\textit{b}~\cite{Aaij:2015yra} & --- & $0.336 \pm 0.027 \pm 0.030$\\
    \arrayrulecolor{black!100}\midrule
    HFAG average\tablefootnote{The HFAG average does not include the most recent Belle measurement~\cite{Hirose:2016wfn}.}~\cite{Amhis:2012bh} & $0.397 \pm 0.040 \pm 0.028$ & $0.316 \pm 0.016 \pm 0.010$ \\
    Our average & --- & $0.311 \pm 0.016$ \\
    \arrayrulecolor{black!30}\midrule
    SM prediction & $0.299 \pm 0.011$~\cite{Lattice:2015rga} & $0.252 \pm 0.003$~\cite{Tanaka:2012nw} \\
    \arrayrulecolor{black!100}\bottomrule
  \end{tabular}
  \caption{A summary of results associated with $b \to c \tau \nu$. Our average
    includes the most recent Belle measurements of $R_{D^*}$, it is calculated
    by taking an error-weighted mean after summing statistical and systematic
    uncertainties in quadrature.}
  \label{tbl:rdrdstartable}
\end{table}

A common origin for $R_{D^{(*)}}$ and the anomalous $b\rightarrow s$ data is
suggested naturally if the former is explained by the effects of the operator
$(\bar{c} \gamma_\mu P_L b)(\bar{\tau} \gamma^\mu P_L \nu)$, related in its
general structure by $\mathrm{SU}(2)_L$ invariance to the aforementioned
four-fermion effective operator accounting for the $b \rightarrow s$ anomalies.
A number of models exploring this idea have been suggested in the
literature~\cite{Alonso:2015sja, Bauer:2015knc, Becirevic:2016oho,
  Becirevic:2016yqi, Boucenna:2016wpr, Boucenna:2016qad, Calibbi:2015kma,
  Crivellin:2017zlb, Deppisch:2016qqd, Deshpand:2016cpw, Fajfer:2015ycq,
  Feruglio:2016gvd, Feruglio:2017rjo, Megias:2017ove, Popov:2016fzr} (along with
many others addressing one or the other anomaly,
\textit{e.g.}~\cite{Becirevic:2015asa, Becirevic:2017jtw, Buras:2013qja,
  Freytsis:2015qca, Gauld:2013qba, Glashow:2014iga, Gripaios:2014tna,
  Hiller:2014ula, Hiller:2014yaa, Mahmoudi:2014mja, Megias:2016bde, Pas:2015hca,
  Sahoo:2015fla, Sahoo:2015qha, Sakaki:2013bfa, Sierra:2015fma,
  Varzielas:2015iva, deBoer:2015boa}) and among these minimal explanations the
Bauer--Neubert (BN) model~\cite{Bauer:2015knc} is one of notable simplicity and
explanatory power: a TeV-scale scalar leptoquark protagonist mediating
$\bar{B}\rightarrow D^{(*)}\tau \bar{\nu}$ at tree-level and the $b \rightarrow
s$ decays through one-loop box diagrams. The leptoquark transforms under the SM
gauge group like a right-handed down-type quark and its pattern of couplings to
SM fermions can also reconcile the measured and predicted values of the
anomalous magnetic moment of the muon, another enduring tension.

Taken together, these measurements paint a picture of new physics interacting
more strongly with the second and third generations of SM fermions, introducing
lepton flavor non-universality and FCNC interactions at energies not
significantly higher than the electroweak scale. Interestingly, many of these
phenomenological motifs arise naturally in radiative models of neutrino mass,
hinting towards the attractive possibility of a common explanation for both
phenomena.

The disparity in scale between the masses of the charged fermions and the sub-eV
neutrinos is a well-established shortcoming of the SM. A distinguishing feature
is that the neutrinos may be Majorana fermions whose mass term can be generated
from suitable lepton number violating effective operators when the high-scale
physics is integrated out. Effective operators that violate lepton number by two
units ($\Delta L = 2$) have been categorized and studied in the
literature~\cite{Babu:2001ex, deGouvea:2007qla}, and a diverse landscape of
models emerges by considering different completions of these in the ultraviolet
(UV). The process of opening up the operators and developing renormalizable
models of neutrino mass has been formalized into a minimal model building
prescription~\cite{Angel:2012ug} from which the canonical seesaw models and
popular radiative scenarios emerge naturally. Previous work has also considered
radiative neutrino mass models whose particle content addresses
$R_{K}$~\cite{Pas:2015hca, Cheung:2016fjo, Cheung:2017efc, Cheung:2016frv, Popov:2016fzr},
$R_{D^{(*)}}$~\cite{Deppisch:2016qqd, Popov:2016fzr} and $(g-2)_\mu$~\cite{Babu:2010vp,
  Cheung:2016fjo, Cheung:2017efc, Cheung:2016frv, Popov:2016fzr}. In Refs.~\cite{Pas:2015hca,
  Deppisch:2016qqd} the flavor anomalies are explained through two light scalar
or vector leptoquarks whose couplings to the SM Higgs doublet and fermions
prohibit a consistent assignment of lepton number to the leptoquarks such that
the symmetry is respected. Thus $\mathrm{U}(1)_L$ is explicitly broken by two
units and the neutrinos gain mass at the one-loop
level~\cite{AristizabalSierra:2007nf}, apart from the imposition of any
additional symmetries\footnote{Mass generation in Ref.~\cite{Babu:2010vp} occurs
  at the two-loop level because the Yukawa couplings of one of the leptoquarks
  to the left-chiral fermions is turned off.}. A general feature of such models
is that large amounts of fine-tuning are required to suppress the neutrino mass
to the required scale with at least one set of leptoquark--fermion couplings
sizeable enough to explain the anomalies.

Our aim in this work is twofold: (i) to study the scalar leptoquark model in the
context of some previously unconsidered constraints and comment more definitely
on its viability as an explanation of both $R_{D^{(*)}}$ and $R_{K^{(*)}}$; and
(ii) to build on previous work by considering a two-loop neutrino mass model
(first presented in Ref.~\cite{Angel:2013hla}) whose particle content includes
the TeV-scale scalar leptoquark present in the BN scenario. In doing so we hope
to establish the explanatory power of this simple extension of the SM,
emphasizing the simplicity with which it can be embedded into a radiative model
of Majorana neutrino mass. We find that the two-loop scheme heavily alleviates
the fine-tuning present in the one-loop models, and we expect this result to be
general for all two-loop topologies.

The remainder of this work is structured as follows.
Section~\ref{sec:thescalarleptoquarkmodel} outlines the scalar leptoquark model
in which the phenomenological analysis of
Section~\ref{sec:phenomenologicalanalysis} takes place. Within this analysis, we
present the regions of parameter space interesting for the flavor anomalies in
Section~\ref{sec:signals}, relevant constraints for the model in
Section~\ref{sec:constraints} and a general discussion of our results in
Section~\ref{sec:resultsanddiscussion}. Finally, in Section~\ref{sec:mv} we
incorporate the scalar leptoquark into a representative two-loop neutrino mass
model.

\section{The scalar leptoquark model}
\label{sec:thescalarleptoquarkmodel}

The leptoquark $\phi$ that features in the BN model transforms under the SM
gauge group as $\phi \sim (\mathbf{3}, \mathbf{1}, -1/3)$, corresponding to the
leptoquark $S_1$ in the nomenclature of Ref.~\cite{Dorsner:2016wpm}. These
transformation properties lead to generalized Yukawa couplings of the leptoquark
to SM quarks and leptons as well as baryon number violating diquark couplings
which we choose to turn off to avoid destabilizing the proton\footnote{This can
  be achieved through the imposition of an appropriate symmetry.}. The part of
the Lagrangian relevant to $\phi$ is\footnote{The correspondence between our
  Yukawa couplings and those of Ref.~\cite{Bauer:2015knc} is $\hat{x}_{ij} =
  -\lambda^L_{ji}$ and $\hat{y}_{ij} = {\lambda^R_{ji}}^{*}$.}
\begin{equation} 
  \label{eq:Lagra} 
  \mathscr{L}_{\phi} = (D_\mu \phi)^\dagger (D^\mu \phi) + m_\phi^2 |\phi|^2 - \kappa |H|^2 |\phi|^2 + \hat{x}_{ij} {\hat{L}_L}^i {\hat{Q}_L}^j \phi^\dagger
  + \hat{y}_{ij} {\hat{e}_R}^i {\hat{u}_R}^j \phi +
  \text{h.c.},
\end{equation}
where $H$ is the SM Higgs doublet, $i,j \in \{1,2,3\}$ are generational indices,
interaction eigenstate fields are hatted and $\chi \psi = \overline{\chi^c}
\psi$ for spinor fields, while $\mathrm{SU}(2)_L$ indices have been suppressed.
We move from the interaction to the charged-fermion mass basis through the
unitary transformations
\begin{equation}
  \begin{split}
    \hat{u}_L^i = (L_u)^{ij} u_L^j, \quad \hat{d}_L^i = (L_d)^{ij} d_L^j, \quad \hat{u}^i_R = (R_u)^{ij} u^j_R,\\
    \hat{e}_L^i = (L_e)^{ij} e_L^j, \quad \hat{\nu}_L^i = (L_e)^{ij} \breve{\nu}_L^j, \quad \hat{e}_R^i = (R_e)^{ij} e_R^j,
  \end{split}
\end{equation}
where $\mathbf{V} = \mathbf{L}_u^\dagger \mathbf{L}_d$ is the
Cabibbo--Kobayashi--Maskawa (CKM) matrix and the
Pontecorvo--Maki--Nakagawa--Sakata (PMNS) matrix $\mathbf{U}$ rotates the
neutrino weak-eigenstate fields $\breve{\nu}_L^i$ into the mass basis: $\nu_L^i
= U^{ij} \breve{\nu}_L^j$. Applying these transformations, the pertinent parts
of the Lagrangian can be written
\begin{equation}
  \begin{split}
    \label{eq:Lagray} 
    \mathscr{L}_\phi &\supset x_{ij} \breve{\nu}_L^i d_L^j \phi^\dagger - [\mathbf{x} \mathbf{V}^\dagger]_{ij} e_L^i u_L^j \phi^\dagger + y_{ij} e_R^i u_R^j \phi + \text{h.c.}\\
    &\equiv x_{ij} \breve{\nu}_L^i d_L^j \phi^\dagger - z_{ij} e_L^i u_L^j \phi^\dagger + y_{ij}
    e_R^i u_R^j \phi + \text{h.c.}
  \end{split}
\end{equation}
where the Yukawa couplings to the left-handed fermions are related through
\begin{equation} \label{eq:mixing}
  \mathbf{z} = \mathbf{x}\mathbf{V}^\dagger  \; .
\end{equation}
The $x_{ij}$ and $y_{ij}$ are free parameters in our model, with the $z_{ij}$
fixed through Eq.~\eqref{eq:mixing}. The Yukawa couplings of the leptoquark to
the first generation of SM fermions are heavily constrained by a number of
processes we discuss in Section~\ref{sec:constraints}. In general, constraints
from processes involving the down-quark are more severe for this leptoquark, and
for the sake of simplicity we therefore take
\begin{equation} \label{eq:freeparams}
  \mathbf{x} = \begin{pmatrix} 0 & 0 & 0 \\ 0 & x_{22} & x_{23} \\ 0 & x_{32} & x_{33} \end{pmatrix}
\end{equation}
throughout this work. Note that in our notation $x_{22} = x_{\nu_\mu s}$,
\textit{et cetera}. We emphasize that even with such a texture, non-zero Yukawa
couplings to the up-quark cannot be avoided since they are generated through the
quark mixing of Eq.~\eqref{eq:mixing}.

Approximate bounds on the mass of the $\phi$ can be inferred from collider
searches. After pair-production, the final states of interest for this work are
$\ell\ell j j$, $\ell j j + \slashed{E}_T$ and $jj + \slashed{E}_T$, where $\ell
\in \{\mu, \tau\}$. The current most stringent results from these channels are
presented here. Experimental limits are usually presented in $(m_{\text{LQ}},
\beta)$ space, where $\beta$ represents the branching ratio to the charged
lepton and quark. The CMS collaboration places an upper limit of $1080\text{
  (}760\text{)} \text{ GeV}$ on the mass of second generation scalar leptoquarks
in the $\mu \mu j j$ channel assuming $\beta = 1\text{ (}0.5\text{)}$, while in
the combined $\mu \mu j j$ and $\mu j j + \slashed{E}_T$ channel, the mass
exclusion reach for $\beta < 1$ is improved: for $\beta = 0.5$, for example,
second generation leptoquark masses below $800 \text{ GeV}$ are
excluded~\cite{Khachatryan:2015vaa}. The most stringent limits in the $b b +
\slashed{E}_T$ channel come from ATLAS. Their analysis excludes third generation
leptoquark masses below $625 \text{ GeV}$ at 95\% confidence for $\beta =
0$~\cite{Aad:2015caa}. Ref.~\cite{Dumont:2016xpj} finds a lower bound between
$400 \text{ -- } 640 \text{ GeV}$ for the BN leptoquark, although this range is
specific to certain parameter choices.

\section{Phenomenological analysis}
\label{sec:phenomenologicalanalysis}

The leptoquark $\phi$ supports a rich beyond-the-standard-model phenomenology
which includes FCNC interactions as well as the possibility of lepton flavor
violation and non-universality. The primary motivations for this work are
charged current processes in the up-quark sector and FCNCs in the down-quark
sector, since these are posited to explain the anomalous measurements in
$R_{D^{(*)}}$ and the $b \to s$ transition, respectively. The new physics
essential to explain these anomalies also implies many heavily constrained
exotic processes, whose adverse effects on the parameter space available to the
model are also computed. Throughout this section, we account for the running of
$\alpha_s$ from the leptoquark-mass scale to the scale appropriate to the
process considered.

For notational convenience, we remove the breve from the neutrino
flavor-eigenstate fields, since we work exclusively with these in this section.
We also define
\begin{equation}
  \hat{m}_\phi = \frac{m_\phi}{\text{TeV}}.
\end{equation}

\subsection{Signals}
\label{sec:signals}

Below we study the ways in which the leptoquark can ameliorate the discrepancies
in the charged current processes $\bar{B} \to D \tau \bar{\nu}$ and $\bar{B} \to
D^* \tau \bar{\nu}$ as well as the neutral current decays associated with the
anomalous $b \to s$ data. We also include the leptoquark's contribution to the
anomalous magnetic moment of the muon.

\subsubsection{Charged current processes}
\label{sec:chargedcurrentprocesses}

The leptoquark's role in decays of the form $b \rightarrow c \ell_i \nu_{j}$ can
be parameterized by the effective Lagrangian~\cite{Sakaki:2013bfa}
\begin{equation} \label{eq:CCHam}
  \begin{split}
    \mathscr{L}^{ij}_{\text{CC}} &= -\frac{4 G_F}{\sqrt{2}} V_{cb} \left[
      C_V^{ij}(\bar{c} \gamma^\mu P_L b)(\bar{\ell}_i \gamma_\mu P_L
      \nu_{j}) + C^{ij}_S (\bar{c}P_L b)(\bar{\ell}_i P_L\nu_{j}) \right. \\
    &\quad \left. + C^{ij}_T (\bar{c} \sigma^{\mu \nu} P_L b)
      (\bar{\ell}_i \sigma_{\mu \nu} P_L \nu_{j})\right] + \text{ h.c.},
  \end{split}
\end{equation}
with the vector, scalar and tensor contributions generated after Fierz
transformation, with Wilson coefficients at the leptoquark mass scale given by
\begin{subequations} \label{eq:ccoperators}
  \begin{align} 
    C_V^{ij} &= \frac{1}{2 \sqrt{2} G_F V_{cb}} \frac{z_{i2}^* x_{j3}}{2m_\phi^2} + \delta_{ij},\\
    C_S^{ij} &= \frac{1}{2 \sqrt{2} G_F V_{cb}} \frac{y_{i 2} x_{j 3}}{2m_\phi^2},\\
    C_T^{ij} &= -\frac{1}{4} C_S^{ij}.
  \end{align}
\end{subequations}

The values of these operators required for a good fit to the available $R_D$ and
$R_{D^{*}}$ data have been studied in the literature, e.g.~\cite{Sakaki:2013bfa,
  Bardhan:2016uhr, Freytsis:2015qca, Choudhury:2016ulr, Bhattacharya:2016zcw,
  Bhattacharya:2015ida}, often under the assumption of lepton-flavor
conservation---that is, new physics allowed only in $C_{V,S,T}^{33}$. One of the
best-fit points suggested by Ref.~\cite{Freytsis:2015qca}:
\begin{equation} \label{eq:bfp}
  \frac{z_{32}^* x_{33}}{\hat{m}_\phi^2} \approx 0.35, \quad \frac{y_{32} x_{33}}{\hat{m}_\phi^2} \approx 0,
\end{equation}
is compatible with new physics only in $C_V^{33}$, and this is the benchmark
considered in the original conception of the BN model. The most recent
measurements of $R_{D^*}$~\cite{Sato:2016svk, Hirose:2016wfn} could not have
been included in their analysis.

We use these results to guide our study but proceed more generally. We evaluate
$R_D$ and $R_{D^{*}}$ by taking an incoherent sum over neutrino flavors in the
final state while accounting for the interference between the SM and leptoquark
contributions when the flavors of the charged lepton and neutrino coincide. The
ratio $R_D$ is evaluated using recently calculated form factors from lattice
QCD~\cite{Lattice:2015rga}, and $R_{D^{*}}$ using form
factors~\cite{Amhis:2012bh} extracted from experiments by
BaBar~\cite{Aubert:2007rs, Aubert:2008yv} and Belle~\cite{Abe:2001yf,
  Dungel:2010uk}, since the lattice results are as yet unavailable. We stress
that the $B \rightarrow D^{*}$ form factors are extracted from measurements of
the decays $\bar{B} \rightarrow D^* (\mu, e) \nu$ assuming the SM, and therefore
our calculation becomes unreliable when the leptoquark effects in the muonic
mode are large. We implement the calculation presented in
Ref.~\cite{Bardhan:2016uhr} and refer the reader there for further detail. 

\begin{figure}[t]
  \centering%
\centering \includegraphics[scale=0.455]{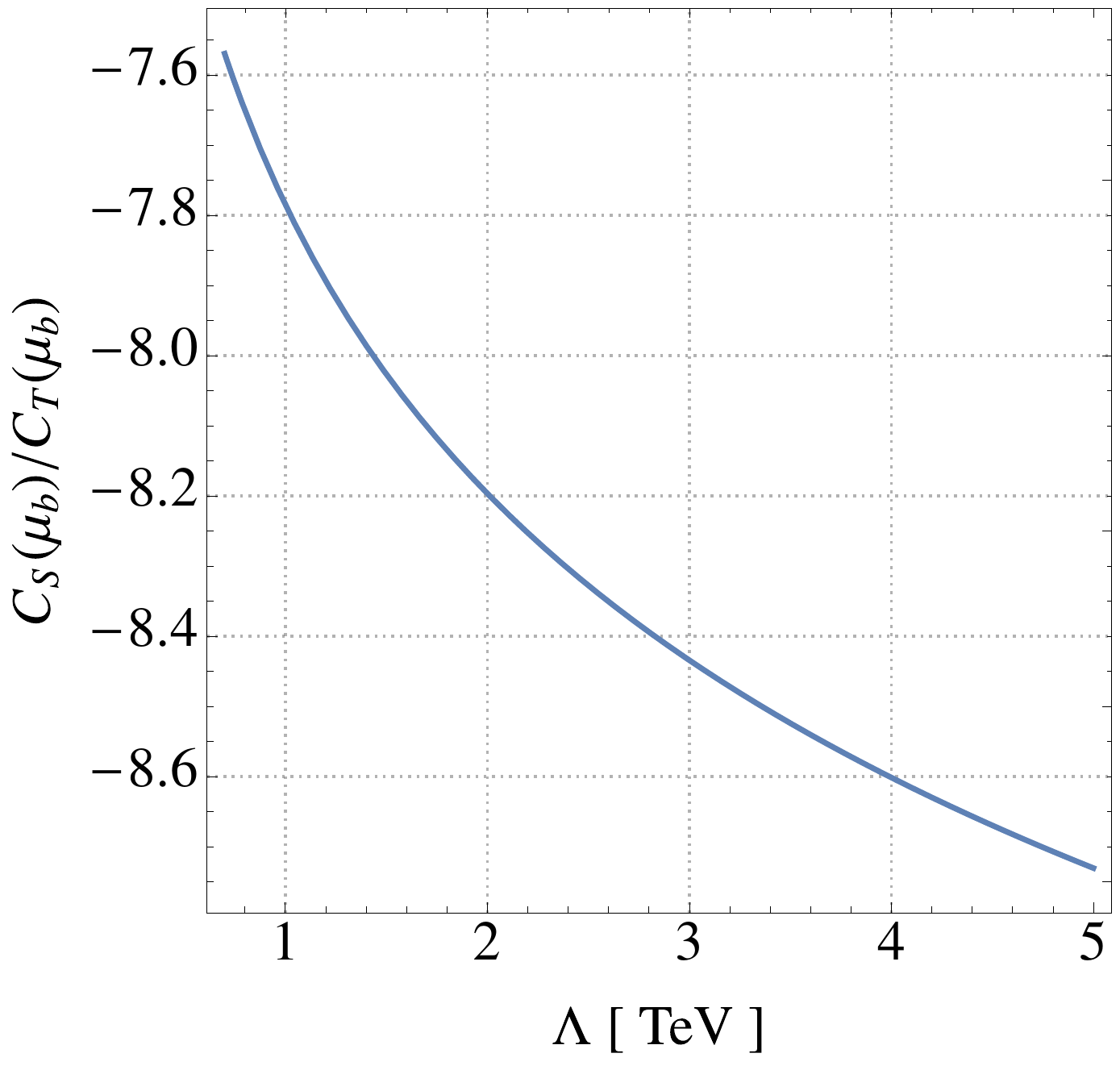}
\caption{The dependence of the ratio of the tensor and scalar Wilson
  coefficients evaluated at $\mu_b$ in $b \to c \ell \nu$ as a function of the
  new-physics scale $\Lambda$, at which the ratio is $-4$. The figure depicts
  the values down to which the ratio $C_S/C_T$ evolves at $\mu_b$. For example,
  running from $1 \text{ TeV}$ to $\mu_b$ implies $C_S/C_T= -7.8$.}
\label{fig:runrat}
\end{figure}

We account for the effects of the running of the strong coupling $\alpha_s$ down
from the high scale ($\Lambda$) to the $b$-quark mass scale ($\mu_b$) for the
scalar and tensor currents. The vector coefficient $C_V$ does not run due to the
Ward identity of QCD. At leading logarithmic order
\begin{subequations} \label{eq:runningrd}
  \begin{align}
    C_S (\mu_b) &= \left[ \frac{\alpha_s(m_t)}{\alpha_s(\mu_b)} \right]^{\frac{\gamma_S}{2\beta_0^{(5)}}} \left[ \frac{\alpha_s(\Lambda)}{\alpha_s(m_t)} \right]^{\frac{\gamma_S}{2\beta_0^{(6)}}} C_S(\Lambda),\\
    C_T (\mu_b) &= \left[ \frac{\alpha_s(m_t)}{\alpha_s(\mu_b)} \right]^{\frac{\gamma_T}{2\beta_0^{(5)}}} \left[ \frac{\alpha_s(\Lambda)}{\alpha_s(m_t)} \right]^{\frac{\gamma_T}{2\beta_0^{(6)}}}C_T(\Lambda),
  \end{align}
\end{subequations}
where $\gamma_S = -8$, $\gamma_T = 8/3$ and $\beta_0^{(n_f)} = 11 -
2n_f/3$~\cite{Dorsner:2013tla}. We use the \textit{Mathematica} package
\texttt{RunDec}~\cite{Chetyrkin:2000yt} to run $\alpha_s$ from $\Lambda \sim
\text{TeV}$ to $\mu_b = \overline{m_b} = 4.2 \text{ GeV}$. This results in a
modification of the relation between the scalar and tensor Wilson coefficients:
$C_T(\Lambda) = -\frac{1}{4} C_S(\Lambda)$. Although most of the running occurs
at the low scale (between $\mu_b$ and $m_t$), the relationship between these
coefficients still depends non-negligibly on the chosen high scale. To
illustrate this dependence, we plot the ratio $C_S(\mu_b)/C_T(\mu_b)$ against
$\Lambda$ in Fig.~\ref{fig:runrat}. Running down to $\mu_b$ from higher scales
increases the magnitude of the scalar coefficient relative to the tensor one.

\begin{figure}[t]
  \centering
  \begin{minipage}[t]{0.45\linewidth}
    \centering \includegraphics[scale=0.55]{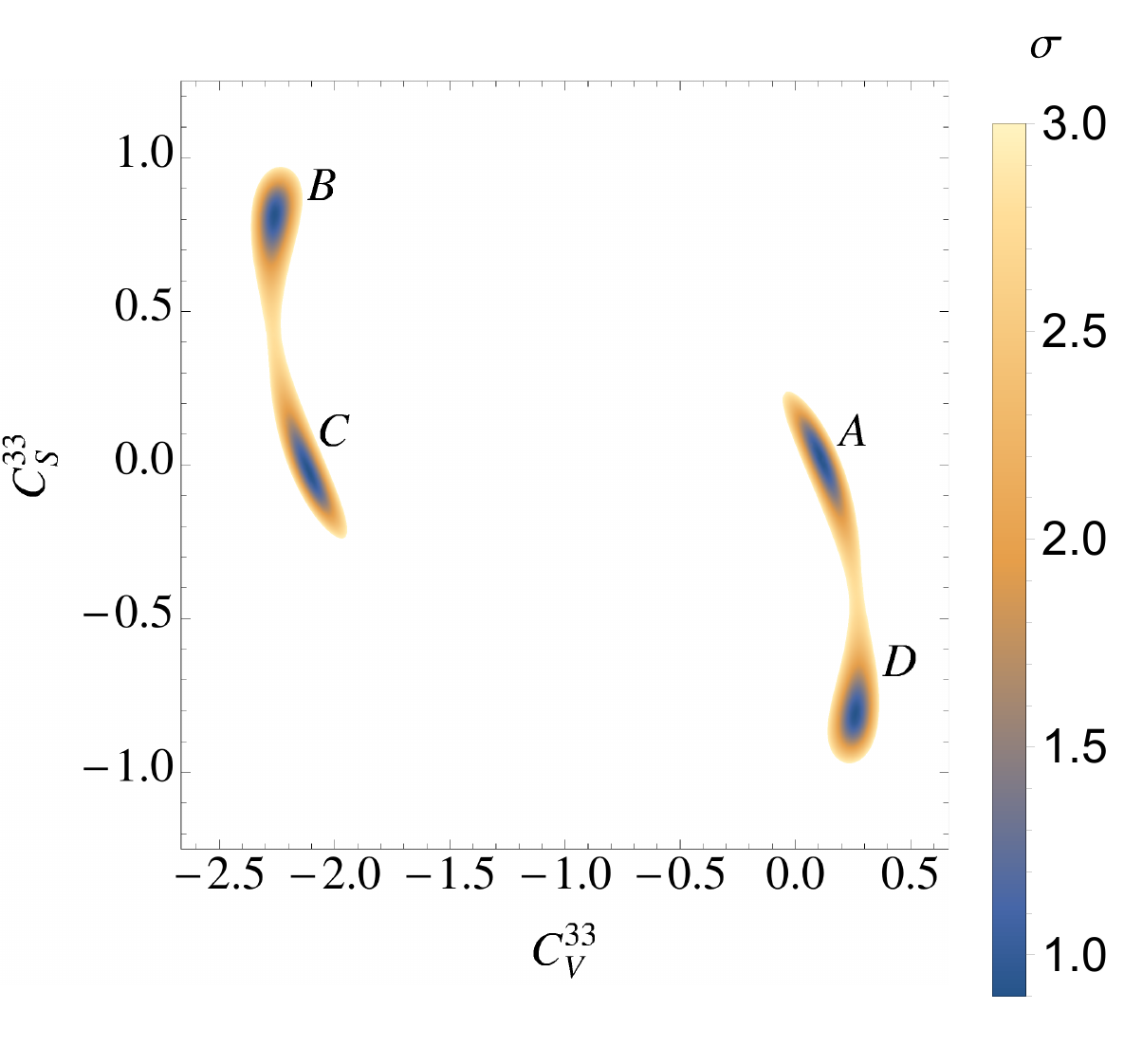}
  \end{minipage}
  \hfill
  \begin{minipage}[t]{0.45\linewidth}
    \centering \includegraphics[scale=0.55]{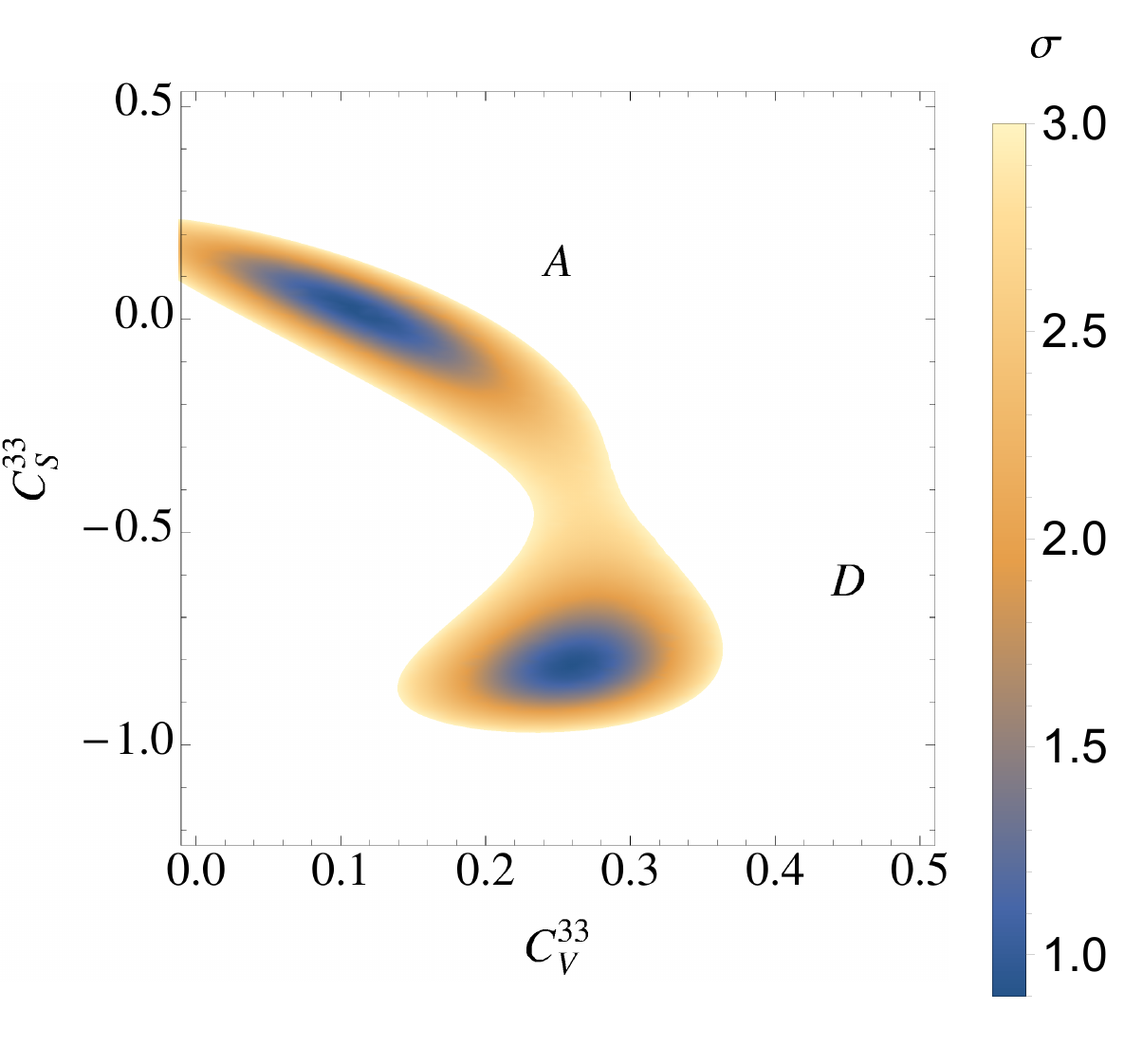}
  \end{minipage}
  \caption{The values for $C_V^{33}$ and $C_S^{33}$ corresponding to a good fit
    to the $R_D$ and $R_{D^*}$ data at $\Lambda = 1 \text{ TeV}$. The colors
    indicate the $\sigma$ values of our fit. The right plot is zoomed to the
    area around regions $A$ and $D$.}
  \label{fig:rdrdstarregions}
\end{figure}

In Fig.~\ref{fig:rdrdstarregions} we present the results of our $\chi^2$ fit to
the measured values of $R_D$ and $R_{D^{*}}$ in $C^{33}_V$--$C^{33}_S$ space to
elucidate the regions of interest. Our fit includes experimental uncertainties,
but none from the theory side. We fit to our own experimental average: $0.311
\pm 0.016$ (an error-weighted mean calculated by adding statistical and
systematic uncertainties in quadrature) which includes the most recent Belle
measurements. For simplicity we assume that the phase of the operators is
aligned with the SM contribution and we do not account for the experimental
correlation between the measurements of $R_D$ and $R_{D^{*}}$. There exist four
regions which provide a good fit to the data for $\Lambda = 1 \text{ TeV}$, the
most easily accessible has best-fit point $(C_V^{33}, C_S^{33}) \approx (0.11,
0.034)$, corresponding to the same region as that surrounding the point given in
Eq.~\eqref{eq:bfp}. Our results in this region are thus in good agreement with
those of Ref.~\cite{Freytsis:2015qca}. The best-fit points of the four regions
are summarized in Table~\ref{tbl:bfpts}.
\begin{table}[t]
  \centering
  \begin{tabular}{cc}
    \toprule
    Region & best-fit point ($C_V^{33}$, $C_S^{33}$)\\
    \midrule
    A & $(0.11, 0.034)$ \\
    B & $(-2.25, 0.81)$ \\
    C & $(-2.12, -0.015)$ \\
    D & $(0.26, -0.81)$ \\
    \bottomrule
  \end{tabular}
  \caption{The best-fit points for our $\chi^2$ fit to the $b \to c \tau \nu$
    data for $\Lambda = 1 \text{ TeV}$. Four distinct regions emerge from our
    analysis, of which region $A$ is the most convenient to attain in a UV
    complete model, since it involves small values of the Wilson coefficients
    and guarantees perturbative Yukawa couplings.}
  \label{tbl:bfpts}
\end{table}

\subsubsection{Neutral current processes}
\label{sec:neutralcurrentprocesses}

\begin{figure}[t]%
  \centering
  \includegraphics{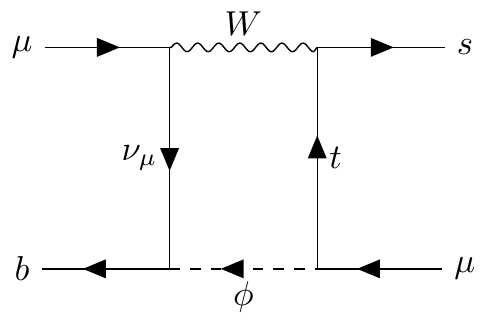}
  \qquad
  \includegraphics{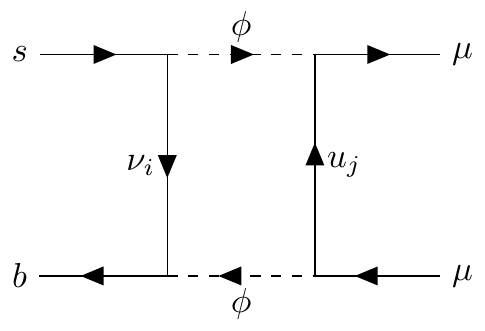}
  \caption{The box diagrams contributing to $C_{LL}^\phi$ and $C_{LR}^\phi$ in
    this scalar leptoquark model.}
  \label{fig:boxes}
\end{figure}

The physics underlying the neutral current $b \rightarrow s$ transitions in the
model can be described by the effective Lagrangian $\mathscr{L}_{\text{NC}}$:
\begin{equation}
  \mathscr{L}_{\text{NC}} = \frac{4 G_F}{\sqrt{2}} V_{tb} V^*_{ts} \frac{\alpha}{4\pi} \sum_{IJ} C^\mu_{IJ}  \mathscr{O}^{\mu}_{IJ},
\end{equation}
where $I,J \in \{L, R\}$ and the operators in this chiral basis are defined
below in terms of $\mathscr{O}_{9,10}$. Following the matching procedure
performed in Ref.~\cite{Misiak:1992bc}, we find that the field $\phi$ generates
the operators
\begin{equation}
  \begin{split}
  \mathscr{O}_{LL}^\mu \equiv
  \frac{1}{2}(\mathscr{O}_9^\mu - \mathscr{O}_{10}^\mu)
  &= (\bar{s}\gamma^\mu P_L b)(\bar{\mu} \gamma_\mu P_L \mu),\\
  \mathscr{O}_{LR}^\mu  \equiv
  \frac{1}{2}(\mathscr{O}_9^\mu + \mathscr{O}_{10}^\mu)
  &= (\bar{s}\gamma^\mu P_L
  b)(\bar{\mu}\gamma_\mu P_R \mu)
  \end{split}
\end{equation}
at the one-loop level with coefficients~\cite{Bauer:2015knc}
\begin{subequations} \label{eq:cllclreqs}
  \begin{align}
    C_{LL}^{\phi,\mu} &= \frac{m_t^2}{8 \pi \alpha m_\phi^2}|z_{23}|^2 -
                         \frac{\sqrt{2}}{64\pi \alpha G_F m_\phi^2 V_{tb}V^*_{ts}}\sum_i x_{i 3}
                         x_{i 2}^* \sum_j |z_{2 j}|^2 , \label{eq:cll}\\
    \begin{split}
      C_{LR}^{\phi,\mu} &= \frac{m_t^2}{16 \pi \alpha m_\phi^2}|y_{2 3}|^2
      \left[ \ln \frac{m_\phi^2}{m_t^2} - f \left( \frac{m_t^2}{m_W^2} \right)
      \right]\\ &\quad - \frac{\sqrt{2}}{64\pi \alpha G_F m_\phi^2
        V_{tb}V^*_{ts}}\sum_i x_{i 3} x_{i 2}^* \sum_j |y_{2 j}|^2, \label{eq:clr}
    \end{split}
  \end{align}
\end{subequations}
where
\begin{equation}
  f(x) = 1 - \frac{3}{x - 1} + \frac{3}{(x - 1)^2} \ln x.
\end{equation}
For the rest of the discussion we remove the $\mu$ superscript from the Wilson
coefficients and operators associated with $b \to s \mu \mu$, since we only
consider new-physics effects in the muonic mode. The dominant contributions are
from the box diagrams shown in Fig.~\ref{fig:boxes}. There are additional lepton
flavor universal contributions from $\gamma$ and $Z$ penguins, however these are
subdominant: the $Z$ penguins are suppressed by small neutrino masses and only
the small short-range contribution from the $\gamma$ penguins contributes to
$C^\phi_9$.

\begin{figure}[t]
\centering \includegraphics[scale=0.755]{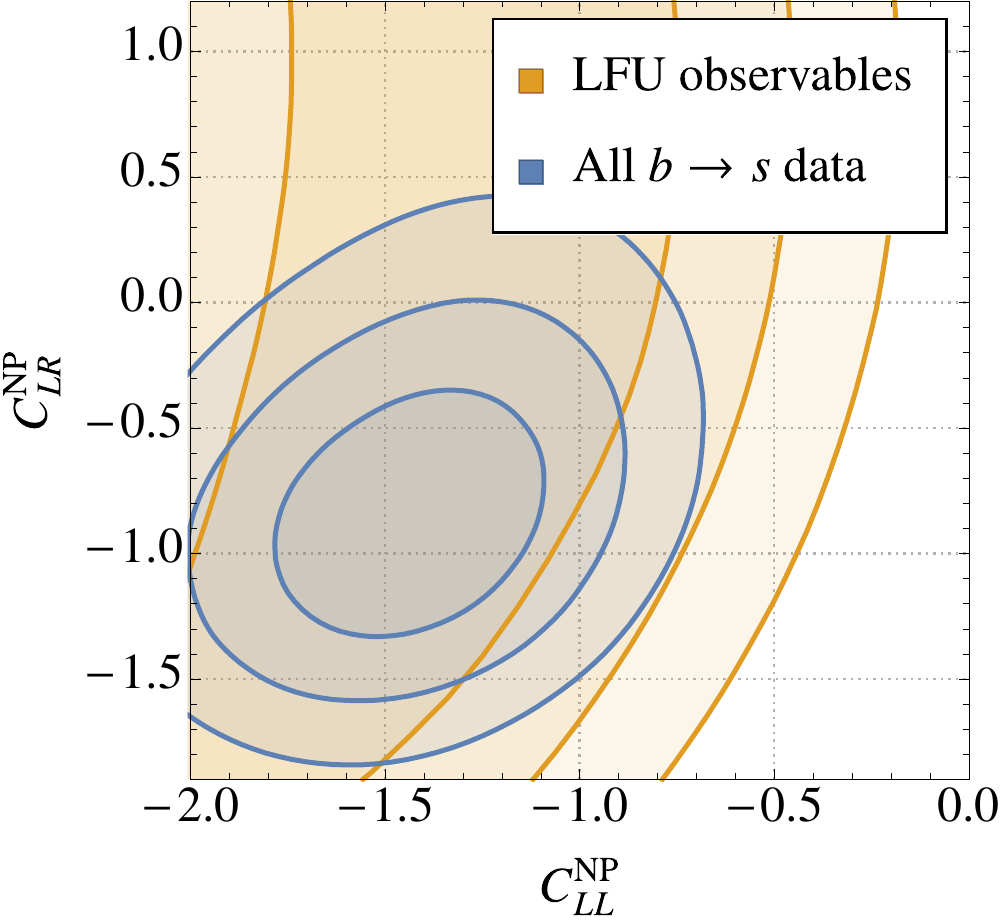}
\caption{ Fig.~1 in Ref.~\cite{Altmannshofer:2017yso} translated into the chiral
  basis. The figure shows the allowed $1$, $2$ and $3\sigma$ contours in the
  $C_{LL}^{\text{NP}}$--$C_{LR}^{\text{NP}}$ plane. The orange contours
  represent the fit to only LFU observables while the blue contours take into
  account all $b\to s$ observables including the branching ratios of $B_s\to
  \mu^+\mu^-$ and the BaBar measurement of $B\to X_s e^+
  e^-$~\cite{Lees:2013nxa}.}
\label{fig:cllclr}
\end{figure}

The authors of Refs.~\cite{Altmannshofer:2017yso, Capdevila:2017bsm,
  Ciuchini:2017mik, Geng:2017svp, DAmico:2017mtc} conduct a global fit of all
available experimental data on the $b\rightarrow s$ decays. They find a good fit
to the data for the chiral coefficients generated by the leptoquark for
\begin{equation} \label{eq:cll12clr0}
  C_{LL}^{\text{NP}} \approx -1.2 \ \text{ and } \ C_{IJ}^{\text{NP}} \approx 0 \ \text{ otherwise},
\end{equation}
where new physics is assumed to significantly alter only the muonic mode and the
fit is performed for $C_{IJ} \in \mathbb{R}$. This choice of coefficients
eliminates the tensions in $R_{K^{(*)}}$ and results in a significantly improved
fit to all of the $b \to s$ data, with a total $4.2\sigma$ pull from the
SM~\cite{Altmannshofer:2017yso}. Although a better fit to all of the data can be
achieved for $C_{LR}^{\text{NP}} < 0$, the choice $C_{LR}^{\text{NP}}\approx 0$
allows slightly smaller values of $C_{LL}^{\text{NP}}$ to explain the
$R_{K^{(*)}}$ anomalies. We translate the top plot of Fig.~1 in
Ref.~\cite{Altmannshofer:2017yso} into the chiral basis relevant for our
leptoquark in Fig.~\ref{fig:cllclr} to elucidate the regions of interest. A good
fit to the measurements of the LFU observables $R_K$ and $R_{K^*}$ is implied
for $-1.8 \lesssim C_{LL}^\phi \lesssim -0.8$ with $C_{LR}^\phi = 0$ and values
close to unity for the mixed-chirality contribution require smaller values for
$C_{LL}^\phi$ to meet the central value of the LFU measurements. The condition
$C_{LR}^\phi \approx 0$ implies a suppression of the Yukawa couplings $y_{2i}$,
while $C_{LL}^\phi \approx -1.2$ requires large leptoquark couplings to the
second and third generation of left-handed quarks for the second term in
Eq.~\eqref{eq:cll}---corresponding to the second diagram in
Fig.~\ref{fig:boxes}---to dominate over the first, since it alone can be
negative.

Throughout the text, we follow Ref.~\cite{Becirevic:2016oho} 
for the calculation of $R_K$.

\subsubsection{Anomalous magnetic moment of the muon}
\label{sec:gminus2}

The leptoquark also mediates one-loop corrections to the $\gamma \mu \mu$ vertex,
contributing to the muon anomalous magnetic moment. In the limit that $m_\phi^2
\gg m_t^2$, the contribution of $\phi$ to $a_\mu = (g - 2)_\mu/2$ is given
by~\cite{Bauer:2015knc, Djouadi:1989md, Chakraverty:2001yg, Cheung:2001ip}
\begin{equation} \label{eq:amu}
  a_\mu^\phi = \sum_{i} \frac{m_\mu m_{u_i}}{4\pi^2 m_\phi^2} \left(
    \frac{7}{4} - \ln \frac{m_\phi^2}{m_{u_i}^2} \right) \text{Re} 
  (y_{2i} z_{2 i}) - \frac{m_\mu^2}{32 \pi^2 m_\phi^2}
  \left[ \sum_i |z_{2 i}|^2 + \sum_i |y_{2 i}|^2 \right],
\end{equation}
and the same-chirality terms are suppressed relative to the mixed-chirality term
by a factor of the muon mass, leading to the requirement of non-vanishing
right-handed couplings for an adequate explanation. We require that the
leptoquark contribution account for the measured anomaly, and thus that
$a_\mu^\phi = (287 \pm 80) \cdot 10^{-11}$~\cite{Davier:2010nc}. 

\begin{figure}[t]
  \centering
\begin{minipage}[t]{.45\linewidth}
\centering \includegraphics[scale=0.47]{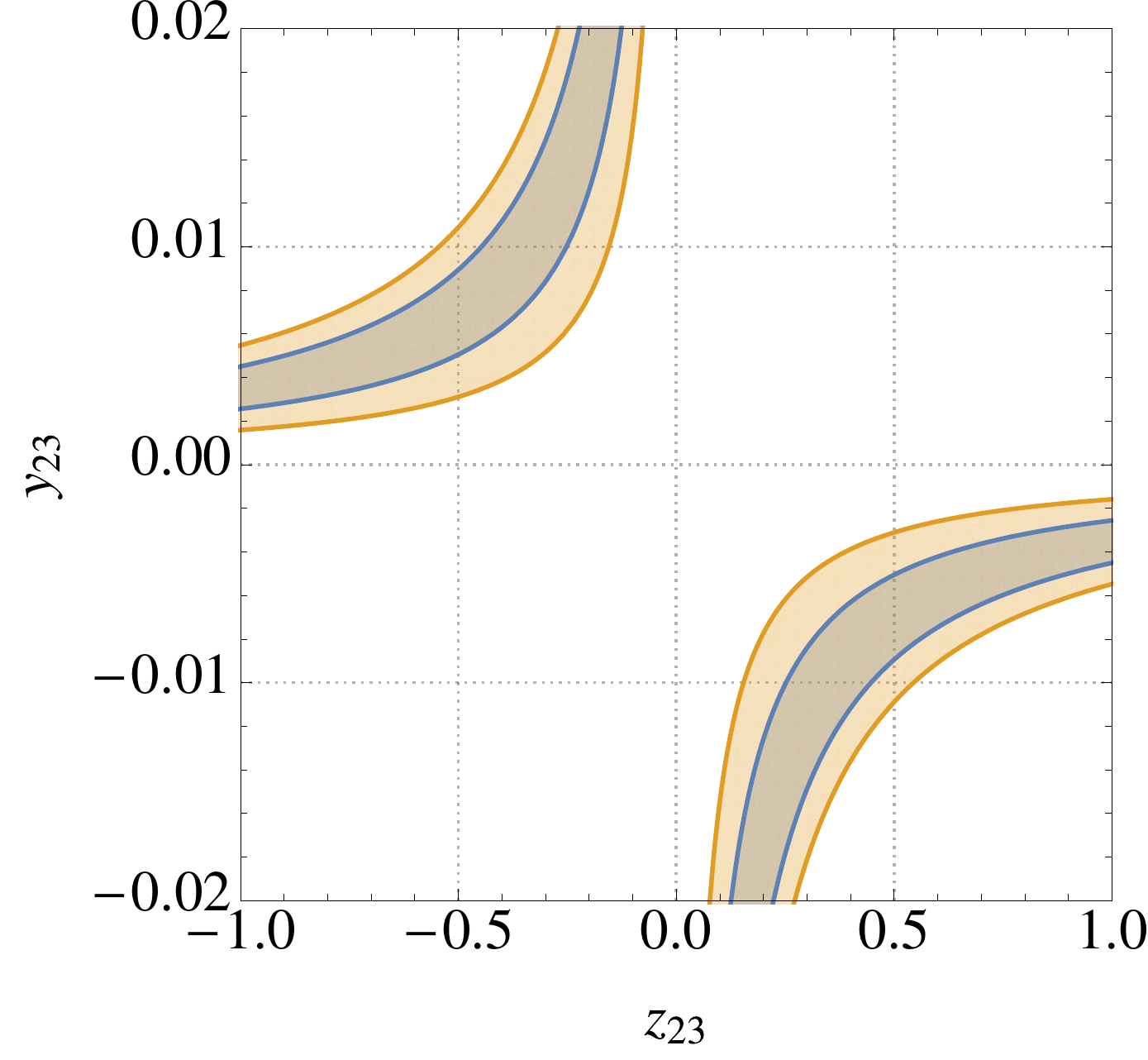}
\phantomsubcaption \label{fig:gm2plot}
\end{minipage}%
\hfill
\begin{minipage}[t]{.45\linewidth}
\centering \includegraphics[scale=0.455]{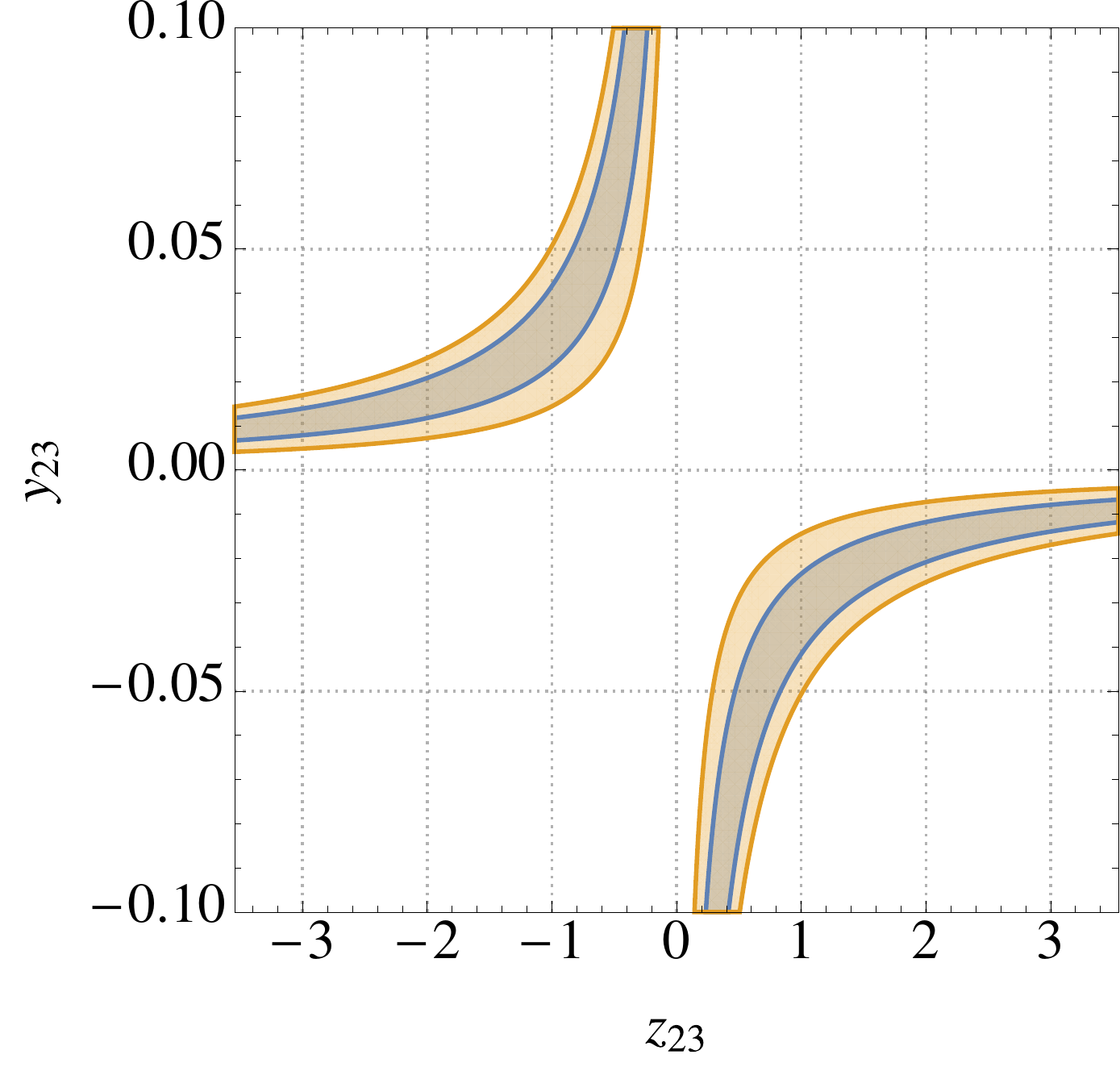}
\phantomsubcaption \label{fig:gm2plot2}
\end{minipage}
\caption{The $1$ and $2\sigma$ allowed regions for $a_\mu$ in the
  $y_{23}$--$z_{23}$ plane for leptoquark masses of $m_\phi = 1 \text{ TeV}$
  (left) and $m_\phi = 5 \text{ TeV}$ (right). The top-mass enhancement in the
  first term of Eq.~\eqref{eq:amu} allows the model to accommodate $a_\mu$ with
  very small values for $|y_{23}|$ with $z_{23} \neq 0$.}\label{fig:gm2plots}
\end{figure}

The top-mass enhancement in the first term makes this the dominant contribution
in this model, and we illustrate the interesting values of $y_{32}$ and $z_{32}$
in Fig.~\ref{fig:gm2plots} for leptoquark mass values of $m_\phi=1 \text{ TeV}$
and $m_\phi=5 \text{ TeV}$. Large $z_{23}$ values assist the model's explanation
of $R_{K^{(*)}}$, hence a combined explanation of this and the $(g-2)_\mu$ anomaly
prefers a small $y_{23}$. Explicitly, the condition~\cite{Bauer:2015knc}
\begin{equation} \label{eq:bnamueq}
  -20.7 (1 + 1.06 \ln\hat{m}_\phi) \text{Re}(y_{23} z_{23}) \approx 0.08 \hat{m}_\phi^2
\end{equation}
must be satisfied to meet the central value of the measurements of $a_\mu$ in
this minimal case.

\subsection{Constraints}
\label{sec:constraints}

We proceed by studying the constraints important for the leptoquark $\phi$ in
the regions of parameter space dictated by the flavor anomalies. This analysis
includes the constraints imposed by $B$, $K$ and $D$ meson decays,
$B_s$--$\bar{B}_s$ mixing, lepton-flavor violating processes and electroweak
measurements.

Many of these processes are studied in the context of an effective-operator
framework. Since much of the nomenclature for four-fermion operator coefficients
is often only based on the Lorentz-structure of each term, keeping the naming
conventions present in the flavor-physics literature for each process can lead
to ambiguity. For this reason we index each effective Lagrangian appearing in
this section and retain the common names for each term, with the Lagrangian's
index prepended to the label. For example, $C_{i,V_L}$ might correspond to the
coefficient of an operator like $(\bar{\phi} \gamma_\mu P_L \chi)(\bar{\psi}
\gamma^\mu P_L \omega)$ in $\mathscr{L}_i$, where $\phi,\chi,\psi,\omega$
represent Fermion fields. For clarity we remind the reader that the coefficients
of $\mathscr{L}_{\textsc{CC}}$ and $\mathscr{L}_{\textsc{NC}}$ from the previous
section are left unindexed.

For the reader's convenience, we signpost the important results
of this section below.

\paragraph{Constraints on the left-handed couplings.} A feature of the BN model
is that the effective operators mediating the $b \to s \mu \mu$ decays are
generated through box diagrams, since $\phi$ only couples down-type quarks to
neutrinos at tree-level. As a consequence, the large Yukawa couplings required
to meet the $b \to s$ measurements will mediate FCNC processes with a neutrino
pair in the final state---processes to which they are related by
$\textrm{SU}(2)_L$ invariance---at tree level. This makes the $b \to s \nu \nu$
decays and $K^+ \to \pi^+ \nu \nu$ very constraining for this model's
explanation of $R_{K^{(*)}}$. The former decay is more important, since it
involves the combination of left-handed couplings present in
Eq.~\eqref{eq:cllclreqs}: $\sum_{i} x_{i3}x^*_{i2}$, and essential to ensure a
negative value for $C^\phi_{LL}$. The leptoquark also contributes to
$B_s$--$\bar{B}_s$ mixing through box diagrams similar to those given in
Fig.~\ref{fig:boxes} with neutrinos running through both internal fermion lines.
We find measurements of $B_s$--$\bar{B}_s$ mixing to be more constraining than
those of FCNC decays for leptoquark masses larger than a few TeV. For small
leptoquark masses, precision electroweak measurements of the $Z\ell \bar{\ell}$
couplings place upper bounds on the sum of the absolute squares of left-handed
couplings, and a relative sign difference between couplings to the
third-generation quarks and those to the second implies the possibility of a
mild cancellation taming these effects. A very important constraint on the
left-handed coupling $|x_{22}|$ can be derived from the meson decay $D^0 \to \mu
\mu$, a large value of which aids the explanation of $R_{K^{(*)}}$ in this
scenario. It has also recently been pointed out~\cite{Becirevic:2016oho} that
the LFU evident in the ratio $R_{D}^{\mu/e} = \Gamma(\bar{B}\rightarrow D \mu
\bar{\nu}) / \Gamma(\bar{B}\rightarrow D e \bar{\nu})$ represents a significant
hurdle to the leptoquark's explanation of $R_{K}$, and we discuss this
constraint below.

Constraints on the product of left-handed couplings discussed above also
frustrate the model's attempts to explain measurements of the ratios $R_D$ and
$R_{D^{*}}$, specifically in those areas of parameter space suggested by
new-physics effects only in $C_V^{ij}$. This implies the need for non-vanishing
right-handed couplings $y_{ij}$.

\paragraph{Constraints on the right-handed couplings.} The right-handed
couplings $y_{ij}$ are generally less constrained in this leptoquark model,
since they mediate interactions involving fewer fermion species. The most
stringent limits come from mixed-chirality contributions to tau decays such as
$\tau \to \mu \mu \mu$ and $\tau \to \mu \gamma$, as well as the precision
electroweak measurements mentioned above. Many right-handed couplings also
feature in the model's contributions to $B$, $D$, and $K$ decays.

\subsubsection{Semileptonic charged current processes}
\label{sec:semileptonicchargedcurrentprocesses}

Leptonic and semileptonic charged current processes are a sensitive probe of the
model we study, since the leptoquark $\phi$ provides tree-level channels for
leptonic pseudoscalar meson decays and semileptonic decays of the tau. In order
to describe these processes, we generalize the Lagrangian presented in
Eq.~\eqref{eq:CCHam} to 
\begin{equation} \label{eq:CCHam2}
  \begin{split}
    \mathscr{L}^{ijkl}_{1} &= -\frac{4 G_F}{\sqrt{2}} V_{u_i d_j} \left[
      C_{1,V}^{ijkl}(\bar{u}_i \gamma^\mu P_L d_j)(\bar{\ell}_k \gamma_\mu P_L
      \nu_{l}) + C^{ijkl}_{1,S} (\bar{u}_i P_L d_j)(\bar{\ell}_k P_L\nu_{l}) \right. \\
    &\quad \left. + C^{ijkl}_{1,T} (\bar{u}_i \sigma^{\mu \nu} P_L d_j)
      (\bar{\ell}_k \sigma_{\mu \nu} P_L \nu_{l})\right],
  \end{split}
\end{equation}
where the vector, scalar and tensor Wilson coefficients at the leptoquark mass
scale now read
\begin{subequations} \label{eq:ccoperators2}
  \begin{align} 
    C_{1,V}^{ijkl} &= \frac{1}{2 \sqrt{2} G_F V_{u_i d_j}} \frac{z_{kj}^* x_{li}}{2m_\phi^2} + \delta_{kl},\\
    C_{1,S}^{ijkl} &= \frac{1}{2 \sqrt{2} G_F V_{u_i d_j}} \frac{y_{k j} x_{l i}}{2m_\phi^2},\\
    C_{1,T}^{ijkl} &= -\frac{1}{4} C_{1,S}^{ijkl}, \label{eq:ccops3}
  \end{align}
\end{subequations}
in analogy with Eqs.~\eqref{eq:ccoperators}. The leptonic decay rate for a
pseudoscalar meson $P_{ij} \sim \bar{u}_i d_j$ is then given
by~\cite{Becirevic:2016oho}
\begin{equation} \label{eq:plnu}
  \begin{split}
    \Gamma(P_{ij} \to \ell_k \nu_{l}) &= \frac{G_F^2 m_P |V_{u_i d_j}|^2}{8\pi} f_P^2 m_{\ell_k}^2 \left( 1 - \frac{m_{\ell_l}^2}{m_P^2} \right)^2\\ &\quad \cdot \left| C_{1,V}^{ijkl} - C_{1,S}^{ijkl} \frac{m_P^2}{m_{\ell_k}(m_{u_i} + m_{d_j})} \right|^2,
  \end{split}
\end{equation}
where $f_P$ is the pseudoscalar meson decay constant. As before, we account for
the effect of the running of $\alpha_s$ from the high scale to the scales
appropriate for each decay for the scalar operator. We take the relevant scale
to be $\mu = \overline{m_c} = 1.27 \text{ GeV}$ for the $D$ meson decays and $\mu = 2
\text{ GeV}$ for the $K$ decays, since this is the matching scale used in
Ref.~\cite{Aoki:2016frl}, from which we take the decay constants. Explicitly,
\begin{equation}
  C_{1,S} (\mu) = \left[\frac{\alpha_s(m_b)}{\alpha_s(\mu)} \right]^{\frac{\gamma_S}{2\beta_0^{(4)}}} \left[\frac{\alpha_s(m_t)}{\alpha_s(m_b)} \right]^{\frac{\gamma_S}{2\beta_0^{(5)}}} \left[\frac{\alpha_s(\Lambda)}{\alpha_s(m_t)} \right]^{\frac{\gamma_S}{2\beta_0^{(6)}}} C_{1,S}(\Lambda),
\end{equation}
while the running for the scalar operator featuring in the $B$ decays is the
same as in Eq.~\eqref{eq:runningrd}. Eq.~\eqref{eq:plnu} is finely sensitive to
the Wilson coefficient $C_{1,S}$ since it has the effect of lifting the chiral
suppression of the SM due to the charged lepton mass in the denominator of the
last term. Recent work~\cite{Becirevic:2016oho} has pointed out the importance
of considering the decays $B \to \ell \bar{\nu}$, $K \to \ell \bar{\nu}$, $D_s
\to \ell \bar{\nu}$ and $B \to D^{(*)} \ell \nu$, to which we also add a
discussion of $\tau \to K \nu$ and $B_c \to \ell \bar{\nu}$ below. In addition,
for each relevant process we calculate a LFU ratio, since in many cases these
are well measured quantities which constitute powerful probes of any new-physics
attempting to explain $R_{D^{(*)}}$ or $R_{K^{(*)}}$. We summarize the limits
and values we take for these decays and their relevant ratios in
Table~\ref{tbl:decays}. All values of the decay constants used throughout this
discussion are taken from Ref.~\cite{Aoki:2016frl}.
\begin{table}[t]
  \centering
  \begin{tabular}{cc}
    \toprule
    Observable & Experimental value \\
    \midrule
    $\text{Br}(K\to\mu\nu)$       & $(63.56 \pm 0.11) \%$  \\
    $\text{Br}(D_s \to \mu \nu)$  & $(0.556 \pm 0.025) \%$ \\
    $\text{Br}(D_s \to \tau \nu)$ & $(5.55 \pm 0.24) \%$   \\
    $\text{Br}(B \to \mu \nu)$    & $< 1.0 \cdot 10^{-6}$  \\
    $\text{Br}(B \to \tau \nu)$   & $(1.09 \pm 0.24) \cdot 10^{-4}$ \\
    $\text{Br}(B_c \to \tau \nu)$   & $\lesssim 30 \%$~\cite{Alonso:2016oyd} \\
    \arrayrulecolor{black!30}\midrule
    $r_K^{e/\mu} = \frac{\Gamma(K \to e\nu)}{\Gamma(K\to \mu\nu)}$ & $(2.488 \pm 0.009) \cdot 10^{-5}$\\
    $R_{K}^{\tau/\mu} = \frac{\Gamma(\tau \to K\nu)}{\Gamma(K\to \mu\nu)}$ & $(1.101 \pm 0.016) \cdot 10^{-2}$\\
    $R_{D_s}^{\tau/\mu} = \frac{\Gamma(D_s \to \tau\nu)}{\Gamma(D_s\to \mu\nu)}$ & $10.73 \pm 0.69^{+0.56}_{-0.53}$~\cite{Zupanc:2013byn}\\
    $R_D^{\mu/e} = \frac{\Gamma(B \to D \mu \nu)}{\Gamma(B \to D e \nu)}$ & $0.995 \pm 0.022 \pm
0.039$~\cite{Glattauer:2015teq}\\
    $R_{D^*}^{e/\mu} = \frac{\Gamma(B \to D^* e \nu)}{\Gamma(B \to D^* \mu \nu)}$ & $1.04 \pm 0.05 \pm 0.01$~\cite{Abdesselam:2017kjf}\\
    \arrayrulecolor{black!100}\bottomrule
  \end{tabular}
  \caption{A table summarizing the experimental values we take for the various
    leptonic branching ratios and LFU ratios considered in this section.
    Measurements quoted without explicit citation are taken from
    Ref.~\cite{Olive:2016xmw}.}
    \label{tbl:decays}
\end{table}

The ratio
\begin{equation} \label{eq:rkemu}
  r_K^{e/\mu} = \frac{\Gamma(K \to e \nu)}{\Gamma(K \to \mu \nu)}
\end{equation}
is one of the most precisely measured quantities in weak hadronic physics. As
such, the consideration of next-to-leading-order corrections becomes important
for our phenomenological analysis of the effects of the leptoquark $\phi$ on
these decays. Electroweak effects contributing to $r_K^{e/\mu}$ have been
calculated to order $e^2p^4$ in chiral perturbation theory,
e.g.~\cite{Cirigliano:2007ga, Finkemeier:1994ev}. Higher order contributions to
the quotient Eq.~\eqref{eq:rkemu} are proportional to the lowest order
contribution: $r_K^{e/\mu, (0)}$, calculated directly from Eq.~\eqref{eq:plnu}.
Including the effects of leading higher-order logarithms through $\Delta_{LL}$,
Eq.~\eqref{eq:rkemu} can be written
\begin{equation}
  r_K^{e/\mu} = r_K^{e/\mu, (0)} \left( 1 + \Delta^K_{e^2 p^2} + \Delta^K_{e^2 p^4} + \cdots \right) \left( 1 + \Delta_{LL} \right)
\end{equation}
and we take $\Delta_{LL} = 0.055 \%$, $\Delta_{e^2 p^2}^K = -3.786 \%$ and
$\Delta_{e^2 p^4}^K = (0.135 \pm 0.011) \%$~\cite{Cirigliano:2007ga} in our
calculation. 

One can extend the study of lepton-flavor universality in leptonic kaon decays
by considering the crossed process $\tau \to K \nu$. More specifically, the
ratio
\begin{equation}
  R_K^{\tau/\mu} = \frac{\Gamma(\tau \to K\nu)}{\Gamma(K \to \mu \nu)}
\end{equation}
can be used to derive constraints on the muon and tau couplings of the
leptoquark $\phi$, and a similar approach has been taken to constrain the
couplings of a vector leptoquark in Ref.~\cite{Fajfer:2015ycq}. For the
numerator, we find
\begin{equation}
  \Gamma(\tau \to K \nu) = \frac{G_F^2 |V_{us}|^2}{8 \pi} f_K^2 m_\tau^3 \left( 1 - \frac{m_K^2}{m_\tau^2} \right)^2 \sum_{i} \left| C_{1,V}^{123i}  - C_{1,S}^{123i} \frac{m_K^2}{m_\tau(m_u + m_s)} \right|^2,
\end{equation}
and the ratio $R_K^{\tau/\mu}$ is required to lie within $2\sigma$ of its
experimental value: $(1.101 \pm 0.016) \cdot 10^{-2}$~\cite{Olive:2016xmw}.


Pion leptonic decays have been well-studied in the context of leptoquark models,
and measurements of the ratio $R_\pi^{\mu/e} = \Gamma(\pi \to \mu
\nu)/\Gamma(\pi \to e \nu)$ demand that leptoquark interactions with the
electron and first-generation quarks are small\footnote{In the most minimal
  case, a non-zero $x_{2 1}$ implies $z_{2 1} \approx x_{2 1}$ and these
  couplings alone are sufficient to mediate the decay $\pi^+ \rightarrow \mu^+
  \nu$.}~\cite{Buchmuller:1986iq, Davidson:1993qk}. The electron and down-quark
couplings play no role in the anomalies we consider in this work, and we only
require that the appropriate couplings are small enough to evade these
constraints.

\begin{figure}[t]
  \centering
  \begin{minipage}[t]{1\linewidth}
    \centering \includegraphics[scale=1]{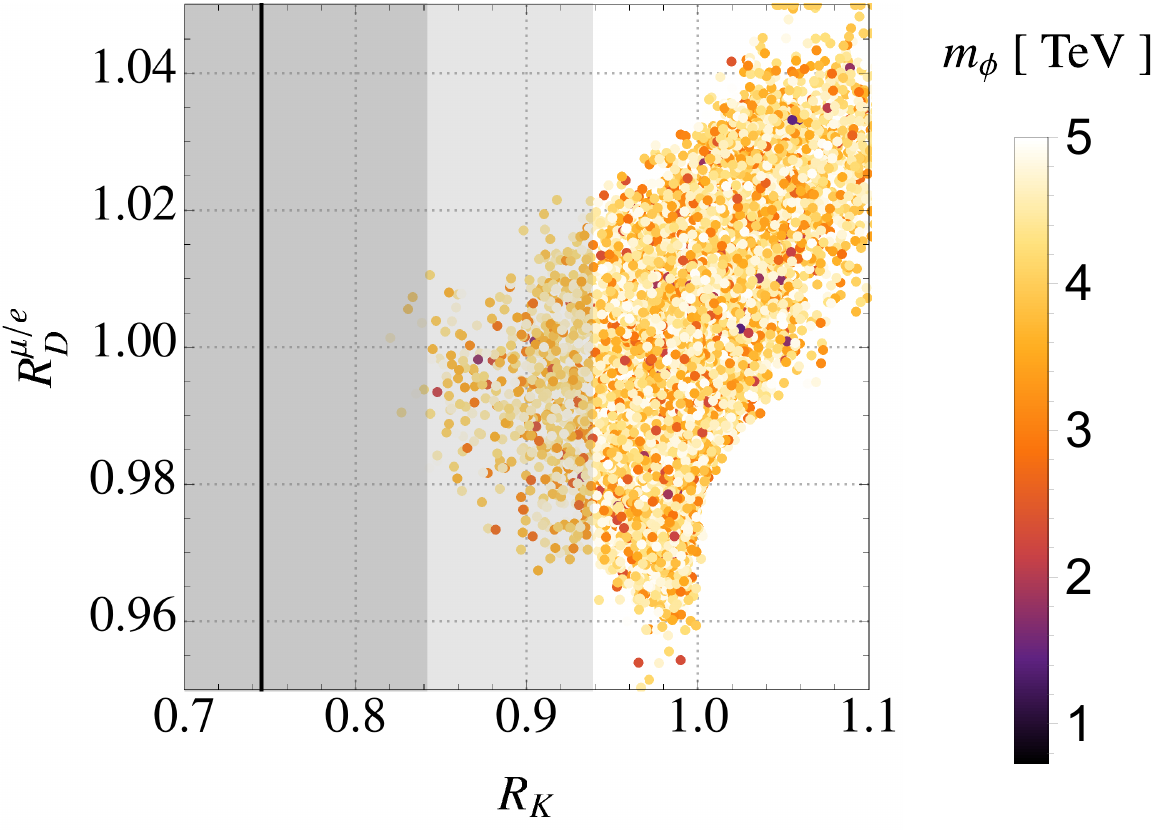}
  \end{minipage}
  \hfill
  \begin{minipage}[t]{1\linewidth}
    \centering \includegraphics[scale=1]{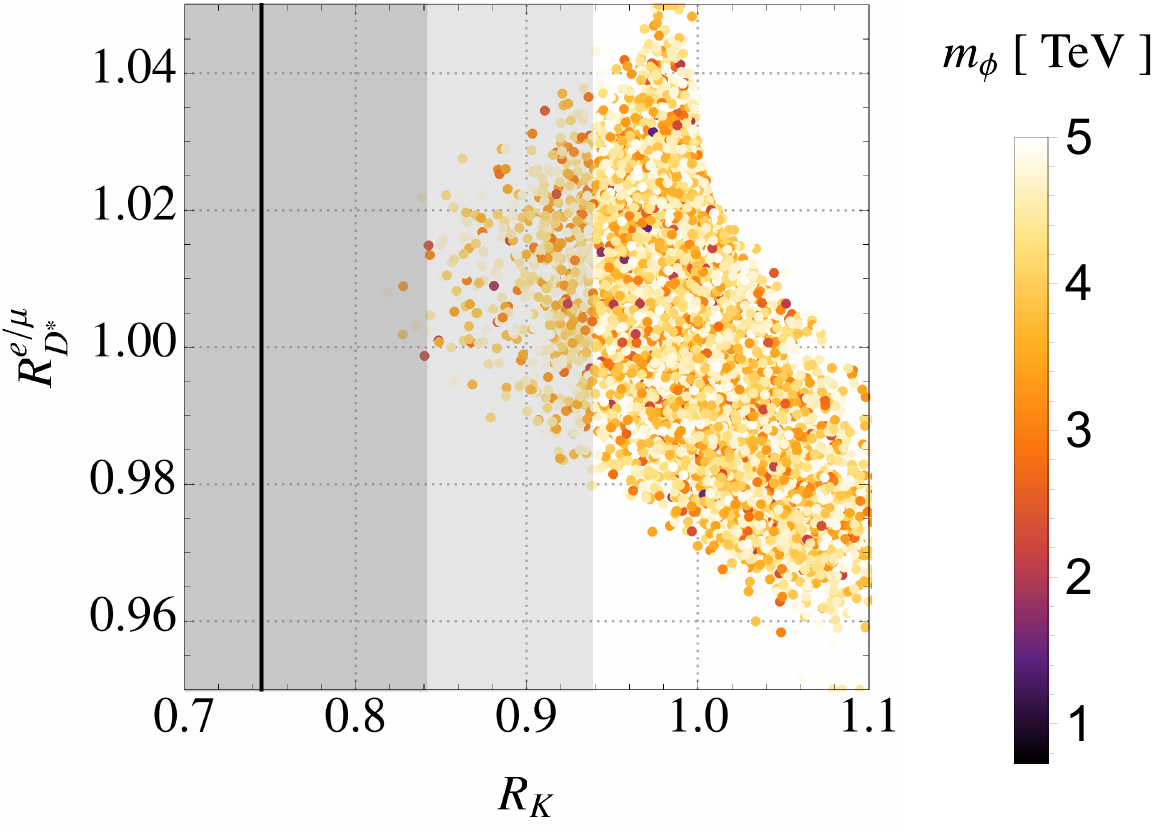}
  \end{minipage}
  \caption{The results of our random scan showing $R_K$ against
      $R_D^{\mu/e}$ (top) and $R^{e/\mu}_{D^*}$ (bottom) for the parameter
      choices detailed in Sec.~\ref{sec:resultsanddiscussion} for `scan I', in
      which the leptoquark mass is allowed to vary to values as large as $5
      \text{ TeV}$. For leptoquark masses between $3$ and $5 \text{ TeV}$, the
      tension in $R_K$ can be significantly resolved while keeping LFU effects
      between electron and muon modes mild.}
  \label{fig:LFUratios}
\end{figure}

\paragraph{Comments on lepton flavor universality in $B \to
  D^{(*)} (e,\mu) \bar{\nu}$.} An additional constraint comes from the
observation that LFU is respected in the ratio of decay rates
\begin{equation}
  R_{D^{(*)}}^{\mu/e} = \frac{\Gamma(\bar{B}\rightarrow
    D^{(*)} \mu \bar{\nu})}{\Gamma(\bar{B}\rightarrow
    D^{(*)} e \bar{\nu})},
\end{equation}
implying a tension with the purported violation in $\mu$--$e$ universality
evident in $R_{K^{(*)}}$. This constraint has been studied in
Ref.~\cite{Becirevic:2016oho}, where it was concluded that the leptoquark model
cannot respect this constraint while explaining the suppression of $R_K$ in the
absence of the right-handed couplings $y_{ij}$. The ratio has been measured to
be $R_D^{\mu/e} = 0.995 \pm 0.022 \pm 0.039$~\cite{Glattauer:2015teq}, while the
reciprocal is presented for the $D^*$ ratio: $R_{D^{*}}^{e/\mu} = 1.04 \pm 0.05
\pm 0.01$~\cite{Abdesselam:2017kjf}. In the case of $R_D^{\mu/e}$, $2\sigma$
consistency with the measurement allows for an approximately $10\%$ deviation
from the SM prediction, a weaker bound than that presented in
Ref.~\cite{Becirevic:2016oho}, while the recent Belle result for
$R_{D^{*}}^{e/\mu}$ permits only a $4\%$ deviation for contributions to the
muonic mode. We find that these constraints become less important for leptoquark
masses larger than $ 1 \text{ TeV}$, permitting sizeable contributions to $R_K$
in this model. We illustrate this point in the top plot of
Fig.~\ref{fig:LFUratios}, where random points passing all of the constraints
presented in our analysis except $R_{D^{*}}^{e/\mu}$ are presented in the
$R_K$--$R_D^{\mu/e}$ plane. The parameters and ranges taken in our scan are the
same as those of scan I in Sec.~\ref{sec:resultsanddiscussion} in which masses
are sampled randomly from the range $[1,5] \text{ TeV}$. The complementary
set-up for $R_{D^{*}}^{e/\mu}$ is shown in the bottom figure of
Fig.~\ref{fig:LFUratios}, \textit{mutatis mutandis}.

\paragraph{Comments on $B_c \to \tau \nu$.} The leptonic decays of the charmed
$B$ meson have not yet been measured---few $B_c$ mesons are produced at $e^+e^-$
$B$-factories and the leptonic mode cannot be reliably reconstructed at
LHC\textit{b}. Despite this, measurements of the $B_c$ lifetime have recently
been shown to imply serious constraints~\cite{Li:2016vvp, Alonso:2016oyd} for
models explaining $R_{D^{(*)}}$ with contributions to the operator $C_S^{3i}$
defined in Eq.~\eqref{eq:CCHam}. Here, we wish to point out that the $B_c \to
\tau \nu$ rate remains SM-like in this leptoquark model due to the presence of
the tensor contribution $C_T^{3i}$, and thus that measurements of the $B_c$
lifetime do not constitute a serious constraint on the model.

\begin{figure}[t]
  \centering \includegraphics[scale=0.75]{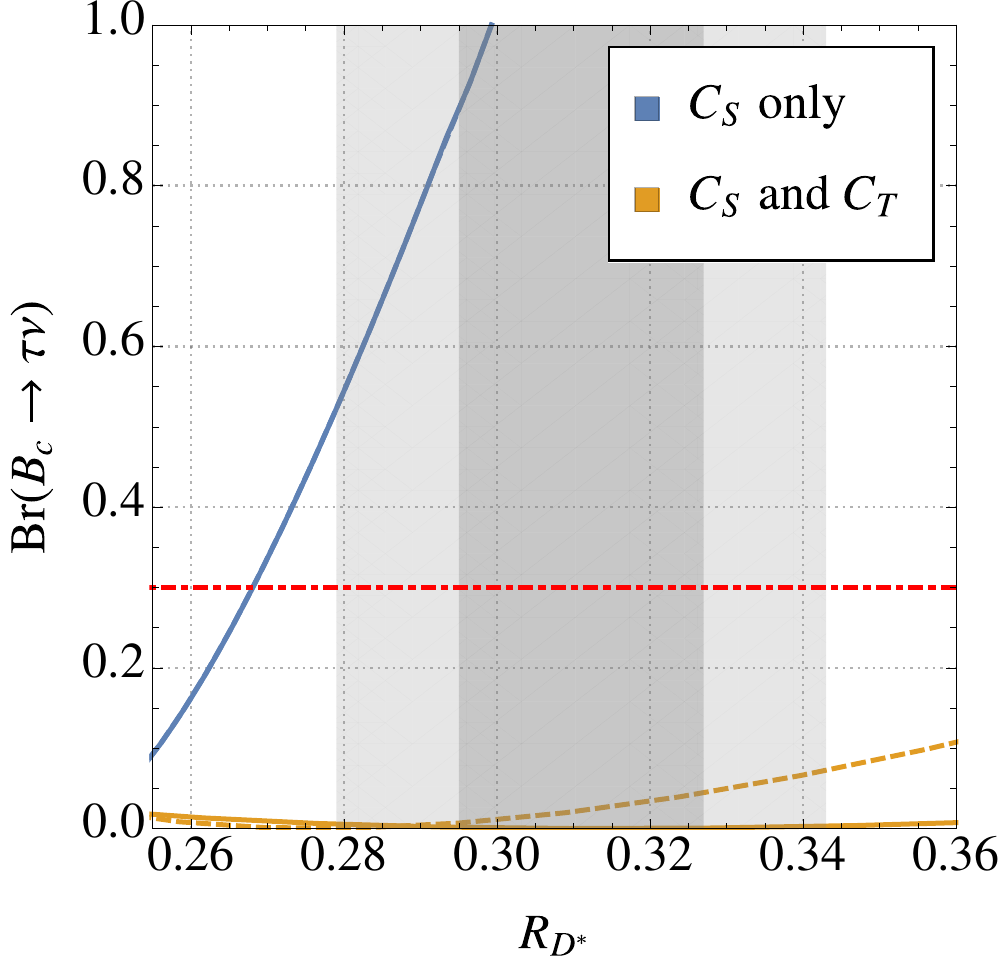}
  \caption{The branching ratio $\text{Br}(B_c \to \tau \nu)$ against $R_{D^*}$
    with new physics only in $C_S$ (solid blue) and new physics in both $C_S$
    and $C_T$ satisfying $C_S / C_T = -4$ (solid orange) and $C_S / C_T = -7.8$
    (dashed orange). The $30\%$ limit is shown in red (dot-dashed). The dark and
    light grey regions represent the $1$ and $2\sigma$ regions for
    $R_{D^{(*)}}$. In this leptoquark model, $\text{Br}(B_c\to \tau \nu)$
    remains SM-like in the region of interest.}
  \label{fig:bctaunu1}
\end{figure}

In Fig.~\ref{fig:bctaunu1} we plot the branching ratio $\text{Br}(B_c \to \tau
\nu)$ in this leptoquark model against interesting values of $R_{D^*}$, in the
spirit of Fig.~1 of Ref.~\cite{Alonso:2016oyd}. The blue curve represents the
contribution from only the Wilson coefficient $C_S$, while the orange curve
represents the contribution from the scalar leptoquark $\phi$ where the scalar
and tensor contributions are related through Eq.~\eqref{eq:ccops3}. The presence
of both the scalar and tensor contributions results renders the branching ratio
SM-like, or slightly suppressed, in the region of interest.

\subsubsection{Lepton flavor violating processes}
\label{sec:leptonflavorviolatingprocesses}

The lepton-flavor symmetries present in the SM are broken by the
Yukawa couplings of the leptoquark to the SM fermions. This implies that $\phi$
can mediate processes that do not conserve lepton flavor, of which those
considered in our analysis are $\ell_i \to \ell_j \gamma$, $\ell_i \to \ell_j
\ell_k \ell_l$ and muon--electron conversion in nuclei: $\mu \ce{^{A}_{Z}N} \to
e \ce{^{A}_{Z}N}$. We use the expressions for these processes found in the
Appendix of Ref.~\cite{Angel:2013hla}, adapted to the case of one leptoquark,
and direct the reader there for more details. We impose the following limits for
the constraints:
\begin{align}
  \text{Br}(\tau \to \mu \gamma) &< 4.4 \cdot 10^{-8}~\text{\cite{Aubert:2009ag}},\\
  \text{Br}(\tau \to \mu \mu \mu) &< 2.1 \cdot 10^{-8}~\text{\cite{Miyazaki:2011xe}},\\
  \text{Br}(\mu \ce{^{197}_{79}Au} \to e \ce{^{197}_{79}Au}) &< 7.0 \cdot 10^{-13}~\text{\cite{Bertl:2006up}}.
\end{align}
In the $\mu \to e$ transition, we only consider muon--electron conversion since
this is the most stringent of the muon's LFV decay modes that the leptoquark can
mediate~\cite{Angel:2013hla, Babu:2010vp, Cai:2014kra}. The tree-level
contributions to muon--electron conversion imply very strong constraints on the
coupling combinations involved. Assuming no accidental cancellation between
terms, the order-of-magnitude bounds~\cite{Angel:2013hla}
\begin{align}
  z_{21}y_{11}^*, y_{21}z_{11}^* &\lesssim \left( 4 \cdot 10^{-9} - 7 \cdot 10^{-8} \right) \frac{m_\phi^2}{m_W^2},\\
  z_{21}z_{11}^*, y_{21}y_{11}^* &\lesssim \left( 10^{-8} - 10^{-7} \right)\frac{m_\phi^2}{m_W^2}.
\end{align}
can be evaded with small electron couplings.

\subsubsection{Rare meson decays}
\label{sec:raremesondecays}

The most important rare meson decays remain to be mentioned. We group them here
and separate their discussion based on the species of lepton in the final state.
The decays studied are: (1) $B \to K \nu \nu$ and $K^+ \to \pi^+ \nu \nu$,
involving neutrinos, and (2) $D^0 \to \mu \mu$ and $D^+ \to \pi^+ \mu \mu$,
involving charged leptons.

\begin{figure}[t]
  \centering \includegraphics[scale=0.85]{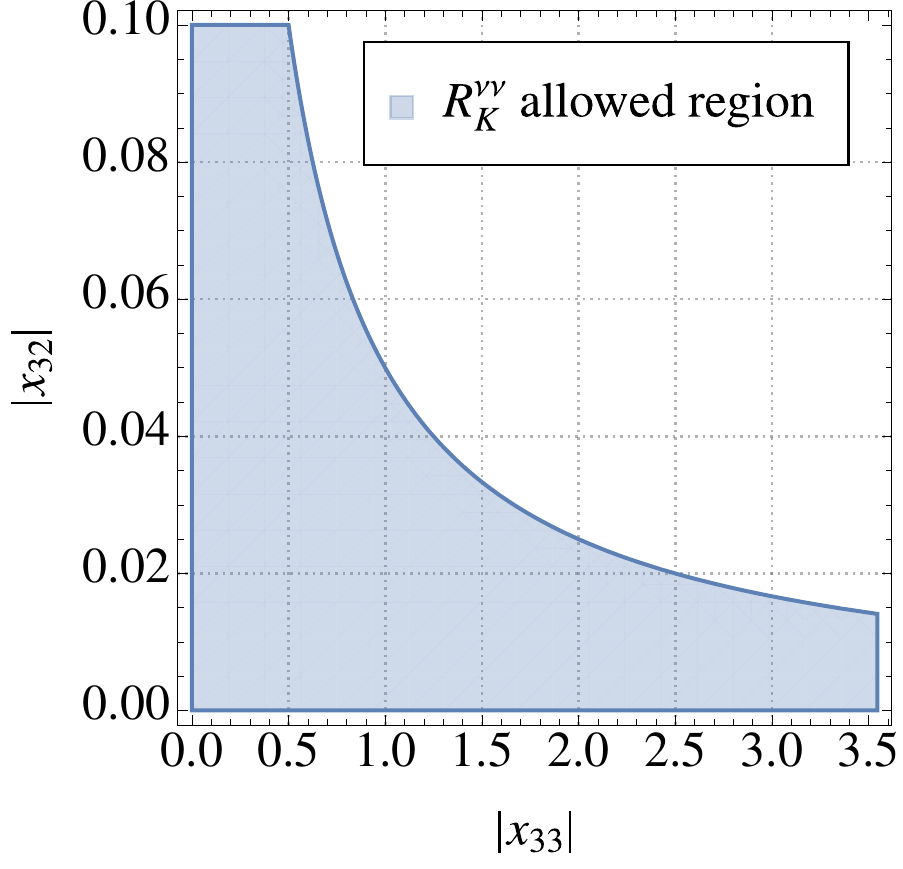}
  \caption{The region allowed by experimental limits on the decay $B \to K \nu
    \nu$ in the $|x_{33}|$--$|x_{32}|$ plane for $m_\phi = 1 \text{ TeV}$. All
    other couplings are switched off. A large value of $|x_{33}|$ is essential
    to explaining $R_{D^{(*)}}$, and the figure implies that such a requirement
    keeps $|x_{32}|$ small.}
  \label{fig:rknunuallowed}
\end{figure}

The decays $B \rightarrow K \nu \nu$ and $K^+ \rightarrow \pi^+ \nu \nu$ heavily
constrain the combination of Yukawa couplings $x_{ij}$ in this model since the
SM contributions proceed at loop-level, while our leptoquark mediates such
neutral current quark decays at tree-level. The physics describing this class of
decays is described by the effective Lagrangian~\cite{Altmannshofer:2009ma,
  Buras:2004uu}
\begin{equation}
  \begin{split}
    \mathscr{L}_2^{i j k l}
    &= \frac{8 G_F}{\sqrt{2}} \frac{e^2}{16 \pi^2} V_{t d_i} V^*_{t d_j} \left[ C^{ijkl}_{2,L} (\bar{d}_i \gamma_\mu P_{L} d_j)(\bar{\nu}_k
      \gamma^\mu P_L \nu_l) \right. \\ &\quad \left. + C^{ijkl}_{2,R} (\bar{d}_i \gamma_\mu P_{R} d_j)(\bar{\nu}_k
      \gamma^\mu P_L \nu_l) \right] + \text{ h.c.}
  \end{split}
\end{equation}
and operator coefficients
\begin{equation}
  C_{2,L}^{ijkl} = -\frac{\sqrt{2}\pi^2}{e^2 G_F m_\phi^2}\frac{x^*_{k j} x_{l i}}{V_{t d_i}V_{t d_j}^*} + C_L^{\text{SM}}\delta_{kl}, \quad C_{2,R}^{ijkl} = 0,
\end{equation}
where $C_L^{\text{SM}} = -X(m_t^2/m_W^2)/s_w^2$. The SM loop function $X(x)$ is
given by~\cite{Buras:2004uu, Altmannshofer:2009ma, Buchalla:1998ba,
  Misiak:1999yg}
\begin{equation}
  X(x) = 
  \frac{x}{8} \left[ \frac{x+2}{x-1} + \frac{3x-6}{(x-1)^2} \ln x \right],
\end{equation}
and the ratio $R_{K}^{\nu\nu} \equiv \Gamma(B\rightarrow K \nu
\nu)/\Gamma(B\rightarrow K \nu \nu)_{\text{SM}}$ is constrained to satisfy
$R_K^{\nu\nu} < 4.3$ at $90\%$ C.L.~\cite{Buras:2014fpa}. We find
\begin{equation} \label{eq:rknunu}
  \begin{split}
    R_K^{\nu\nu} &= \frac{1}{3}\sum_{ij}\frac{|C_{2,L}^{32ij} |^2 }{|C_L^{\text{SM}}|^2}\\
    &= 1 + \frac{a^2}{3m_\phi^4}\sum_{ij}\left|\frac{x^*_{i 2} x_{j 3}}{V_{tb}
        V^*_{ts}}\right|^2 - \frac{2a}{3m_\phi^2} \sum_i \text{Re} \left(\frac{x^*_{i
        2} x_{i 3}}{V_{tb} V^*_{ts}}\right),
  \end{split}
\end{equation}
where $a = \sqrt{2} \pi^2 / (e^2 G_F |C_L^{\text{SM}}|)$. Due to the absence of
right-handed currents, our model predicts $R_K^{\nu\nu}=R_{K^{*}}^{\nu\nu}$
although the bound on $R_{K^{*}}^{\nu\nu}$ is slightly weaker, as is that for
the inclusive decay. A conservative limit on the combination $(\sum x^*_{i 2}
x_{i 3})/\hat{m}^2_\phi$ can be derived using the Schwartz
inequality~\cite{Bauer:2015knc}:
\begin{equation} \label{eq:bnunucond}
  -0.05 \lesssim \frac{[\mathbf{x}^\dagger \mathbf{x}]_{23}}{\hat{m}^2_\phi} \lesssim 0.1,
\end{equation}
where we have assumed $\text{Arg}(x_{i2}^*x_{i3}) = \text{Arg}(V_{tb}V_{ts}^*)$.
We emphasize that this bound represents an insufficient condition for the model
to respect the experimental limits. In Fig.~\ref{fig:rknunuallowed} we present
the allowed region for non-zero $x_{32}$ and $x_{33}$ and $m_\phi = 1 \text{
  TeV}$---a coupling texture interesting for explaining $R_{D^{(*)}}$, although
heavily constrained by $R_K^{\nu\nu}$.

The decay $K^+ \rightarrow \pi^+ \nu \nu$ constitutes the most stringent
constraint on our model from the kaon sector~\cite{Kumar:2016omp}. We
find
\begin{equation}
  \begin{split}
  \text{Br}(K^+\rightarrow \pi^+ \nu \nu) &= \frac{1}{3}\sum_{ij} \kappa_+ \left[ \left(\text{Im} \frac{ V_{ts}V_{td}^* s_w^2 C_{2,L}^{21ij} }{\lambda^5} \right)^2  \right. \\ &\quad \left. +  \left( \text{Re} \frac{ V_{tb}V_{ts}^* s_w^2 C_{2,L}^{21ij}}{\lambda^5} + P_{(u,c)} \delta_{ij} \right)^2 \right],
  \end{split}
\end{equation}
by adapting Eq.~(3.29) of Ref.~\cite{Altmannshofer:2009ma}, where the
factor $\kappa_+ = (5.27 \pm 0.03) \cdot 10^{-11}$ is due mainly to hadronic
matrix elements, $\lambda$ is the CKM Wolfenstein parameter, $P_{(u,c)} = 0.41
\pm 0.05$ accounts for the effects of light-quark loops, and the small
electromagnetic corrections have been neglected. The branching ratio for the
decay has most recently been measured by the E949 collaboration to be
$\text{Br}(K^+ \rightarrow \pi^+ \nu \nu) = (1.73^{+1.15}_{-1.05}) \cdot
10^{-10}$~\cite{Artamonov:2009sz}. A conservative limit can be placed on the
combination of new-physics couplings featuring in $C_{2,L}^{21ij}$ by
considering only same-flavored neutrinos in the final state of the decay. Under
the assumptions that the couplings involved are real and that only one
combination dominates, we find
\begin{equation} \label{eq:kpinunubound}
  -9.1 \cdot 10^{-4} < \frac{[\mathbf{x}^\dagger \mathbf{x}]_{21}}{\hat{m}_\phi^2} < 4.8 \cdot 10^{-4}.
\end{equation}
This bound can be avoided by considering a suppression of the leptoquark
couplings to the first generation of quarks.

In this leptoquark model, the coupling of the $c$-quark to the charged leptons
is essential for the explanation of the $b \to c \tau \nu$ anomalies. Also, as
discussed earlier, the up-quark couplings cannot be entirely avoided due to the
stringency of Eq.~\eqref{eq:kpinunubound} and the mixing of
Eq.~\eqref{eq:mixing}. These factors make the physics of operators of the form
$\mathscr{O}_{ijkl} \sim (u_i \Gamma u_j) (\ell_k \Gamma \ell_l)$ an important
source of constraint on this model. Additionally, in order to ensure
$C_{LL}^\phi \approx -1.2$ in the model's original conception, an ansatz for
$z_{ij}$ was chosen such that $|z_{22}|$ takes $\mathscr{O}(1)$ values.
Constraints from the decays $D^0 \rightarrow \mu \mu$ and $D^+ \rightarrow \pi^+
\mu \mu$ are especially worrying in this case, since the leptoquark mediates
these processes at tree-level. Even within the context of vanishing
first-generation couplings, one cannot avoid inducing new-physics interactions
involving up quarks because of the mixing of Eq.~\eqref{eq:mixing}. The
new-physics contributions to decays of the form $u_i \to u_j \ell_k \ell_l$ can
be contained within the effective Lagrangian
\begin{equation}
  \begin{split}
    \mathscr{L}_3^{i j k l}
    &= \frac{4 G_F}{\sqrt{2}} \bigg[  C^{ijkl}_{3, V_{R}}  (\bar{u}_i \gamma_\mu P_R u_j)(\bar{\ell}_k \gamma^\mu P_R \ell_l) + C_{3, V_{L}}^{ijkl} (\bar{u}_i \gamma_\mu P_L u_j)(\bar{\ell}_k \gamma^\mu P_L \ell_l)\\ &\quad + C_{3, T}^{ijkl} (\bar{u}_i \sigma_{\mu\nu} P_R u_j)(\bar{\ell}_k \sigma^{\mu\nu} P_R \ell_l) + C_{3, S_L}^{ijkl} (\bar{u}_i P_L u_j)(\bar{\ell}_k P_L \ell_l)\\ &\quad + C_{3, S_R}^{ijkl}(\bar{u}_i P_R u_j)(\bar{\ell}_k P_R \ell_l) + \text{h.c.} 
    \bigg],
  \end{split}
\end{equation}
with coefficients $C_{3,i}$ at the leptoquark mass scale given by
\begin{align}
  C_{3,\{V_L,V_R\}}^{ijkl} &= \frac{1}{2\sqrt{2} G_F} \left\{ \begin{matrix} z_{kj}z_{li}^*\\ y^*_{kj} y_{li} \end{matrix}  \right\} \frac{1}{2 m_\phi^2},\\
  C_{3,\{S_L,S_R\}}^{ijkl} &= \frac{1}{2\sqrt{2} G_F} \left\{ \begin{matrix} z_{kj}y_{li}\\ y^*_{kj} z^*_{li} \end{matrix}  \right\} \frac{1}{2 m_\phi^2},\\
  C_{3,T}^{ijkl} &= -\frac{1}{4} C^{ijkl}_{3,S_L}.
\end{align}
For the scalar and tensor operators we account for the running of $\alpha_s$
down to the charm-quark mass scale as in
Sec.~\ref{sec:semileptonicchargedcurrentprocesses}.

For the leptonic decay, we find
\begin{subequations} \label{eq:Dmumu}
  \begin{align}
    \begin{split}
      \Gamma(D^0 \rightarrow \mu \mu) &= \frac{f_D^2 m_D^3 G_F^2}{32\pi}\left(\frac{m_D}{m_c}\right)^2\beta_\mu
      \Bigg[ \left| C_{3,S_L}^{2122}-C_{3,S_R}^{2122}\right|^2 \beta_\mu^2  \\
      & \quad + \left| C_{3,S_L}^{2122}+ C_{3,S_R}^{2122} -\frac{2 m_\mu m_c}{m_D^2} (C_{3,V_L}^{2122}+C_{3,V_R}^{2122})\right|^2 \Bigg] \end{split} \\
    \begin{split}
      &= \frac{f_D^2 m_D^3}{512 \pi m_\phi^4}
      \left( \frac{m_D}{m_c} \right)^2 \beta_\mu \left[ \vphantom{\frac{2 m_\mu
            m_c}{m_D^2}} | y^*_{22} z_{2 1}^* - z_{2 2} y_{2 1} |^2
        \beta_\mu^2 \eta^2 \right. \\ 
      &\quad \left. + \left| \eta(y_{2 2}^* z_{2 1}^* +
          z_{2 2} y_{2 1}) - \frac{2 m_\mu m_c}{m_D^2} (z_{2 2} z_{2 1}^*
          + y_{2 2}^* y_{2 1}) \right|^2 \right] \; ,
    \end{split}
  \end{align}
\end{subequations}
where $\beta_\mu = (1 - 4m_\mu^2/m_D^2)^{1/2} \approx 0.99$, $f_D = 212(2)
\text{ MeV}$~\cite{Aoki:2016frl} and $\eta = C_{3,S_{L}}^{2122}(\overline{m_c}) / C_{3,S_{L}}^{2122}(m_\phi)$. In the limit that the left-handed
contribution dominates, the bound
\begin{equation} \label{eq:zmc}
  |x_{2 2}| < 0.46 \hat{m}_\phi
\end{equation}
can be derived from the experimental upper limit $\text{Br}(D^0 \rightarrow
\mu\mu) < 7.6 \cdot 10^{-9}$~\cite{Aaij:2013cza} assuming $x_{23} \ll x_{22}$.
One can arrange for a mild cancellation between the same- and mixed-chirality
terms in Eq.~\eqref{eq:Dmumu} by allowing the right-handed couplings $y_{2
  (1,2)}$ to take $\mathscr{O}(0.1)$ values, however this creates tensions with
other meson decays such as $D_s \rightarrow \mu \nu$, $K \rightarrow \mu \nu$
and $D^+ \rightarrow \pi^+ \mu \mu$, and we find no overlapping allowed region.

For the decay $D^+ \to \pi^+ \mu \mu$, we implement the calculation of
Ref.~\cite{Fajfer:2015mia}. The branching ratio
\begin{equation}
  \text{Br}(D^+ \to \pi^+ \mu^+ \mu^-) < 8.3 \cdot 10^{-8},
\end{equation}
is measured by extrapolating spectra over the resonant
region~\cite{Aaij:2013sua}, while the bounds on the separate high- and low-$q^2$
bins are
\begin{align}
  \text{Br}(D^+ \to \pi^+ \mu^+ \mu^-)_{q^2 \in [1.56, 4.00]}   &< 2.9 \cdot 10^{-8},\label{eq:Dpimumu1}\\
  \text{Br}(D^+ \to \pi^+ \mu^+ \mu^-)_{q^2 \in [0.0625, 0.276]} &< 2.5 \cdot 10^{-8},\label{eq:Dpimumu2}
\end{align}
where $q^2$ ranges are given in GeV$^2$. Both Eq.~\eqref{eq:Dpimumu1} and
Eq.~\eqref{eq:Dpimumu2} are imposed in our numerical scans.

\subsubsection{Meson mixing}

A complementary constraint on the left-handed couplings can be derived from
$B_s$--$\bar{B}_s$ mixing, providing a stronger bound than $R_K^{\nu\nu}$ for
leptoquark masses larger than a few TeV. The UT\textit{fit} collaboration
determines constraints on $\Delta F = 2$ processes in terms of the quotient of
the meson mixing amplitude and the SM prediction:
\begin{equation}
  C_{B_s} e^{2i\phi_{B_s}} \equiv \frac{\langle B_s | \mathscr{H}^{\Delta F = 2} | \bar{B}_s \rangle}{\langle B_s | \mathscr{H}^{\Delta F = 2}_{\text{SM}} | \bar{B}_s \rangle},
\end{equation}
and the current best fit values for these parameters are $C_{B_s} = 1.052 \pm
0.084$ and $\phi_{B_s} = (0.72 \pm 2.06)^\circ$~\cite{Bona:2007vi}. In the
notation of Ref.~\cite{Bona:2007vi}, our leptoquark only generates the effective
operator $Q^{ij}_1 = C^{bs}_1(\bar{q}^\alpha_i \gamma_\mu P_L
q_j^\alpha)(\bar{q}^\beta_i \gamma^\mu P_L q_j^\beta)$, where $\alpha$ and
$\beta$ are color indices, through box diagrams with neutrinos and leptoquarks
in the loop. The relevant operator coefficient, defined at the high scale
$\Lambda$, is
\begin{equation} \label{eq:Q1}
  C^{bs, \phi}_{1}(\Lambda) = \frac{1}{128 \pi^2} \left(\sum_i \frac{x_{i 3}^* x_{i 2}}{m_\phi}\right)^2,
\end{equation}
in the limit of vanishing SM fermion masses. The SM processes involve similar
box diagrams with top quarks and $W$ bosons in the loop, inducing the Wilson
coefficient (see e.g.~\cite{Fleischer:2008uj})
\begin{equation}
  C^{bs, \text{SM}}_{1} = \frac{G_F^2 m_W^2}{4\pi^2}(V_{tb}^*V_{ts})^2 S_0(m_t^2 / m_W^2),
\end{equation}
where $S_0(x)$ is the well-known Inami-Lim function~\cite{Inami:1980fz}:
\begin{equation}
  S_0(x) = \frac{x^3 -11x^2 + 4x}{4(x-1)^2} - \frac{3x^3}{2(x-1)^3} \ln x.
\end{equation}
We account for the effect of the running of $\alpha_s$ down to $m_W$ for the
coefficient $C^{bs, \phi}_{1}$ to compare with the SM result
using~\cite{Aebischer:2017gaw}
\begin{equation}
  C_1^{bs, \phi}(m_W) =  \left[\frac{\alpha_s(m_t)}{\alpha_s(m_W)}\right]^{\frac{\gamma}{2\beta_0^{(5)}}} \left[\frac{\alpha_s(\Lambda)}{\alpha_s(m_t)}\right]^{\frac{\gamma}{2\beta_0^{(6)}}} C_1^{bs, \phi}(\Lambda),
\end{equation}
where $\gamma = 4$ and $\beta_0^{(n_f)} = 11 - 2 n _f / 3$. The combination of
left-handed couplings in Eq.~\eqref{eq:Q1} is thus required to satisfy
\begin{equation} \label{eq:BsBsbar}
  C_{B_s} e^{2i\phi_{B_s}} = 1 + \frac{1}{32 G_F^2 m_W^2 S_0(m_t^2/m_W^2)} \left(  \frac{\eta^\prime}{V_{tb}^* V_{ts}} \sum_i \frac{x_{i 3}^* x_{i 2}}{m_\phi} \right)^2,
\end{equation}
where $\eta^\prime = C_1^{bs, \phi}(m_W) / C_1^{bs, \phi}(m_\phi)$.

\subsubsection{Precision electroweak measurements}

The Yukawa interactions of the leptoquark with both left- and right-handed SM
fermions give corrections to many electroweak observables. Precision
measurements of these have been translated into bounds on dimension-six
operators in the literature, and we proceed by applying the results of a recent
fit to the electroweak precision data~\cite{Ciuchini:2013pca}. Specifically, we
consider the way in which the couplings $x_{ij}$ and $y_{ij}$ are constrained by
precision electroweak measurements of the $Z \ell \bar{\ell}$ couplings $g_L$
and $g_R$. These receive corrections from leptoquark loops in our
model~\cite{Bauer:2015knc}:
\begin{equation} \label{eq:neccond}
  \begin{split}
  \delta g_I^{\ell_i} &= (-1)^{\delta_{IR}} \frac{3}{32 \pi^2} \frac{m_t^2}{m_\phi^2} \left( \ln \frac{m_\phi^2}{m_t^2} - 1 \right) |\lambda^I_{i 3}|^2 \\ &\quad - \frac{1}{32 \pi^2}\frac{m_Z^2}{m_\phi^2}\sum_{j}^2 |\lambda^I_{i j}|^2 \left[ \left( \delta_{IL} - \frac{4 s_w^2}{3} \right) \left( \ln\frac{m_\phi^2}{m_Z^2} + i\pi + \frac{1}{3} \right) - \frac{s_w^2}{9}\right],
  \end{split}
\end{equation}
where $I \in \{L,R\}$, $\lambda^L_{ij} = z_{ij}$ and $\lambda^R_{ij} = y_{ij}$.
From Eq.~(3.28) and Table 10 of Ref.~\cite{Ciuchini:2013pca}, we calculate the
conservative constraints
\begin{equation} \label{eq:eqZll}
  \text{Re}\delta g_L^{\ell_i} \in [-8.5, 12.0] \cdot 10^{-4}, \quad \text{Re}\delta g_R^{\ell_i} \in [-5.4, 6.7] \cdot 10^{-4}
\end{equation}
at $95\%$ confidence from the fit results obtained using the large-$m_t$
expansion. The expressions in Eq.~\eqref{eq:eqZll} are conservative since we do
not account for correlations between different operators but this does not
affect our results in an important way. The results of the fit are sensitive to
the interference between the SM and leptoquark contributions, hence only the
real part of the $\delta g_I^{\ell_i}$ is constrained.

\section{Results and discussion}
\label{sec:resultsanddiscussion}

Below we study the extent to which the experimental anomalies in $R_{D^{(*)}}$,
$R_{K^{(*)}}$ and $(g-2)_\mu$ can be accommodated in light of the constraints presented
in Sec.~\ref{sec:constraints}. We first consider each anomaly separately and
then present the combined parameter space.

For all of the random scans in this section our Monte Carlo strategy proceeds as
follows. We sample random real values of the free parameters $x_{ij}$ for $i,j
\neq 1$ and leptoquark masses in the range $\hat{m}_\phi \in [0.6, 5]$. Values
are sampled from the region described in Eq.~\eqref{eq:bnunucond}---a necessary
condition for the $x_{ij}$ to respect the bound from $B \to K\nu\nu$, discussed
in Sec.~\ref{sec:raremesondecays}---and the perturbativity bound $|x_{ij}| \leq
\sqrt{4\pi}$ is imposed at sampling. The values chosen for the right-handed
couplings $y_{ij}$ depend on the process studied, although we find that only the
$y_{2i}$ and $y_{32}$ are important for our analysis. Two scans are performed,
here labelled I and II. Scan I explores the parameter space associated with
$R_{K^{(*)}}$ and thus only contains the couplings featuring in
Eq.~\eqref{eq:cllclreqs}, while scan II is intended to elucidate the parameter
space associated with both $R_{K^{(*)}}$ and $R_{D^{(*)}}$, hence $y_{32}$ is
included. An important difference between scans I and II is that the former
allows $C_{LR}^\phi \neq 0$, although this comes at the expense of fewer points
passing all of the constraints since the couplings $y_{22}$ and $y_{23}$ are
heavily constrained by semileptonic charged current processes discussed in
Sec.~\ref{sec:semileptonicchargedcurrentprocesses}. Explicitly, the parameters
and respective ranges over which they are scanned are as follows.
\begin{description}
\item [Scan I.] $6 \cdot 10^6$ points sampled from the region in
  Eq.~\eqref{eq:bnunucond} subject to
  \begin{itemize}
  \item $\hat{m}_\phi \in [0.6,5]$,
  \item $|x_{ij}| \leq \sqrt{4\pi}$ for ${i,j \neq 1}$,
  \item $|y_{22}|, |y_{23}| \leq \sqrt{4\pi}$,
  \item All other couplings are set to zero.
  \end{itemize}
  Of the $6 \cdot 10^6$ points, only $\sim 5 \cdot 10^3$ pass all of the
  constraints.
\item [Scan II.] $6 \cdot 10^6$ points sampled from the region in
  Eq.~\eqref{eq:bnunucond} subject to
  \begin{itemize}
  \item $\hat{m}_\phi \in [0.6,5]$,
  \item $|x_{ij}| \leq \sqrt{4\pi}$ for ${i,j \neq 1}$,
  \item $|y_{23}| \leq 0.05$, $|y_{32}| \leq \sqrt{4\pi}$,
  \item All other couplings, including $y_{22}$, are set to zero.
  \end{itemize}
  We will see from the results of scan I that $y_{22} \approx 0$ is preferred
  for $R_{K^{(*)}}$, hence we take it to vanish in scan II. The range $|y_{23}|
  \leq 0.05$ is motivated \textit{a posteriori} by the fit to $(g - 2)_\mu$ and
  the avoidance of a number of constraints. These relaxed requirements on the
  $y_{2i}$ mean that, of the $6 \cdot 10^6$ generated points, $\sim 3.7 \cdot
  10^4$ pass all of the constraints.
\end{description}
For each of the points the relevant observables and operators $R_D$, $R_{D^*}$,
$C_{LL}^\phi$ and $C_{LR}^\phi$ are calculated and then the associated coupling
constants are filtered through the constraints considered, including
$R_K^{\nu\nu} < 4.3$.

Our analysis mainly focuses on answering the following questions: (1) To what
extent can the present leptoquark model explain $R_{K^{(*)}}$ while maintaining
a SM-like $R_{D^{(*)}}$? (2) To what extent can it explain $R_{D^{(*)}}$ with a
SM-like $R_{K^{(*)}}$? (3) How well can all of the anomalies be explained
together? (4) Can neutrino masses be explained in the regions relevant for the
flavor anomalies? Questions (1)--(3) are addressed below in that order, while
(4) is addressed in Sec.~\ref{sec:nm}. Throughout this discussion, the relative
ease with which this leptoquark model can explain the tension in $(g-2)_\mu$ is
exploited to simplify our study. We do not include its calculation in our
numerical scans, since the values of $x_{23}$ and $y_{23}$ required---namely,
those satisfying Eq.~\eqref{eq:bnamueq}---are such that no constraints are
encountered.

\subsection{Flavor anomalies}

\begin{figure}[t]
  \centering
  \begin{minipage}[t]{0.99\linewidth}
    \centering \includegraphics[scale=0.15]{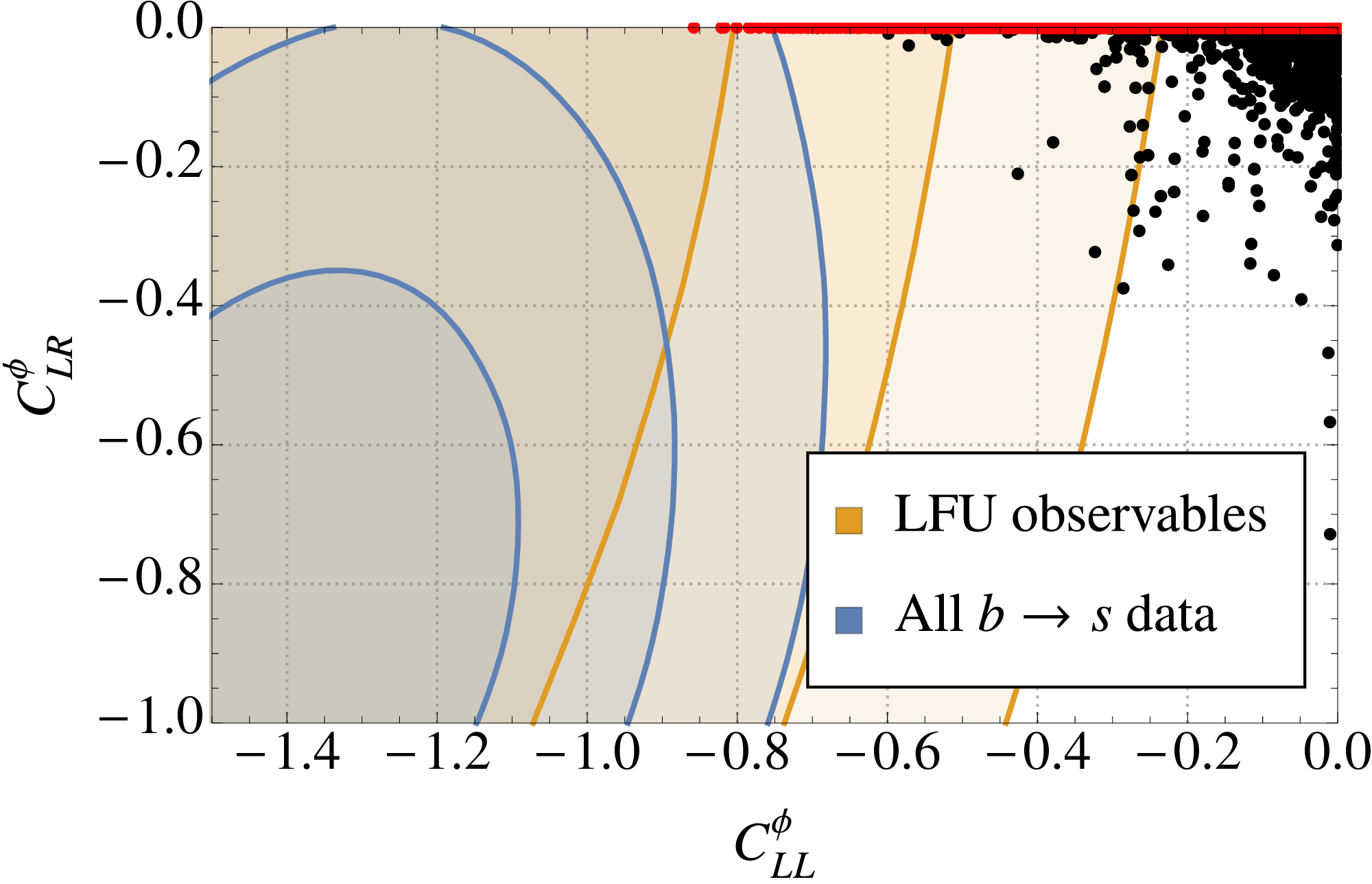}
    \subcaption{The results of scan I (black) and scan II (red) projected onto
      the $C_{LL}^\phi$--$C_{LR}^\phi$ plane. The colored contours correspond to
      those in Fig.~\ref{fig:cllclr}: the orange represent the fit to only LFU
      observables while the blue take into account all $b\to s$ observables. The
      model can alleviate the tensions in LFU observables to just within the
      $1\sigma$ region, a significant improvement on the SM. In this region,
      $C_{LR}^\phi \approx 0$, implying a suppression of the $y_{2i}$. Agreement
      with all of the $b \to s$ data is not as good.\\}
    \label{fig:cllclrpts1}
  \end{minipage} 
  \begin{minipage}[t]{.45\linewidth}
    \centering \includegraphics[scale=0.085]{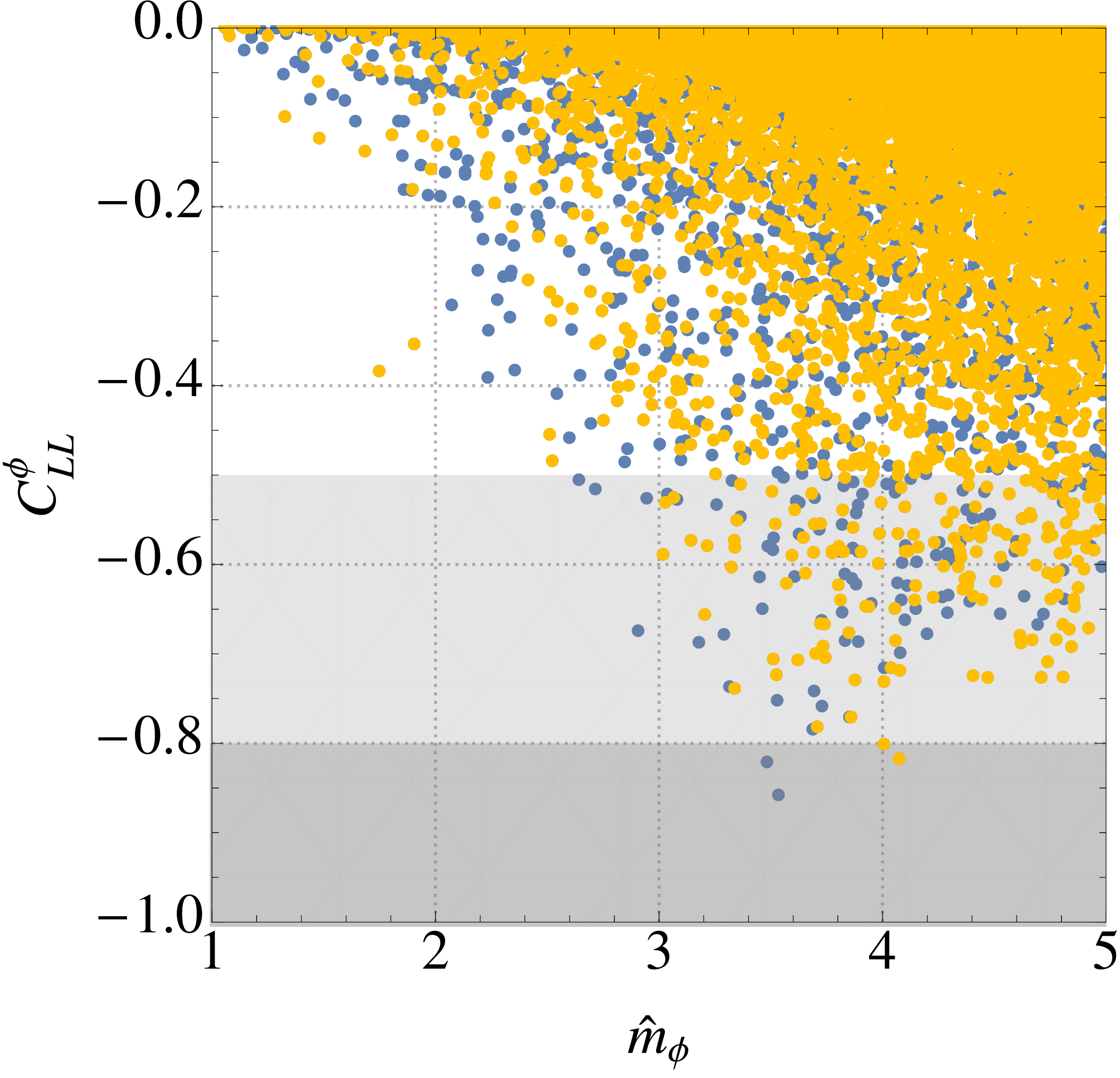}
    \subcaption{A scatter plot showing the results of scan II projected onto the
      $C_{LL}^\phi$--$\hat{m}_\phi$ plane. Yellow points imply SM-like values
      for $R_D$ and $R_{D^{*}}$. The constraints imposed by $D^0 \to \mu \mu$,
      $D^+ \to \pi^+ \mu \mu$ and $Z \to \mu \bar{\mu}$ disfavor light
      leptoquark masses.}
    \label{fig:cllm}
  \end{minipage}
  \hfill
  \begin{minipage}[t]{.45\linewidth}
    \centering \includegraphics[scale=0.085]{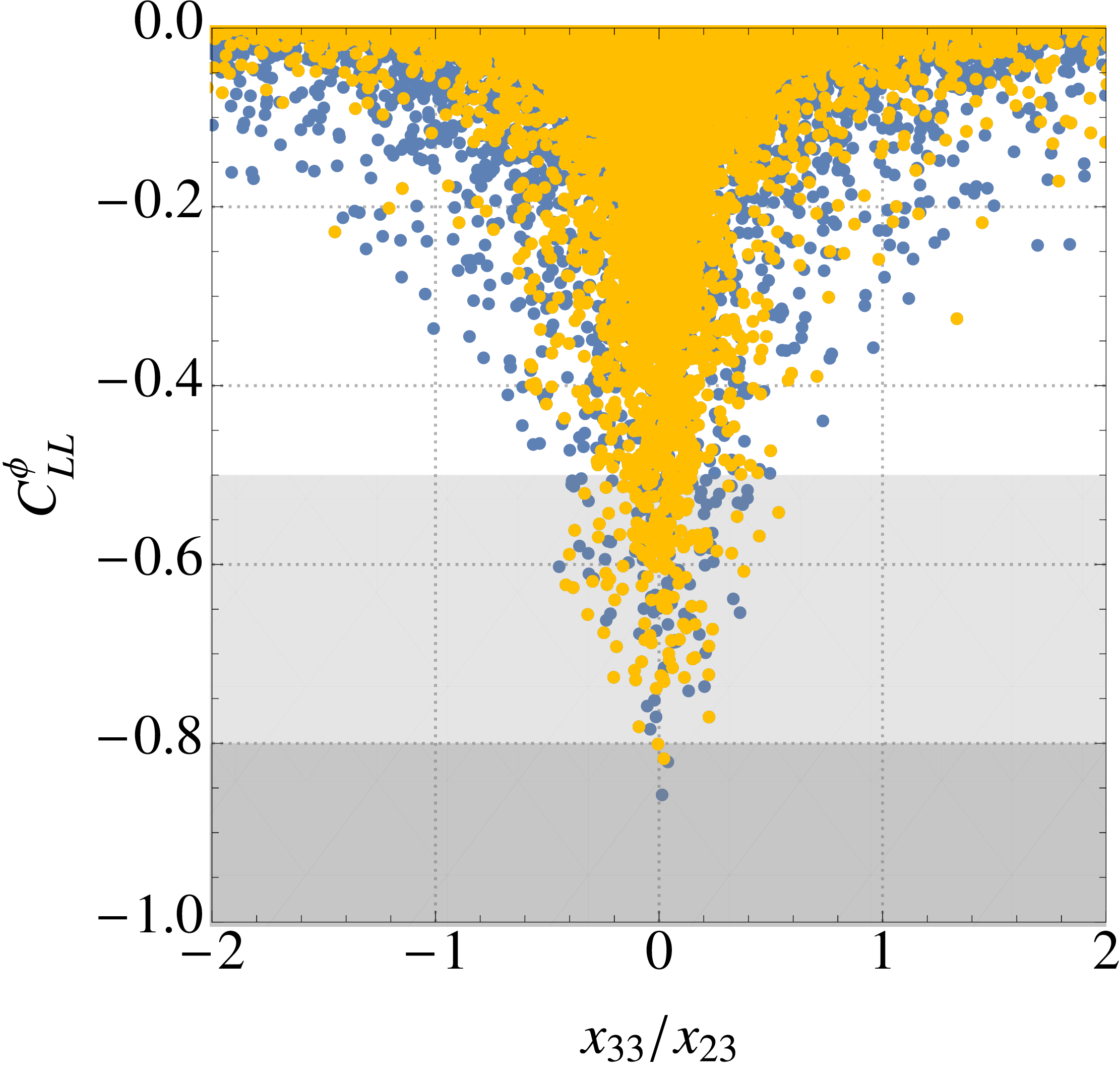}
    \subcaption{A scatter plot of $C_{LL}^\phi$ against the ratio
      $x_{33}/x_{23}$ for parameters subject to scan II. Again, yellow points
      correspond to SM-like $R_D$ and $R_{D^{*}}$. A large, negative value for
      $C_{LL}^\phi$ requires $|x_{23}| > |x_{33}|$ to keep LFV $\tau \to \mu$
      observables at bay.}
    \label{fig:x33x32rat}
  \end{minipage}
  \caption{The key results probing the extent to which the model can explain the
    tensions in $R_{K^{(*)}}$. Significant improvement from the SM is possible
    for leptoquark masses between $3$ and $5 \text{ TeV}$, $|x_{23}| > |x_{33}|$
    and suppressed $y_{2i}$. The grey areas in (b) and (c) are the $1$ and
    $2\sigma$ allowed regions for $R_{K^{(*)}}$.}
  \label{fig:cllclrpts}
\end{figure}

\begin{figure}[!ht] 
    \includegraphics[width=0.40\textwidth]{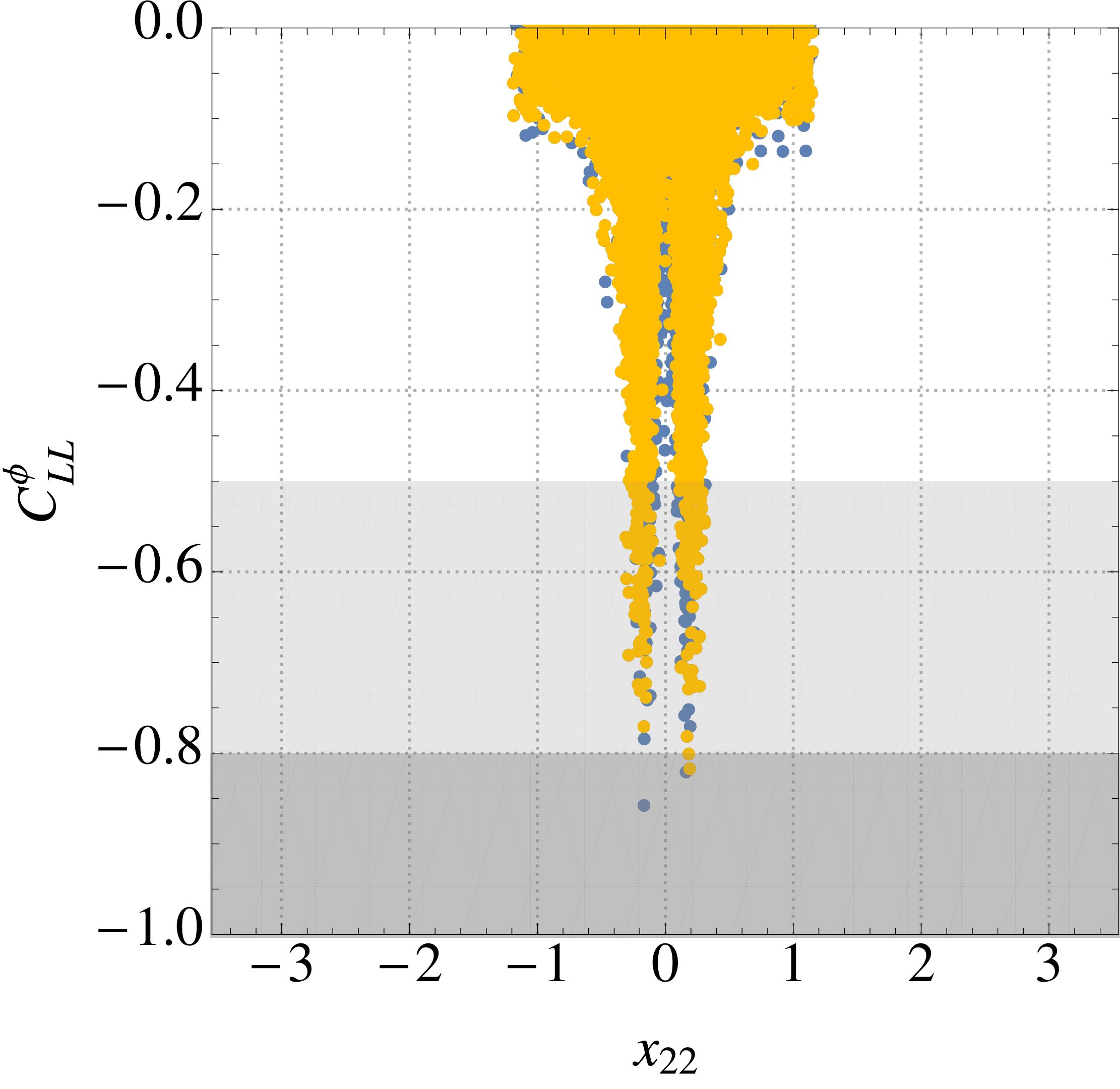} \hfill
    \includegraphics[width=0.40\textwidth]{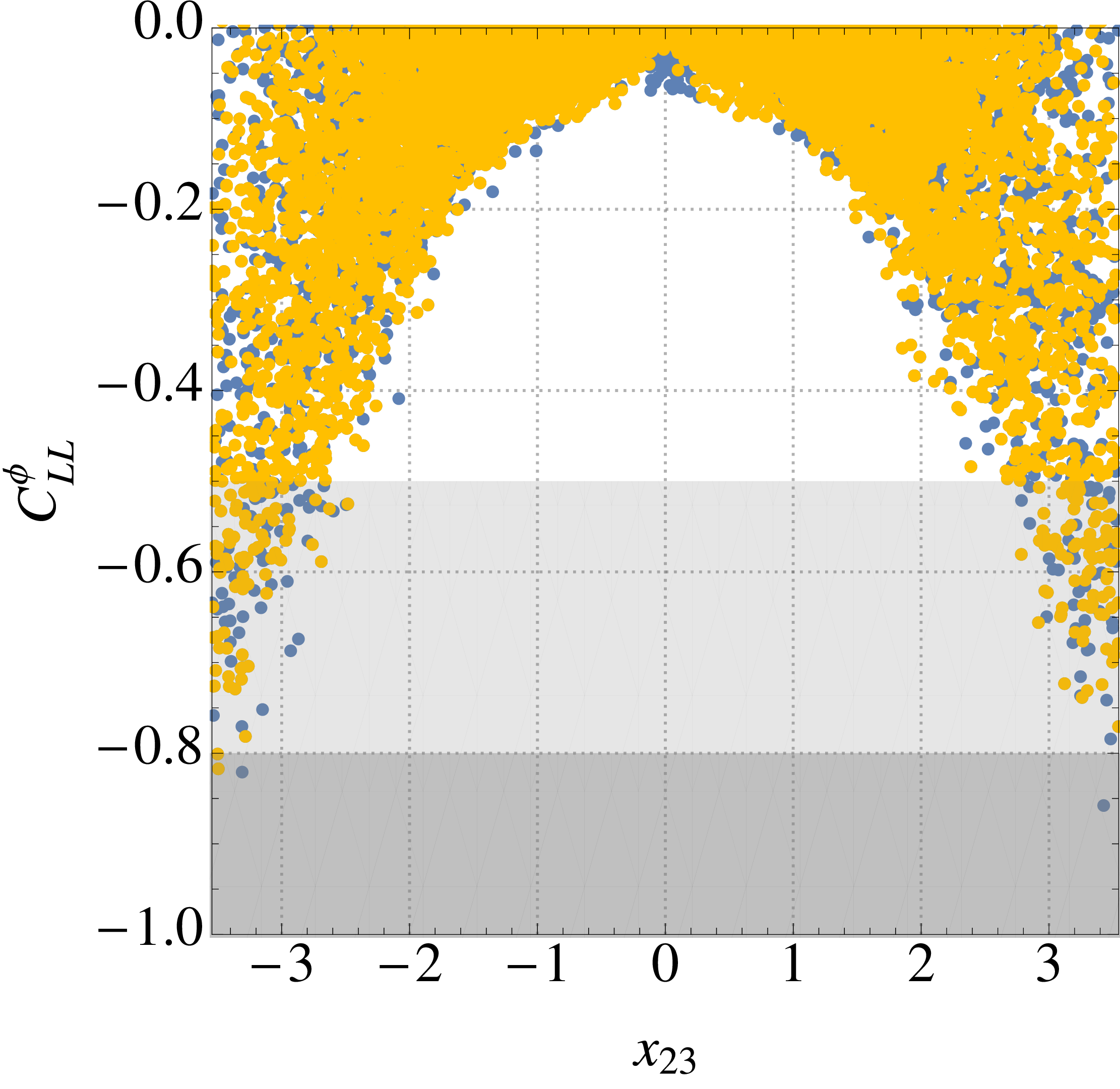}\\ 
    \includegraphics[width=0.40\textwidth]{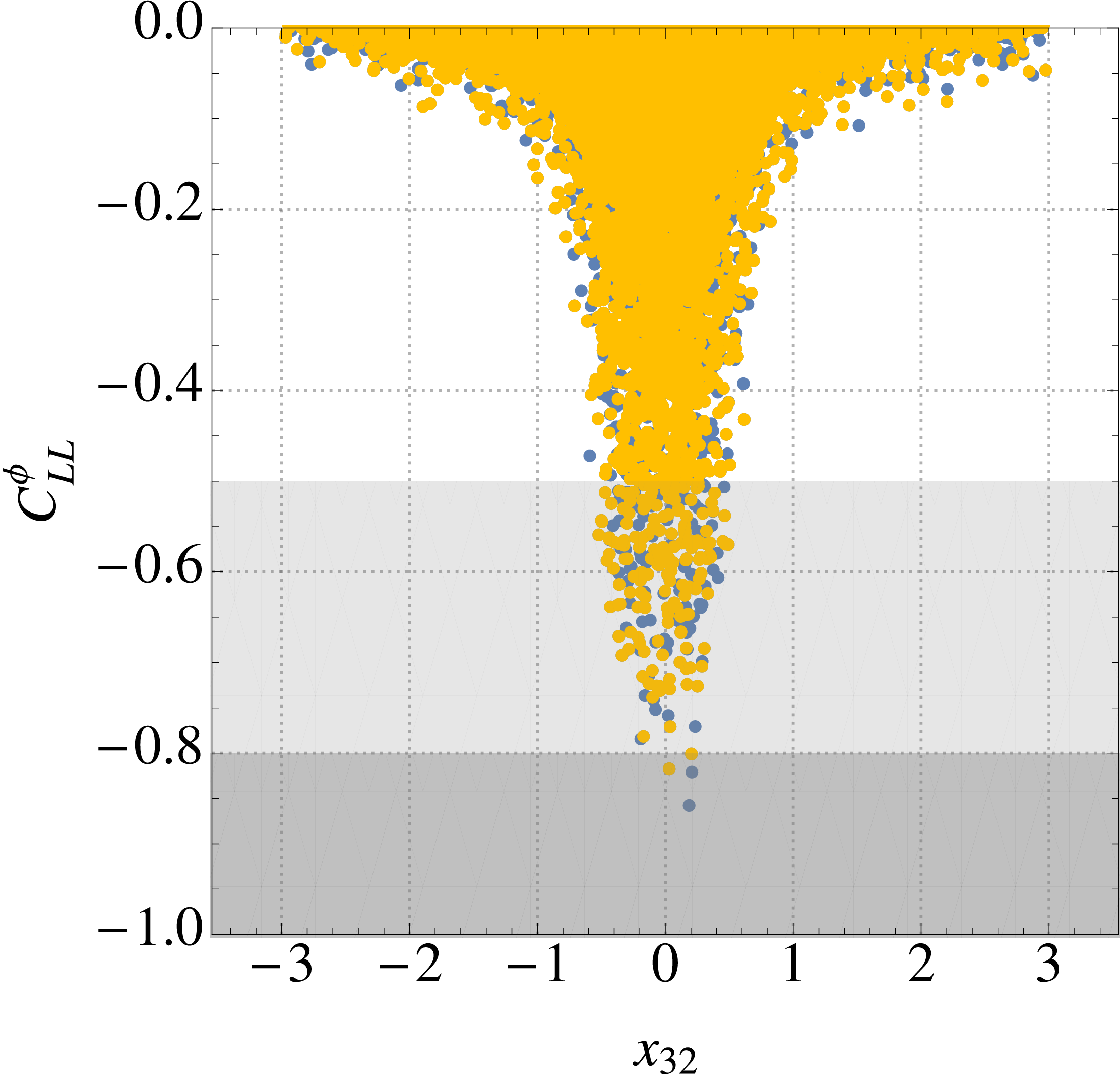} \hfill
    \includegraphics[width=0.40\textwidth]{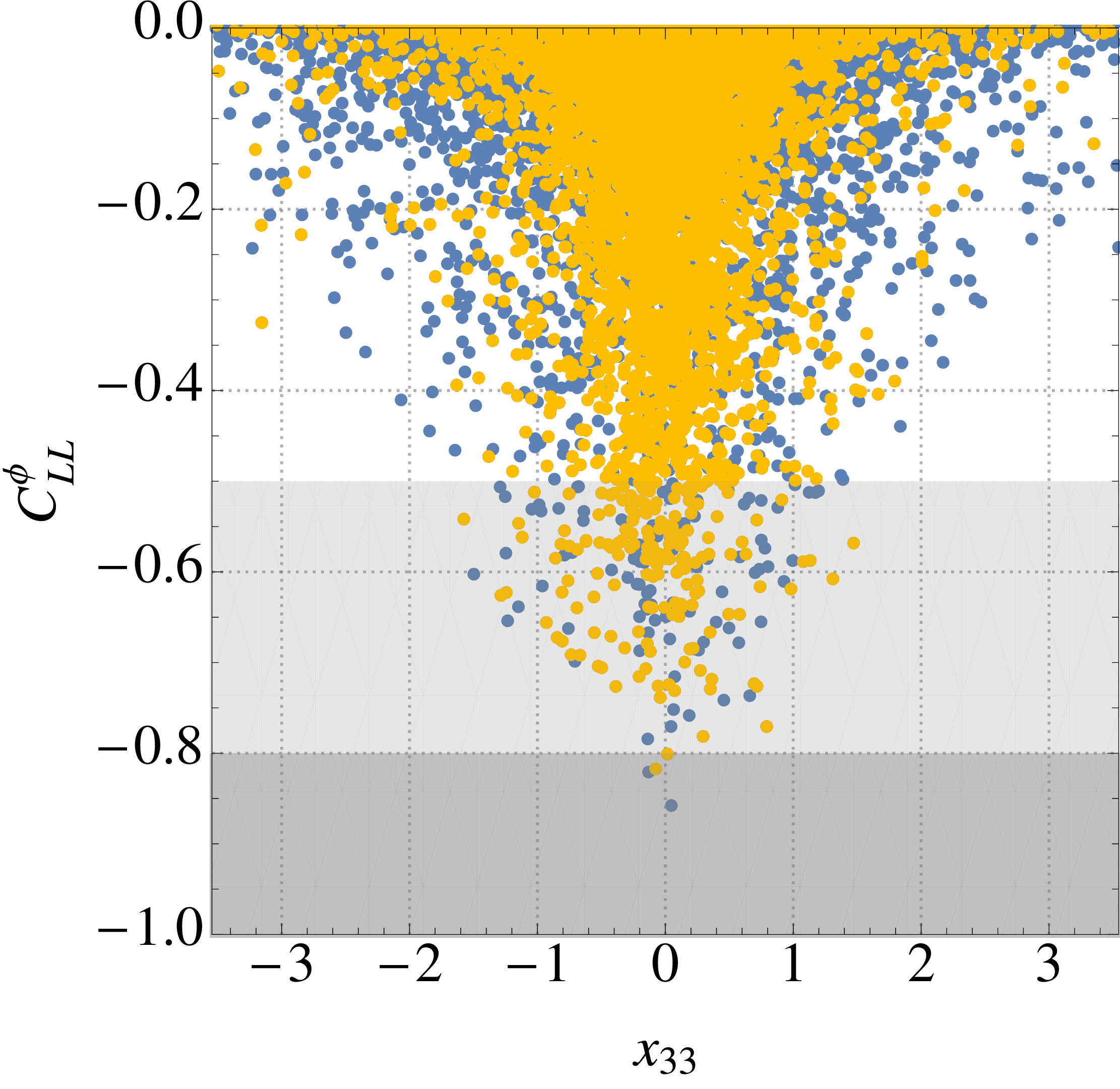} \\
    \includegraphics[width=0.40\textwidth]{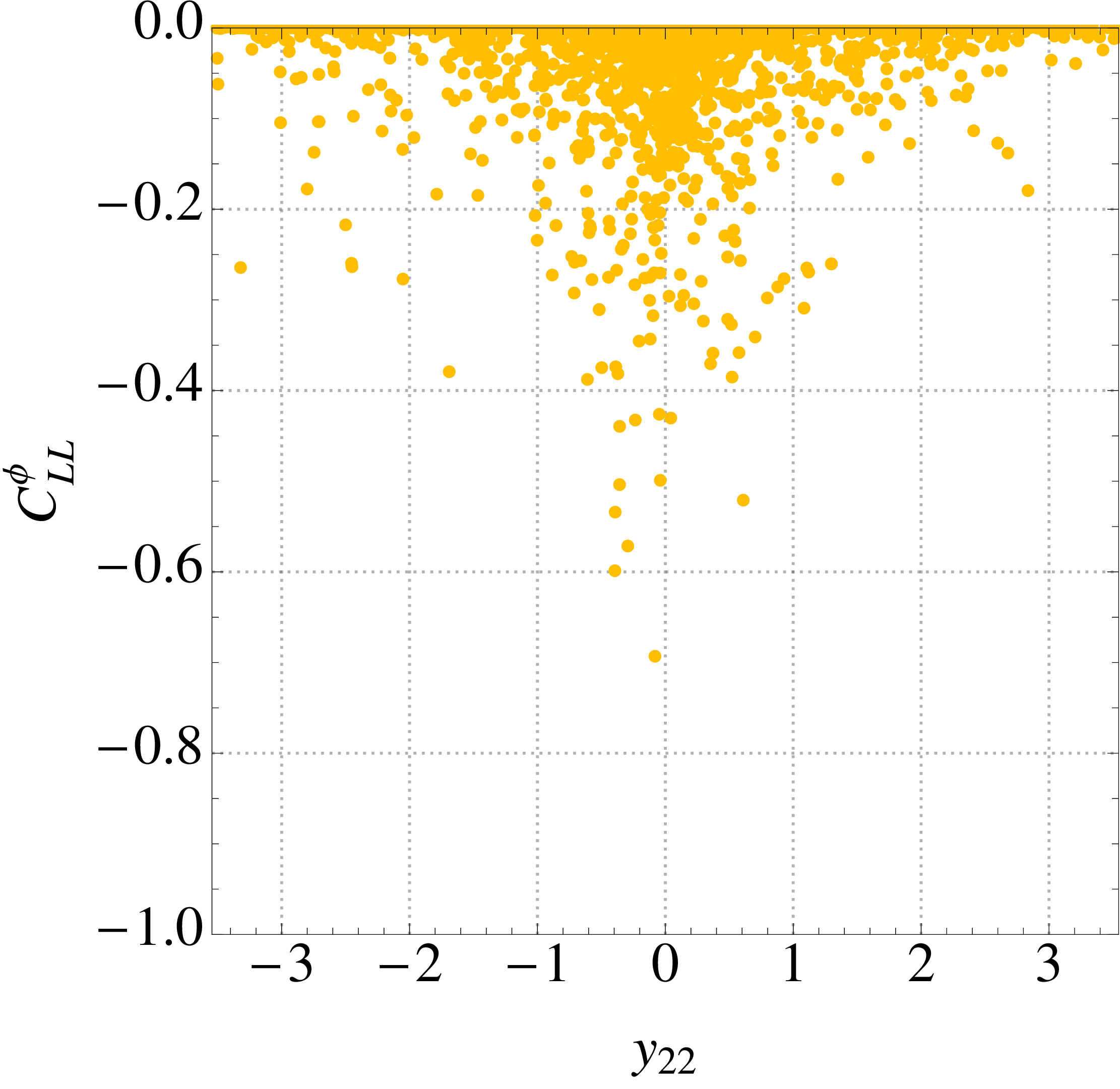} \hfill
    \includegraphics[width=0.40\textwidth]{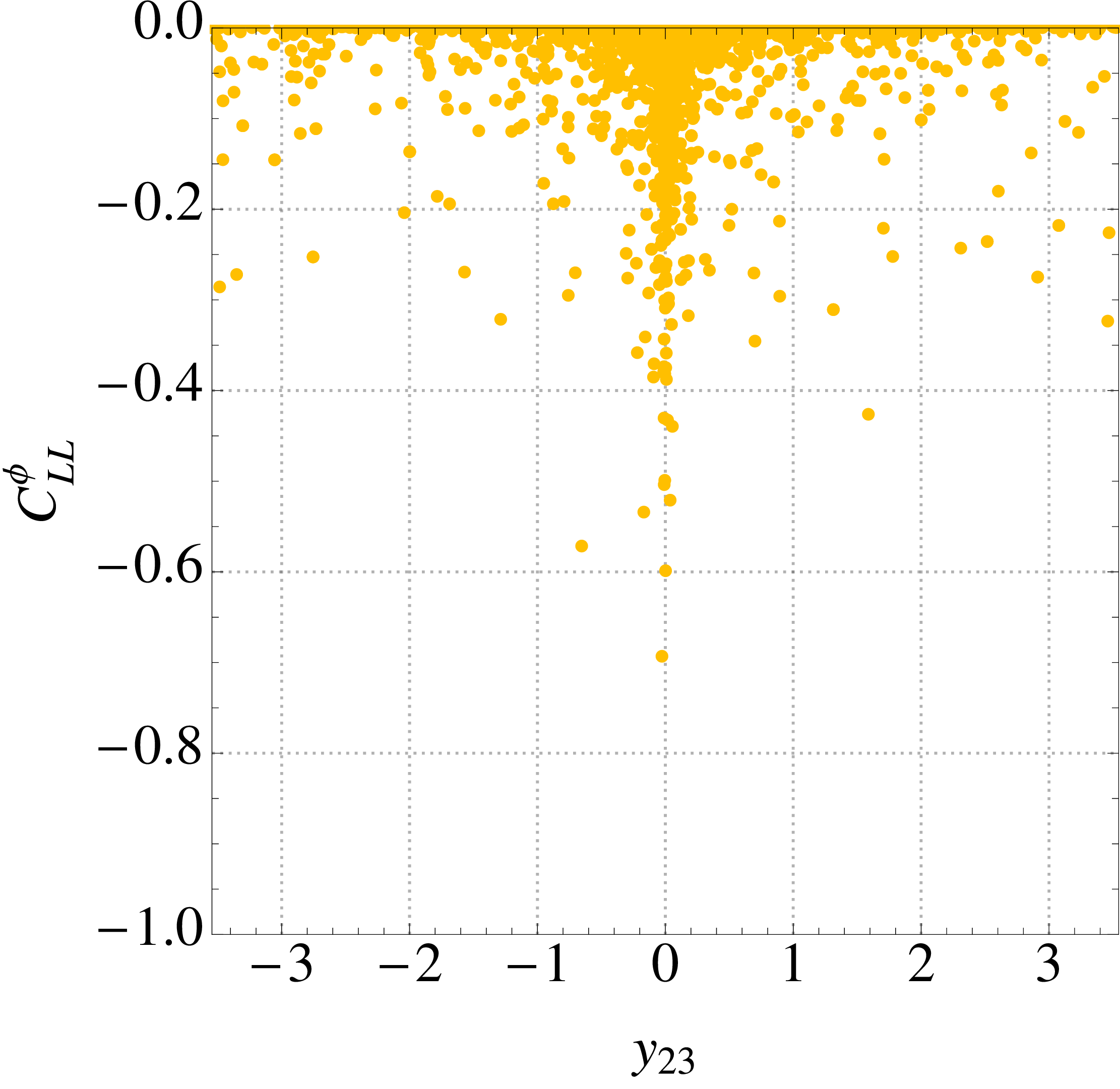} 
    \caption{Slices through the parameter space investigated through scans I and
      II. The value of $C_{LL}^\phi$ is plotted against each Yukawa coupling
      scanned over. Plots agains the $x_{ij}$ contain points from scan II and
      hence $1$ and $2\sigma$ regions for $R_{K^{(*)}}$ can be specified since
      $C_{LR} = 0$, these are shaded grey, and points implying SM-like
      $R_{D^{(*)}}$ values are shown in yellow. Plots against the $y_{2i}$ are
      from scan I, for which all points predict SM-like $R_{D^{(*)}}$ since
      $y_{32} = 0$. Large values of $x_{23}$ are essential for an adequate
      explanation of the $b \to s$ data in this model, while small, but
      non-zero, values for $x_{22}$ are necessary to allow $C_{LL}^\phi$ to be
      negative. The values of $x_{23}$ required to explain the LFU observables
      to $2\sigma$ begin to impinge on the perturbativity constraint $|x_{23}| <
      \sqrt{4\pi}$.}
  \label{fig:rkscans} 
\end{figure}

\paragraph{Explaining $R_{K^{(*)}}$.} In order for the leptoquark model to
explain the measured tensions in the $b \to s$ transition the left-handed
couplings of $\phi$ to the second and third generation of quarks are necessary
to ensure a non-vanishing $C_{LL}^{\phi}$, a parameter space very heavily
constrained by the limits from rare meson decays discussed in
Sec.~\ref{sec:raremesondecays}. The necessary condition Eq.~\eqref{eq:bnunucond}
imposed by the bound on $R_K^{\nu\nu}$ can be combined with Eq.~\eqref{eq:cll}
to give~\cite{Bauer:2015knc}
\begin{equation} \label{eq:rknncond}
  \begin{split}
    \sum_{i=1}^2 |z_{2i}|^2 + \left( 1 - \frac{0.77}{\hat{m}_\phi^2}
    \right) |z_{23}|^2 &\approx \left(|V_{us}|^2 + 1\right) |z_{22}|^2 +
    \left( 1 - \frac{0.8}{\hat{m}_\phi^2} \right) |z_{23}|^2\\ &\gtrsim - 6
    C_{LL}^\phi.
  \end{split}
\end{equation}
It follows that $\mathscr{O}(1)$ couplings to the muon are necessary for the
model to meet the benchmark $C_{LL}^\phi \approx -1.2$. For small leptoquark
masses the model prefers a large $|z_{22}|$ since the top contribution is
suppressed through destructive interference between the box diagrams in
Fig.~\ref{fig:boxes}, however the limit from $D^0 \to \mu \mu$ [see
Eq.~\eqref{eq:zmc}] prohibits such a scenario. Indeed, the analysis of
Ref.~\cite{Becirevic:2016oho} indicates that the constraint following from the
LFU evident in $R_D^{\mu/e}$ also constitutes a very serious stumbling-block for
the model's explanation of the $b \to s$ data for $m_\phi \lesssim 1 \text{
  TeV}$. We make progress by performing a random scan in which the leptoquark
mass is allowed to vary up to $5 \text{ TeV}$---such large masses have the
effect of lifting the suppression on the last term in Eq.~\eqref{eq:rknncond}
and permitting larger values for $z_{22}$ according to Eq.~\eqref{eq:zmc}. In
addition to the $x_{ij}$, we turn on the $y_{2i}$ with $i \neq 1$ in order to
study the extent to which $C_{LR}^\phi$ can contribute. These define the
parameters of scan I, introduced above, and we present the results of this scan
along with those of scan II, for which $C_{LR} = 0$, in Fig.~\ref{fig:cllclrpts}
and Fig.~\ref{fig:rkscans}. We highlight those points for which the
$R_{D^{(*)}}$ observables remain SM-like, that is, within twice the theoretical
error associated with the SM predictions we cite in
Table~\ref{tbl:rdrdstartable}. Consistent with our comments in
Section~\ref{sec:semileptonicchargedcurrentprocesses}, we find that any
phenomenologically viable explanation of the anomalous $b \to s$ data in this
leptoquark model requires $m_\phi \gtrsim 2.5 \text{ TeV}$. Additionally,
constraints from the $\tau \to \mu$ flavor-changing observables require
$|x_{32}| > |x_{33}|$ for large $|x_{32}|$. Although the benchmark value
$C_{LL}^\phi \approx -1.2$ is unattainable in light of the constraints we have
considered for a perturbative $x_{23}$, the model can reduce the tension in
$R_{K^{(*)}}$ to within $1 \sigma$, a significant improvement on the SM. Points
in parameter space implying such large, negative values for $C_{LL}^\phi$ also
entail a vanishing $C_{LR}$, although even in this region agreement with all of
the $b \to s$ data can be slightly better than $3\sigma$.

\begin{figure}[t]
  \centering
  \begin{minipage}[t]{0.45\linewidth}
    \centering \includegraphics[scale=0.7]{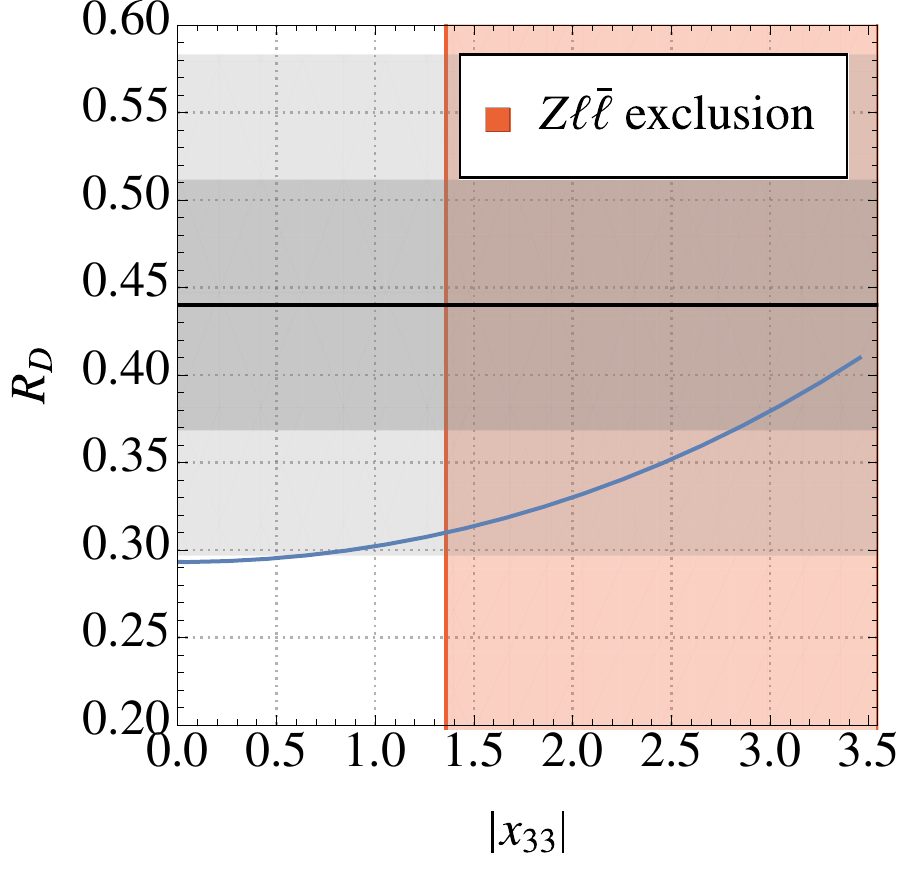}
  \end{minipage}
  \hfill
  \begin{minipage}[t]{0.45\linewidth}
    \centering \includegraphics[scale=0.7]{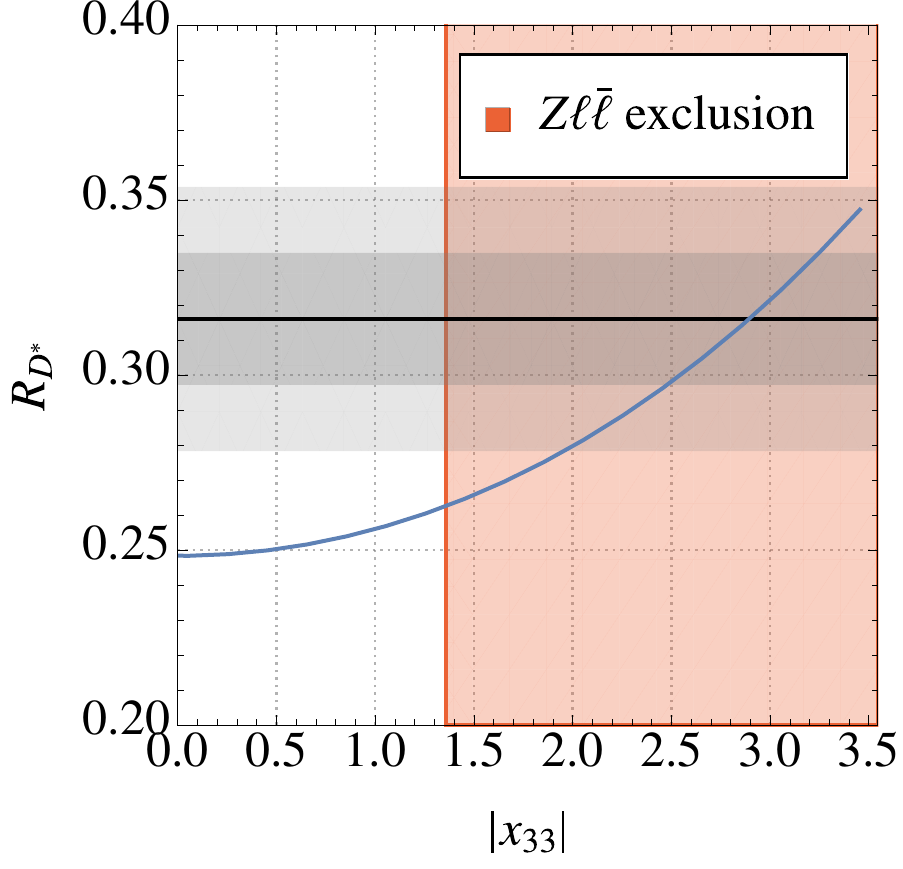}
  \end{minipage}
  \caption{The solid blue lines represents the dependence of $R_D$ (left) and
    $R_{D^{(*)}}$ (right) on $|x_{33}|$ when all other couplings are set to zero
    and $m_\phi = 1 \text{ TeV}$. A non-zero $x_{33}$ generates a small $z_{32}$
    through the quark mixing of Eq.~\eqref{eq:mixing}, although the $|x_{33}|$
    values required to meet the anomalies become large enough to dangerously
    modify the $Z \to \tau\tau$ rate. The values of $|x_{33}|$ excluded by
    measurements of the $Z\tau\tau$ coupling are shaded red. The solid
    black line represents the central values of the measurements for $R_D$ and
    $R_{D^{(*)}}$, and the grey areas are the $1$ and $2\sigma$ regions.}
  \label{fig:rDRes}
\end{figure}

\paragraph{Explaining $R_{D^{(*)}}$.} We move on to consider the extent to which
the leptoquark can explain the anomalies in the $b\to c$ transition. The fit
presented in Fig.~\ref{fig:rdrdstarregions} suggests two scenarios for
explaining the measured tensions in the $b \to c \ell \nu$ decays in the region
$A$: (i) new physics only in $C_V^{33}$, or (ii) new-physics in $C_V^{33}$ along
with contributions from $C_S^{33}$ and $C_T^{33}$. Possibility (i) is consistent
with the best-fit value, and this is the region of parameter space considered in
the model's original form. However, we emphasize that the conditions presented
in Eq.~\eqref{eq:bnunucond} and Eq.~\eqref{eq:eqZll} are sufficient to preclude
that effects in $C_V^{33}$ alone could be responsible for the enhancement of the
$R_{D^{(*)}}$ ratios. The product $x_{32}^* x_{33}$ is heavily constrained from
$R_K^{\nu\nu}$ and $B_s$--$\bar{B}_s$ mixing, as indicated in
Fig.~\ref{fig:rknunuallowed}. One could consider generating $z_{23}$, and
therefore $C_V^{33}$, through quark mixing, thus making do only with a non-zero
$x_{33}$ and avoiding these constraints. This set-up, however, requires
excessively large values of $x_{33}$ to explain $R_{D^{(*)}}$, causing the
leptoquark's contributions to the $Z\tau\bar{\tau}$ coupling to exceed current
experimental limits---a result we illustrate in Fig.~\ref{fig:rDRes}.
In addition, we find the effects of lepton-flavor violation to be subdued, since
such contributions add incoherently to the $W$-mediated SM processes, and thus
entail couplings large enough to conflict with measurements of
$B_s$--$\bar{B}_s$ mixing and precision electroweak observables. Scenario (ii)
involves new physics in $C_S^{ij}$ and $C_T^{ij}$. The most minimal approach
here is to turn on only the bottom--tau-neutrino interaction $x_{33}$ and the
right-handed tau--charm coupling $y_{32}$. A non-zero $x_{33}$ will generate
$C_V^{33}$ through quark-mixing. We find the coupling $y_{32}$ to be weakly
constrained by the precision electroweak measurements discussed earlier: in the
limit $|y_{32}| \gg |y_{3 (1,3)}|$, the bound
\begin{equation}
  |y_{32}| < \frac{3.69 \hat{m}_\phi}{\sqrt{1 + 0.39 \ln \hat{m}_\phi}},
\end{equation}
follows from Eq.~\eqref{eq:eqZll}. In addition, small values of the muon--top
coupling $y_{23}$ will allow sizeable contributions to $(g-2)_\mu$ in the
presence of $x_{23} \neq 0$ because of the top-mass enhancement in the
mixed-chirality term of Eq.~\eqref{eq:amu}. This minimal texture involving only
third-generation couplings to left-handed quarks comes with the additional
benefit that the leptoquark can evade the constraints from measurements of
$R_K^{\nu\nu}$ and $B_s$--$\bar{B}_s$ mixing. In fact, the only serious
constraint is that arising from the modification of the $Z\tau\bar{\tau}$
coupling from a large $x_{33}$, a situation that can be remedied for $y_{32}
\sim \mathscr{O}(1)$, allowing a good fit to the $R_{D^{(*)}}$ data for slightly
smaller values of $x_{33}$. A sizeable $y_{32}$ is thus a necessary requirement
for this leptoquark model to explain the experimental anomalies in $R_D$ and
$R_{D^*}$. For example, the parameter choices $m_\phi = 1 \text{ TeV}$, $x_{33}
= 1.3$ and $y_{32} = 0.3$ are sufficient to explain $R_{D^{(*)}}$ to within
$1\sigma$, and this choice of couplings passes all the constraints we impose.
Note also that the measured tension in $(g-2)_\mu$ can be accommodated at the
same time since the couplings involved are unimportant for $b \to c \tau \nu$.
Saturating both $x_{33}$ and $y_{32}$ at the perturbativity bound $\sqrt{4\pi}$,
we find that an explanation of $R_{D^{(*)}}$ loses viability at $\sim 10 \text{
  TeV}$.

\begin{figure}[t]
  \centering \includegraphics[scale=0.1]{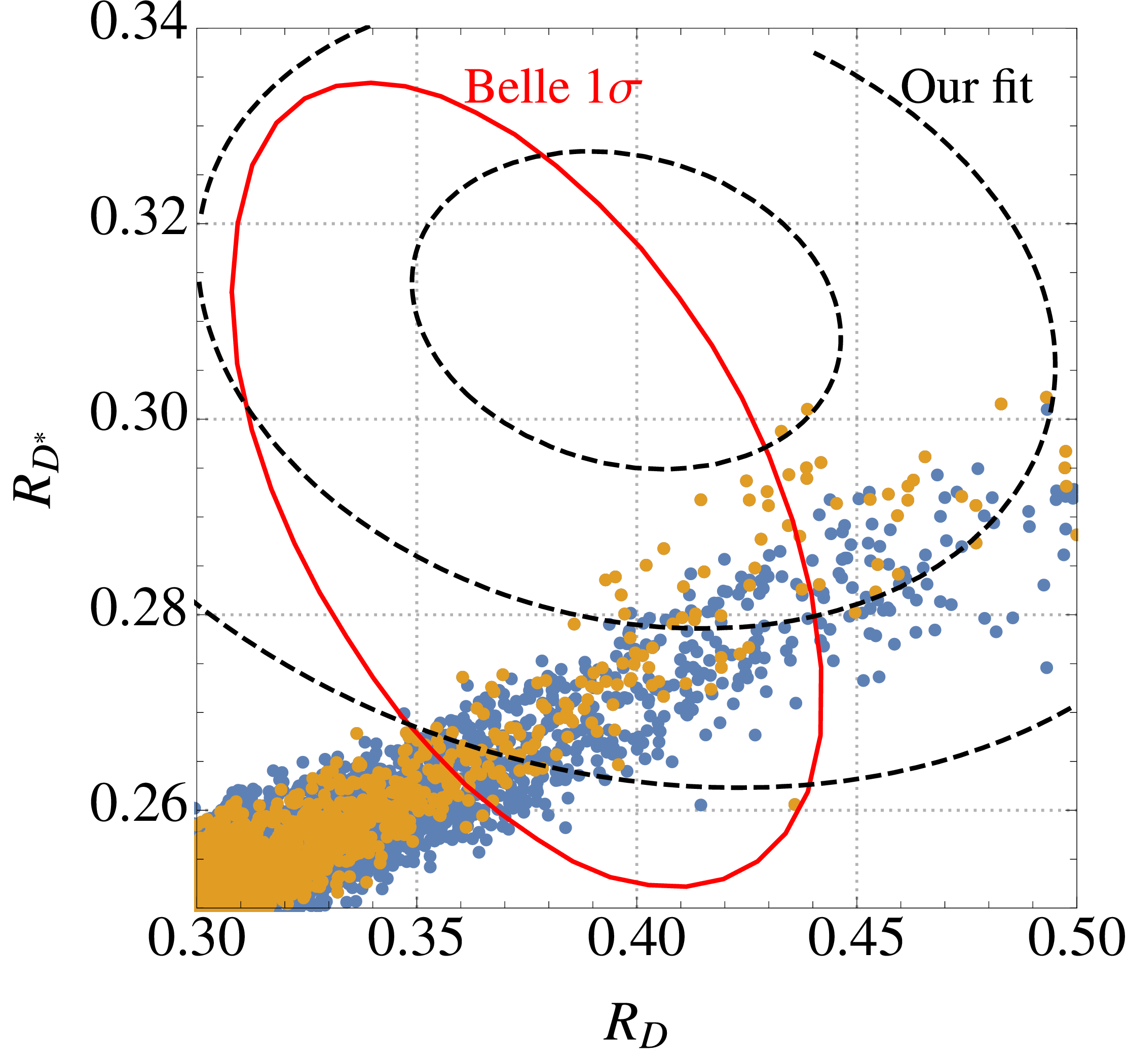} 
  \caption{The results of scan II presented as a scatter plot of $R_D$ against
    $R_{D^*}$. The orange points keep $C_{LL}$ SM-like, while blue points show a
    $> 1\%$ deviation in $C_{LL}$ from the SM prediction. The dashed black
    ellipses represent the $1$, $2$ and $3\sigma$ contours from our fit with an
    assumed correlation $\rho = -0.2$, while the solid red curve indicates the
    $1\sigma$ allowed region implied by the Belle measurement from
    Ref.~\cite{Huschle:2015rga}. The anomalies in $b \to c \tau \nu$ can be
    accommodated in this model, although even small, non-zero values for the
    $x_{i2}$ cause tension with limits from $B \to K \nu \nu$ and measurements
    of $B_s$--$\bar{B}_s$ mixing, causing few points to stray into the $1\sigma$
    region of our fit.}
  \label{fig:money1}
\end{figure}

\begin{figure}[t] 
    \includegraphics[width=0.4\textwidth]{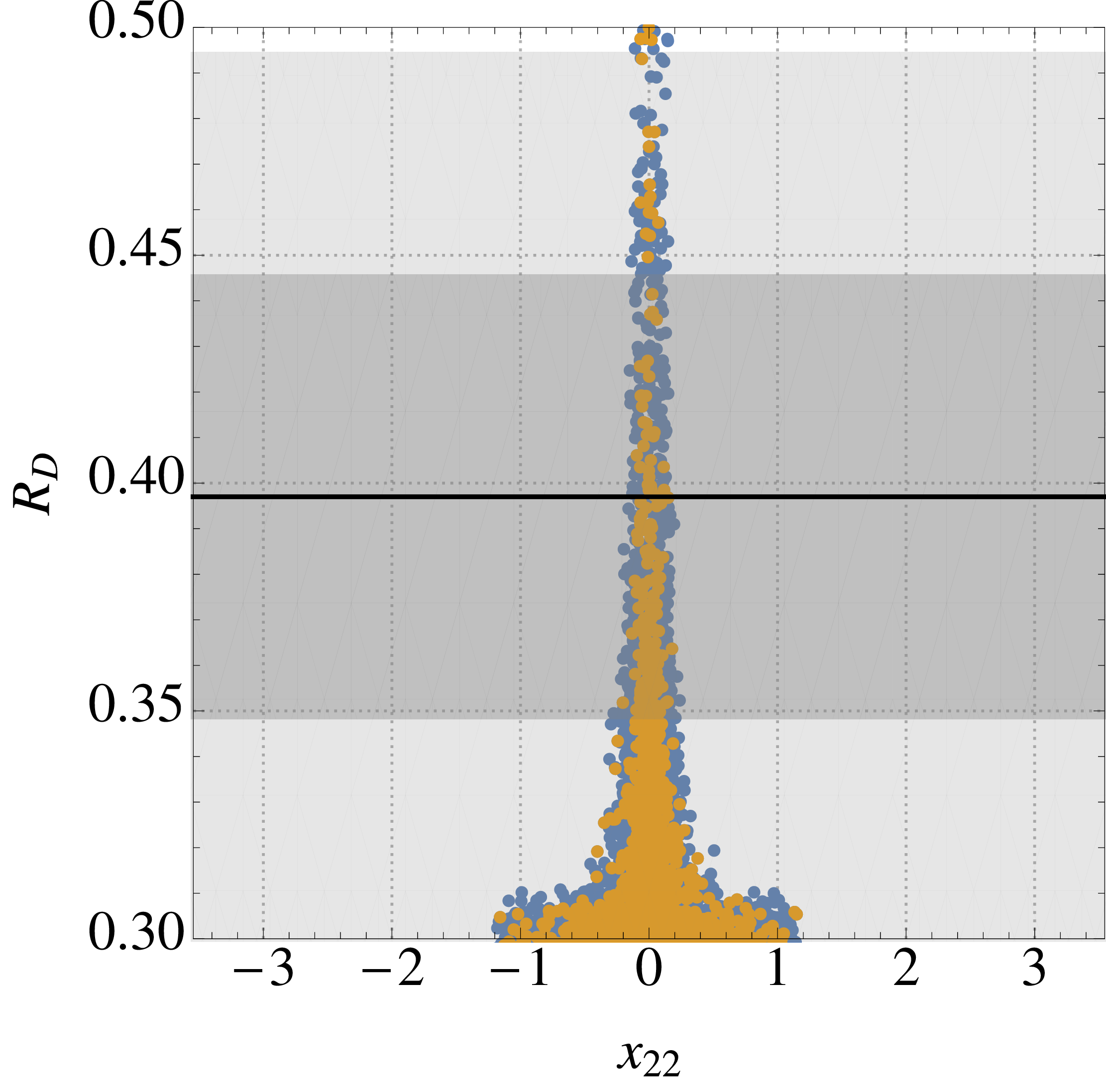} \hfill
    \includegraphics[width=0.4\textwidth]{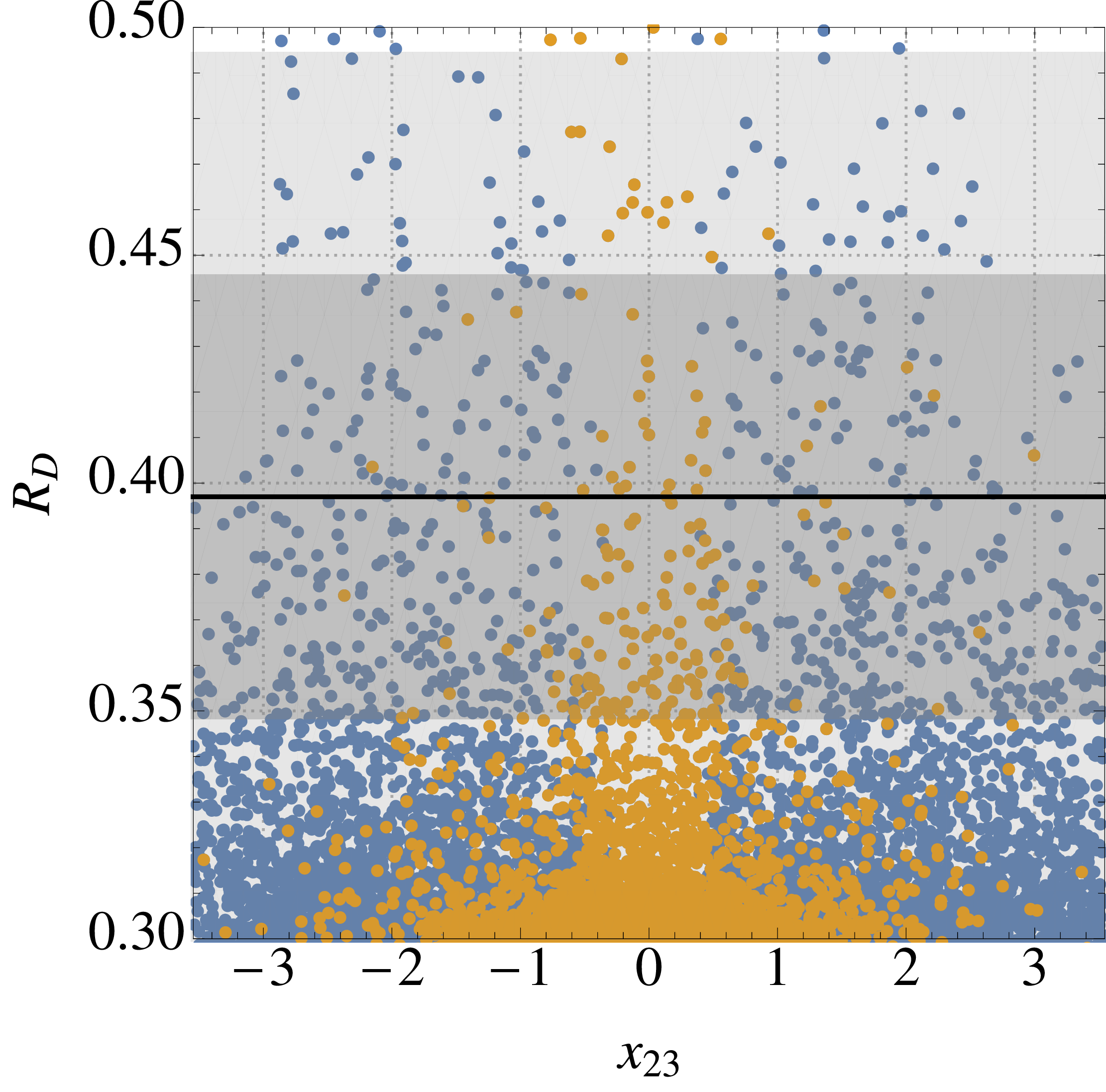} \\
    \includegraphics[width=0.4\textwidth]{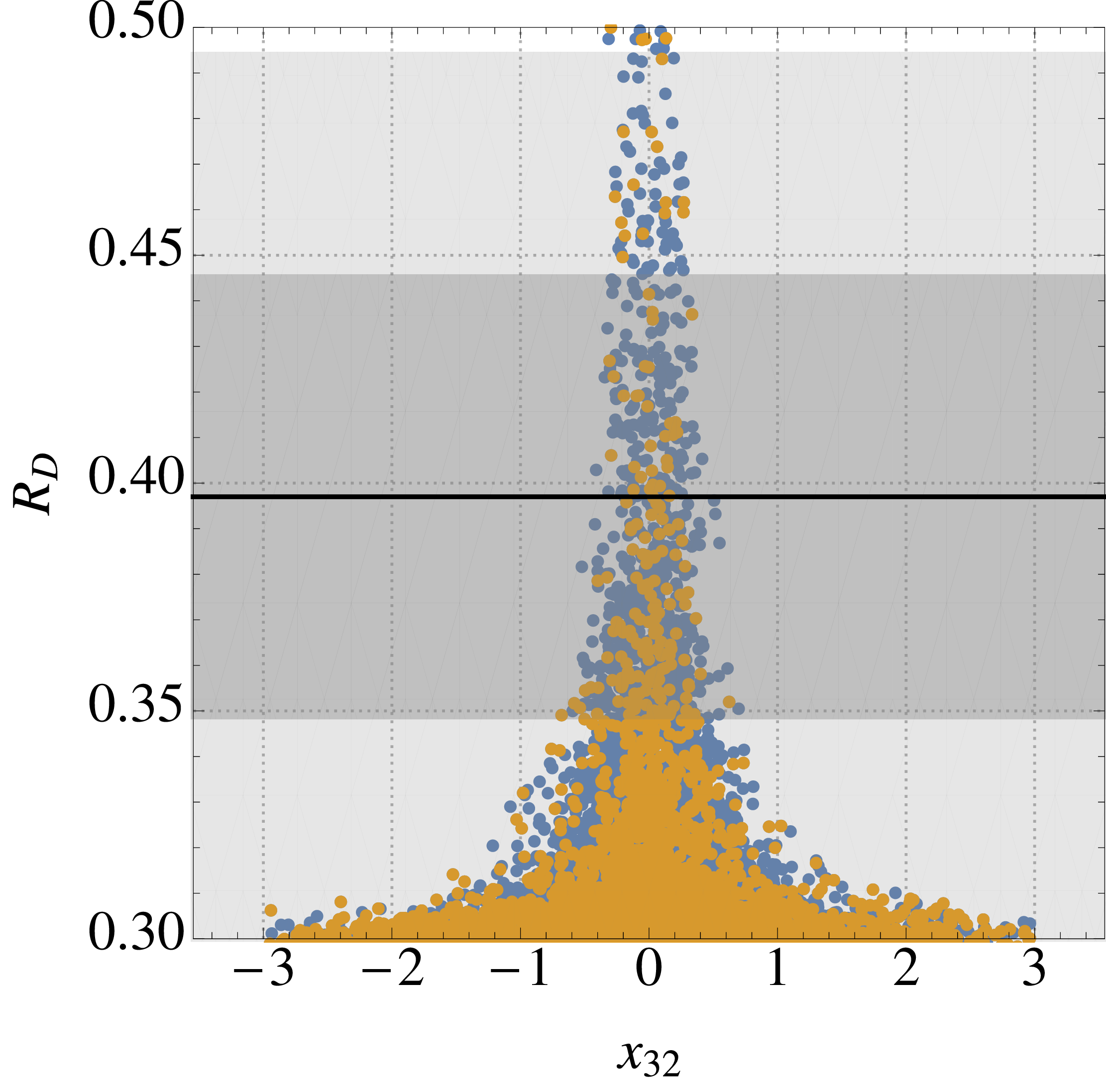} \hfill
    \includegraphics[width=0.4\textwidth]{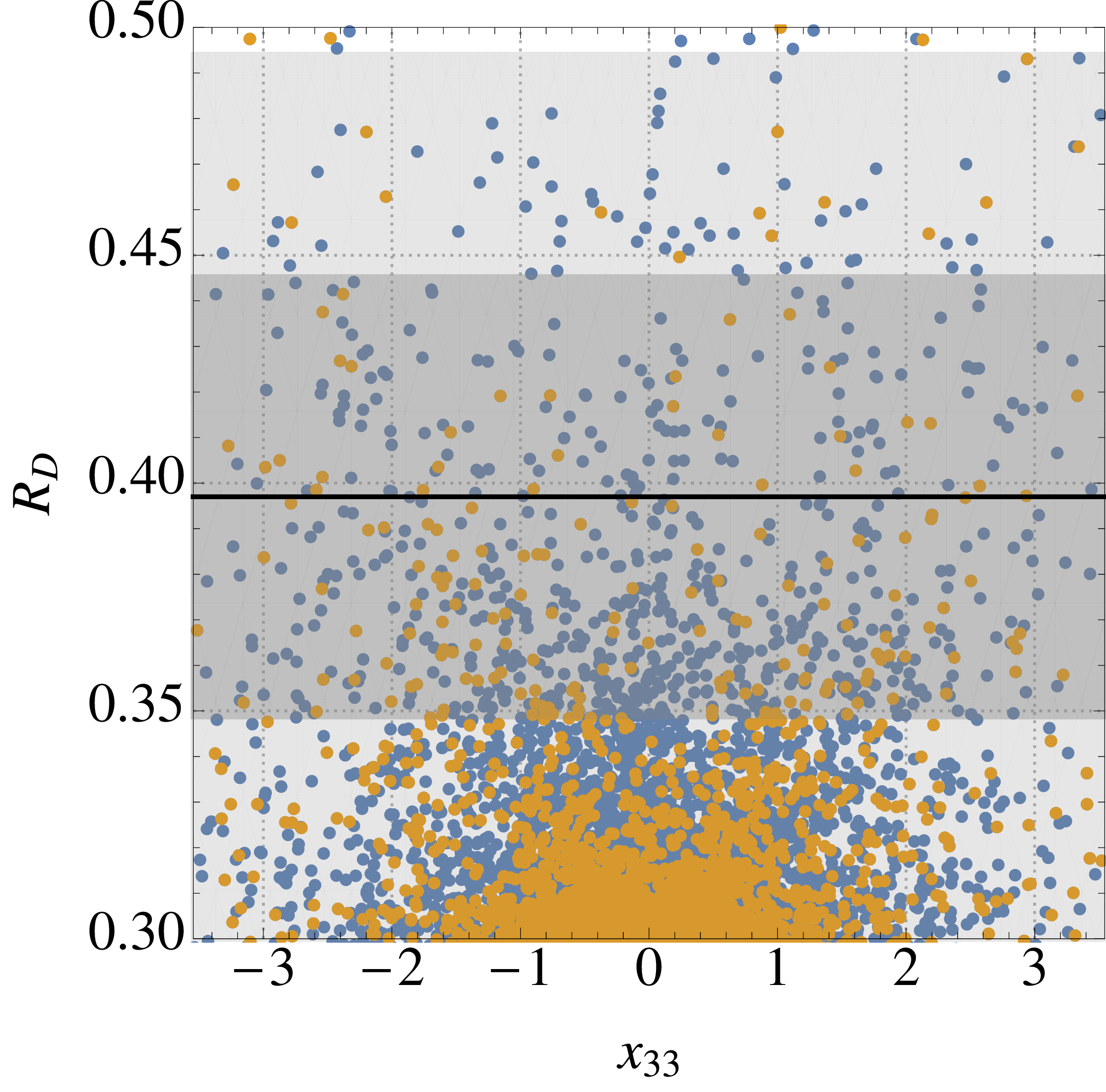} \\
    \includegraphics[width=0.4\textwidth]{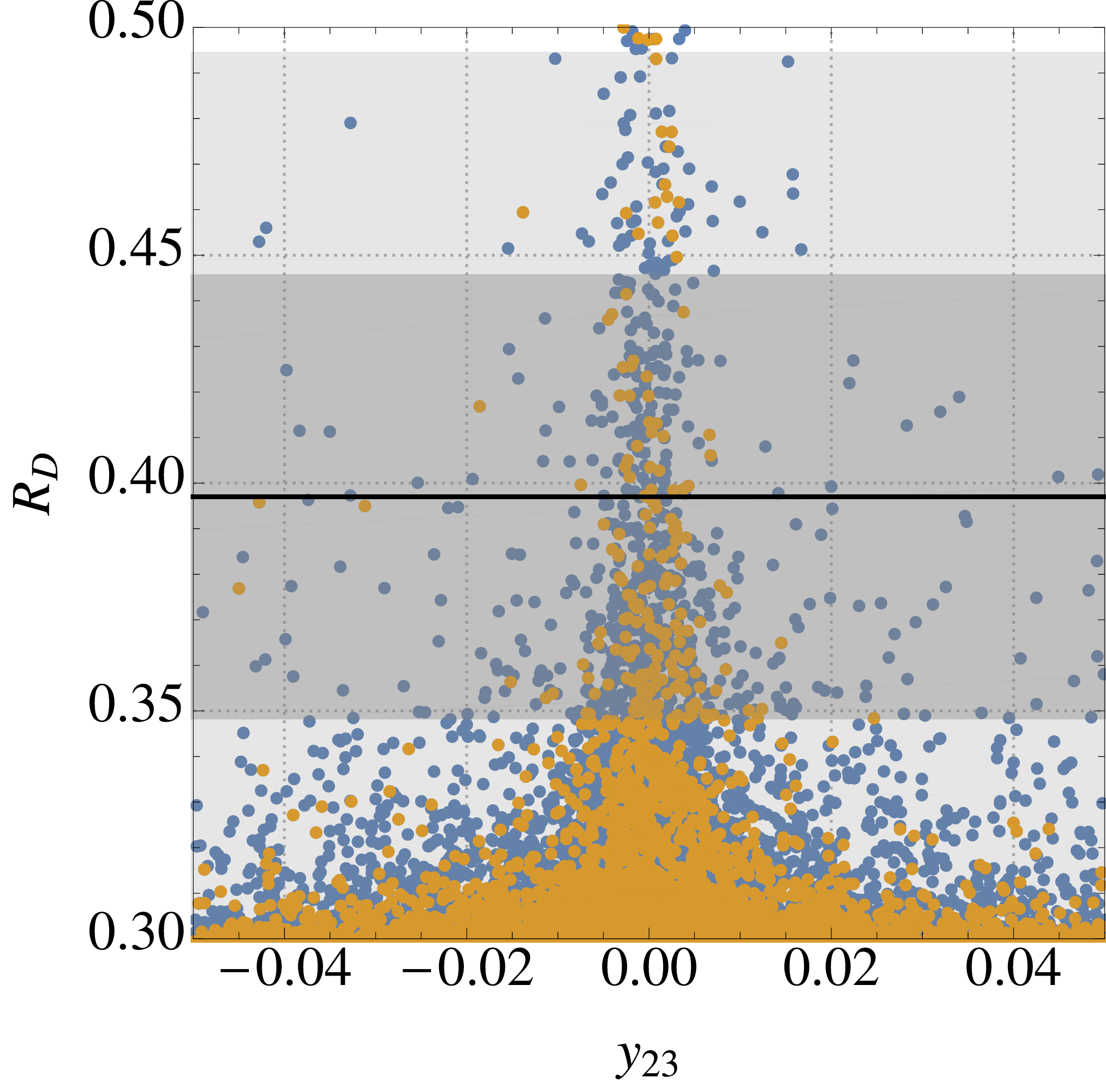} \hfill
    \includegraphics[width=0.4\textwidth]{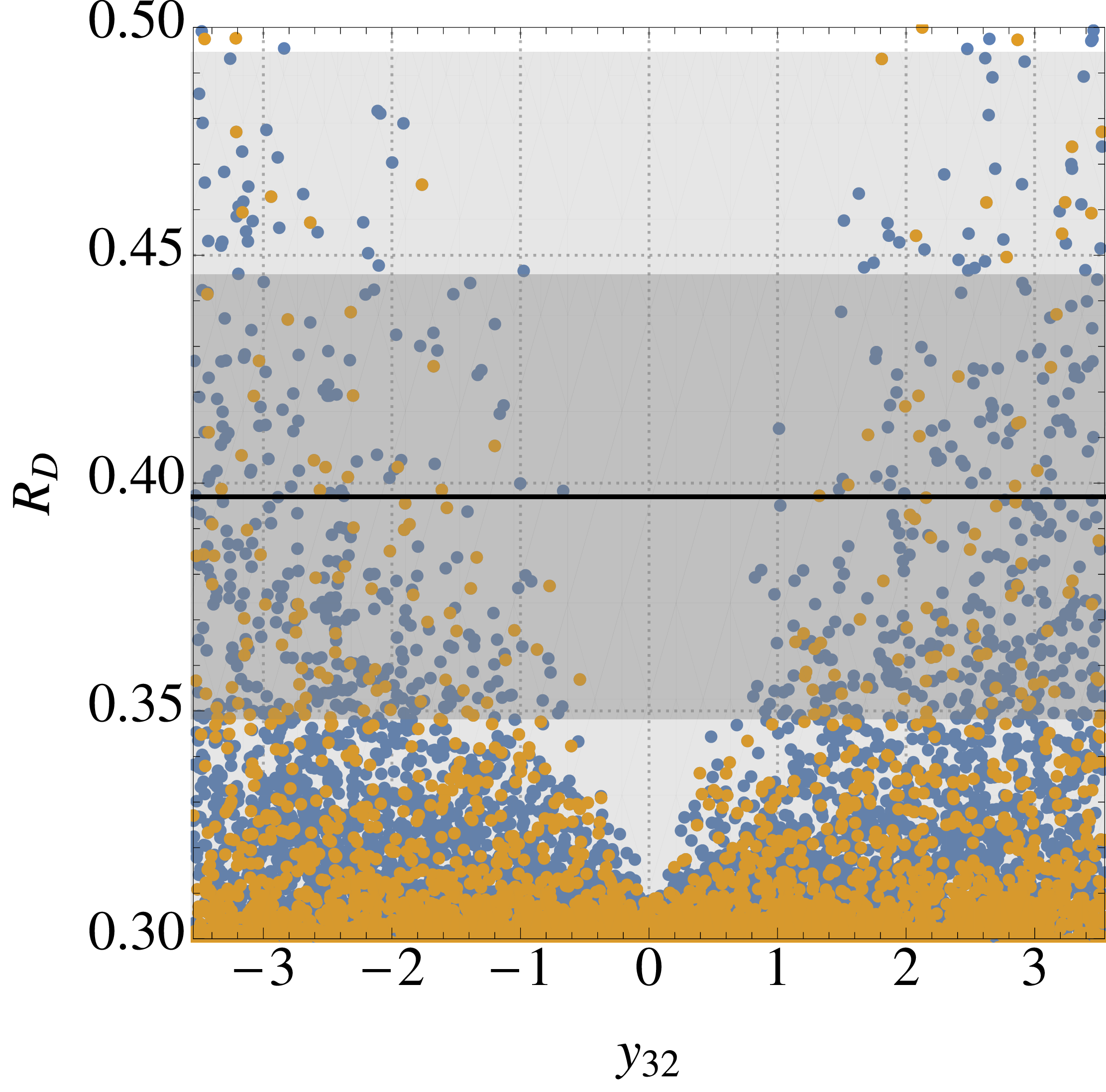} 
    \caption{Slices through the parameter space of scan II. The solid black line
      represents the central value of the $R_D$ measurement, and the grey bands
      correspond to the $1$ and $2\sigma$ regions. The orange points keep
      $C_{LL}$ SM-like, while blue points show $> 1\%$ deviation in $C_{LL}$
      from the SM prediction. Large values $x_{33}$ and $y_{32}$ are necessary
      for an adequate explanation of $R_D$ since these feature in $C_S$ and
      $C_T$. Other left-handed couplings must be small to evade constraints from
      $R_K^{\nu\nu}$ and $B_s$--$\bar{B}_s$ mixing. The results for $R_{D^*}$
      are qualitatively the same.}
  \label{fig:ObsScans} 
\end{figure}

In Fig.~\ref{fig:money1} we present the results of scan II in the
$R_D$--$R_{D^*}$ plane, while Fig.~\ref{fig:ObsScans} displays the values of the
Yukawa couplings from the same scan that lead to interesting $R_D$ values.
Limits on the $B \to K \nu \nu$ rate and measurements of $B_s$--$\bar{B}_s$
mixing constrain the $x_{i2}$ to be small, while large values for $x_{33}$ and
$y_{32}$ are necessary since their product appears in the expressions for $C_{ V
}^{33}$, $C_{ S }^{33}$ and $C_{ T }^{33}$. As discussed above, these large
$x_{33}$ values imply dangerous contributions to $Z \to \tau \tau$, causing few
points to stray into the $1\sigma$ region. The model can, however, significantly
reduce the tension in the $b\to c \tau \nu$ measurements in a large region of
parameter space. Agreement with the Belle result from
Ref.~\cite{Huschle:2015rga} is better than the combined fit, since this model
predicts slightly smaller values of $R_{D^*}$ than those suggested by the BaBar
and LHC\textit{b} measurements.


\paragraph{Explaining both $R_{K^{(*)}}$ and $R_{D^{(*)}}$.} In order to
establish the full power of the model to explain both $R_{D^{(*)}}$ and
$R_{K^{(*)}}$, we perform a complete scan over the 7-dimensional parameter space
spanned by the leptoquark mass and the couplings $x_{ij}$ for $i,j \neq 1$,
$y_{23}$ and $y_{32}$---the parameters of scan II. Results from this scan have
been presented above in the context of explaining one or the other anomaly
separately, although in this case the blue points of Figs.~\ref{fig:cllm},
\ref{fig:x33x32rat}, \ref{fig:rkscans}, \ref{fig:money1}, \ref{fig:ObsScans} and
red points of Fig.~\ref{fig:cllclrpts} are relevant. In addition to these, we
present the results of scan II in $C_{LL}$--$R_D$--$R_{D^*}$ space, where color
is used as the third axis, in Fig.~\ref{fig:ObsScans}. This plot demonstrates a
mild tension between $R_{K^{(*)}}$ and $R_{D^{(*)}}$ in this leptoquark model:
points lying within the $1\sigma$ region for $R_{K^{(*)}}$ keep $R_{ D^* }$
SM-like, while those breaching the $1\sigma$ boundary for $R_D$ and $R_{D^*}$
imply $C^\phi_{LL} \approx 0$. This can be attributed to the behavior evident in
Fig.~\ref{fig:x33x32rat}: large, negative values of $C_{LL}^\phi$ require
$x_{33} \approx 0$, but $x_{33}$ is essential to this model's explanation of
$R_{D^{(*)}}$, since it features in $C^{33}_{V,S,T}$. At best, we find that the
model can explain all of the discrepant measurements to within $2\sigma$, a
striking level of consistency with all constraints and anomalies. In both cases
the $(g-2)_\mu$ anomaly can also be accommodated.

\begin{figure}[t]
    \centering \includegraphics[scale=0.14]{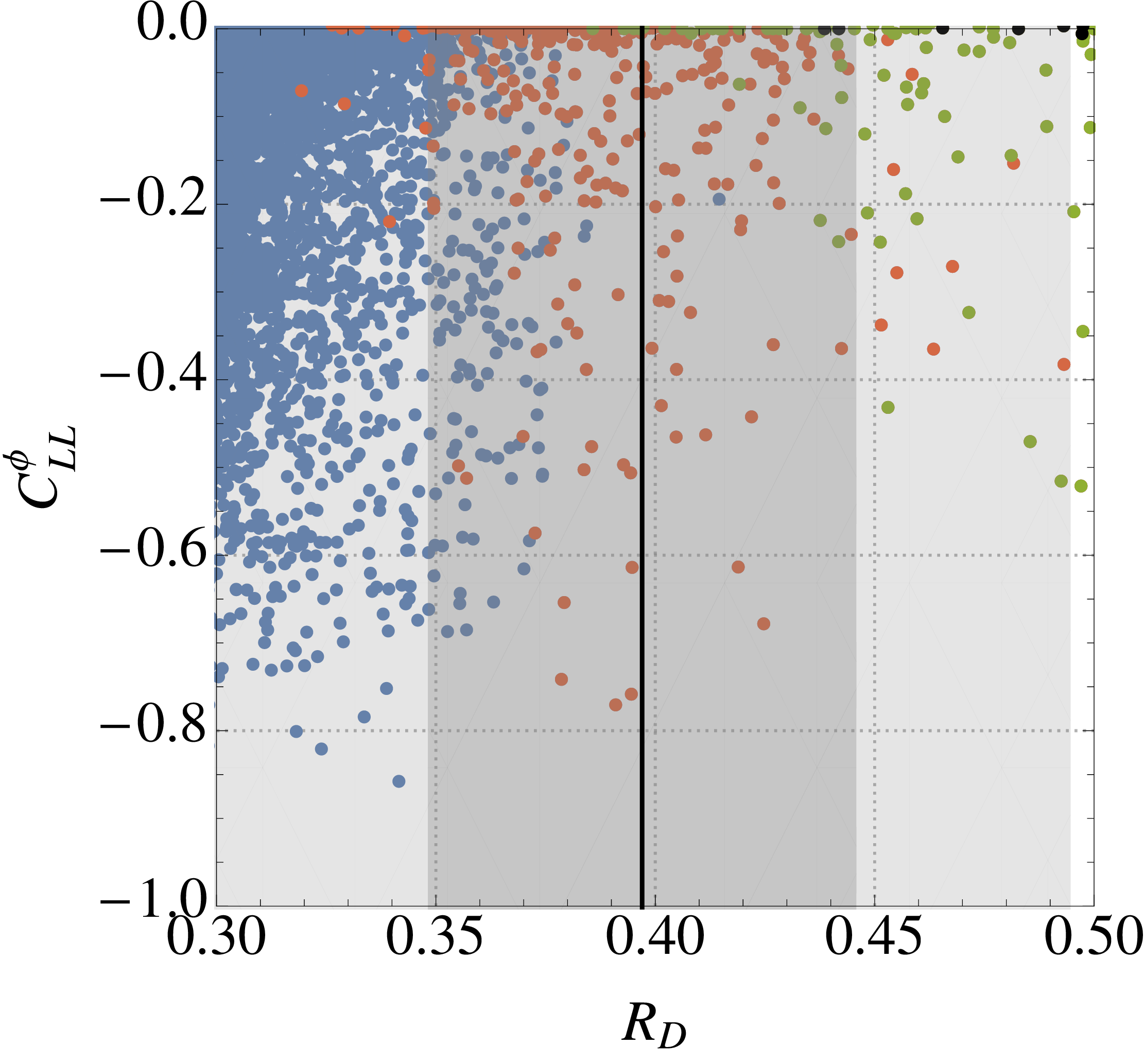}
    \caption{A projection of the random points subject to scan II onto the
      $R_D$--$C_{LL}^\phi$ plane, with colors corresponding to $\sigma$-regions
      of the fit to $R_{D^*}$: black, green, red and blue points lie in the $1$,
      $2$, $3$ or $>3\sigma$ region for $R_{D^*}$. The solid black line
      represents the central value of the $R_D$ measurement, and the grey bands
      correspond to the $1$ and $2\sigma$ regions. Parameter choices leading to
      the required large, negative value for $C_{LL}^\phi$ tend to compromise
      agreement with measurements of $R_{D^*}$. A combined explanation of
      $R_{D^{(*)}}$ and $R_{K^{(*)}}$ is only possible at the $2\sigma$ level
      for both anomalies. This represents a significant improvement on the SM.}
  \label{fig:money}
\end{figure}

\subsection{A representative neutrino mass realization}
\label{sec:mv}

\label{sec:nm}
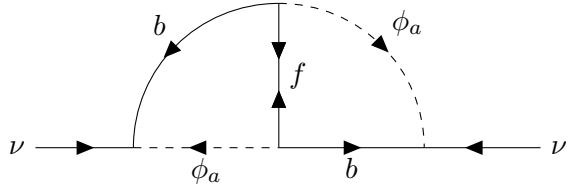
\begin{figure}[t]
  \centering
  \begin{tikzpicture}
    \begin{feynman}
      \vertex (a) {$\nu$};
      \vertex [right=4em of a] (b);
      \vertex [right=5em of b] (c);
      \vertex [right=5em of c] (d);
      \vertex [right=4em of d] (e) {$\nu$};
      \vertex [above=5em of c] (v);
      \diagram* {
        (a) -- [fermion] (b) -- [anti charged scalar, edge label'=$\phi_a$] (c) -- [fermion, edge label'=$b$] (d) -- [anti fermion] (e),
        (b) -- [anti fermion, quarter left, edge label=$b$] (v) -- [charged scalar, quarter left, edge label=$\phi_a$] (d),
        (c) -- [majorana, edge label'=$f$] (v),
      };
    \end{feynman}
  \end{tikzpicture}
  \caption{Two loop neutrino mass generation in the model of
    Ref.~\cite{Angel:2013hla}. For simplicity we consider the case where the
    leptoquark $\phi$ couples significantly only to the third generation of
    quarks. At least two flavors of $\phi$ are required to meet the neutrino
    data.}
  \label{fig:neutrinomass}
\end{figure}
In this section we incorporate the BN leptoquark into the two-loop neutrino mass
model developed and studied in detail in Ref.~\cite{Angel:2013hla}. We summarize
the key features of the model below, and point the reader to the original paper
for more detail.

Following Ref.~\cite{Angel:2013hla} we couple the leptoquark $\phi$ to the
color-octet Majorana fermion $f \sim (\mathbf{8}, \mathbf{1}, 0)$ in order to
introduce the lepton-number violating terms $m_f f f$ and $w_i \bar{d}_i f
\phi$. The dimension-9 $\Delta L = 2$ effective operator $LLQd^cQd^c$ is
generated when the heavy fields $f$ and $\phi$ are integrated out. The neutrino
mass is proportional to the product of down-type quark mass matrices, which is
dominated by the bottom quark mass. We do not consider the case where a strong
hierarchy in the $w_i$ undermines this dominance, and thus only the coupling to
the third generation of quarks ($w_3$) is important for the neutrino mass
generation. For this reason we set $w_{1,2} = 0$ to simplify the calculation of
the neutrino mass. In this limit the neutrino mass matrix will have unit rank
and an additional generation of the leptoquark $\phi$ is needed to satisfy
current oscillation data. Replacing $\phi$ with $\phi_a = (\phi_1, \phi_2)$ in
Eq.~\eqref{eq:Lagra}, small neutrino masses are generated through the two-loop
graph shown in Fig.~\ref{fig:neutrinomass} and the neutrino mass is given by
\begin{equation}
  \label{eq:massformula}
  M_{ij} \approx 4\frac{m_f
    m_b^2}{(2\pi)^8} \sum_{a, b}^2 (x_{i3a} w_{3a}) I_{ab}
  (x_{j3b} w_{3b}),
\end{equation}
where $\mathbf{I}$ is the matrix of loop integrals in the leptoquark-generation
space whose explicit form can be found in Ref.~\cite{Angel:2013hla}. This
expression for the mass matrix can be solved for the $x_{i3a}$ through the
Casas--Ibarra procedure~\cite{Casas:2001sr} to give
\begin{equation}
  \label{eq:ci}
  x_{i3a} = \frac{(2\pi)^4}{2w_{3a}m_b\sqrt{m_f}} U^*_{ij} [\tilde{\mathbf{M}}^{1/2}]_{jk} R_{kb} [\tilde{\mathbf{I}}^{-1/2}\mathbf{S}]_{ba},
\end{equation}
where tildes denote real and positive diagonal matrices and $\mathbf{S}$
diagonalizes the matrix $\mathbf{I}$. We use the best-fit values from the NuFIT
collaboration for the neutrino mixing angles and mass-squared
differences~\cite{Esteban:2016qun, nufitweb}:
\begin{equation}
  \begin{aligned}
    \sin^2 \theta_{12} &= 0.306,   & \Delta m_{21}^2 &= 7.50 \cdot 10^{-5} \text{ eV}^{2},\\
    \sin^2 \theta_{13} &= 0.02166, & \Delta m_{31}^2 &= 2.524 \cdot 10^{-3} \text{ eV}^{2} \text{ (NO)},\\
    \sin^2 \theta_{23} &= 0.441 \text{ (NO)},   & \Delta m_{32}^2 &= -2.514 \cdot 10^{-3} \text{ eV}^{2} \text{ (IO)},\\
    \sin^2 \theta_{23} &= 0.587 \text{ (IO)}. 
  \end{aligned}
\end{equation}
The mass-squared differences fix the elements of $\tilde{\mathbf{M}}$, since the
lightest neutrino in this model is almost massless. In the cases of normal and
inverted neutrino mass hierarchy,
\begin{equation} \label{eq:r}
  \mathbf{R}^{\textsc{NO}} = \begin{pmatrix} 0 & 0\\ \cos\theta & -\sin\theta \\ \sin\theta & \cos\theta\end{pmatrix}, \quad
  \mathbf{R}^{\textsc{IO}} = \begin{pmatrix} \cos\theta & -\sin\theta \\ \sin\theta & \cos\theta \\ 0 & 0 \\\end{pmatrix},
\end{equation}
and $\theta \in \mathbb{C}$ parameterizes the leptoquark--fermion Yukawa
couplings through Eq.~\eqref{eq:ci} in such a way that the correct pattern of
neutrino masses and mixings is produced. Here we consider the region of
parameter space where $m_{\phi_2} , m_{f} \gg m_{\phi_1}$ so that $\phi_1$
comes to be identified as the BN leptoquark, while $\phi_2$ and $f$ are
effectively divorced from the flavor anomalies. For this reason we refer to
$\phi_1$ simply as $\phi$ and suppress the leptoquark-flavor indices for the
remainder of the discussion unless a distinction is necessary. The limit
$m_{\phi_1} \ll m_{\phi_2}$ also allows for a simplification in the matrix
product $\tilde{\mathbf{I}}^{-1/2}\mathbf{S}$ featuring in Eq.~\eqref{eq:ci}:
\begin{equation}
  \tilde{\mathbf{I}}^{-1/2}\mathbf{S} \approx I_{11}^{-1/2} \begin{pmatrix} -i & i/\epsilon \\ 1 & \epsilon \\ \end{pmatrix},
\end{equation}
where $\epsilon \equiv I_{12}/I_{11} \ll 1$. This flavor structure implies that its
contribution to neutrino mixing is small, and thus the PMNS parameters are
principally determined by the Yukawa couplings $x_{i3a}$. We exploit this
relative insensitivity to $m_f$ and $m_{\phi_2}$ to simplify our analysis in
the following.

\begin{figure}[t]
  \centering%
  \begin{minipage}[t]{0.45\linewidth}
    \centering \includegraphics[scale=0.55]{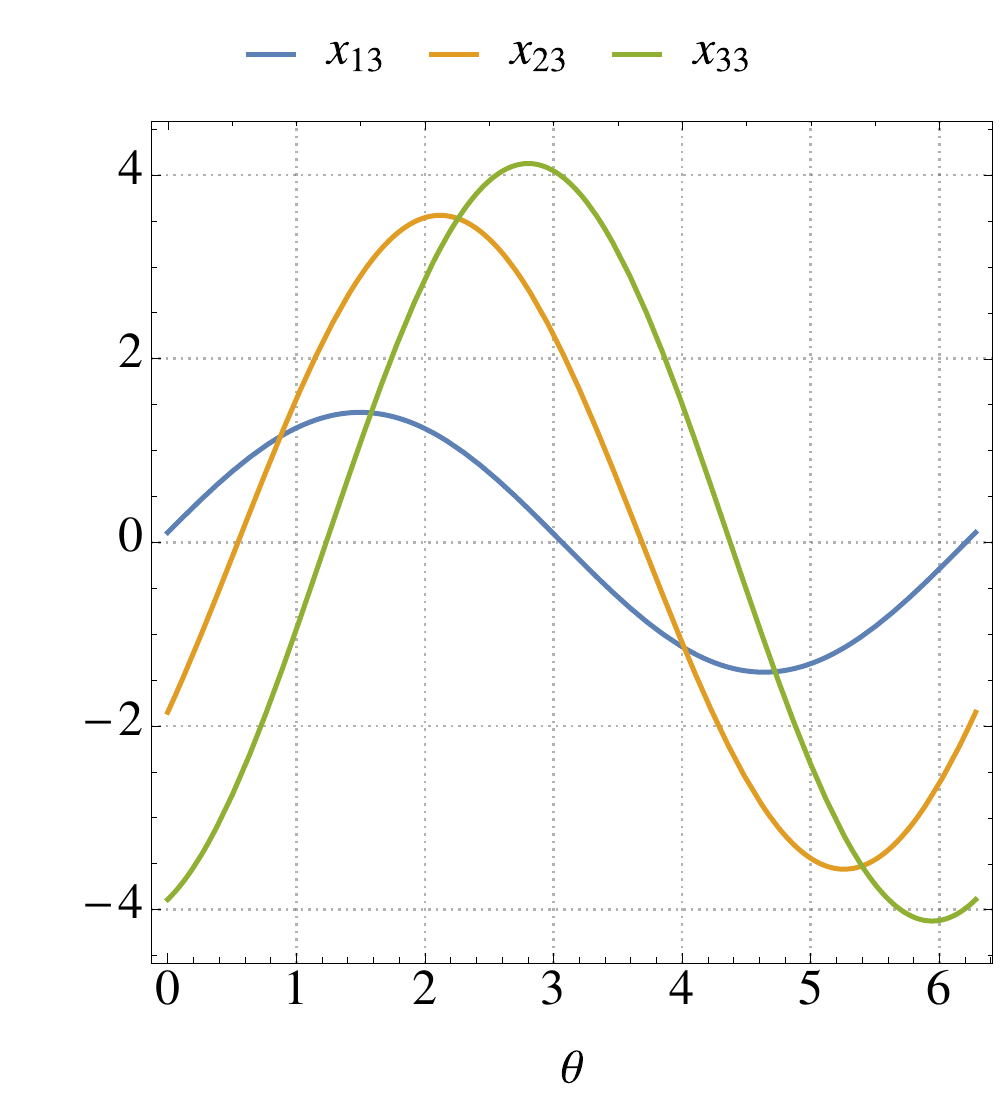}
    \subcaption{Normal neutrino mass hierarchy.}
    \label{fig:noxi3plot}
  \end{minipage}
  \hfill
  \begin{minipage}[t]{0.45\linewidth}
    \centering \includegraphics[scale=0.51]{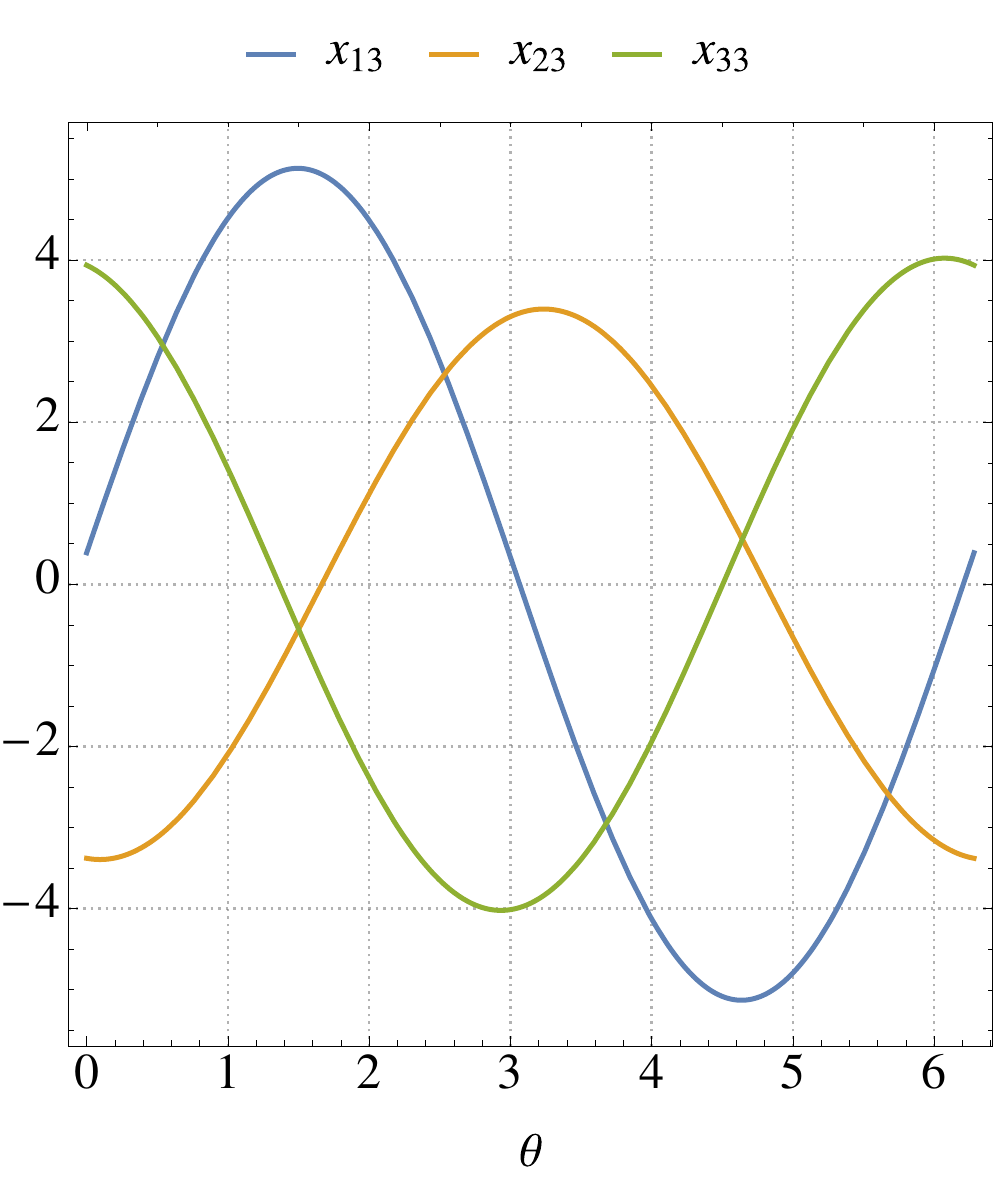}
    \subcaption{Inverted neutrino mass hierarchy.}
    \label{fig:ioxi3plot}
  \end{minipage}
  \caption{Plots of the relative sizes of the couplings of the leptoquark
    $\phi_1$ to the bottom quark and the $i$th neutrino flavor against $\theta$,
    the Casas--Ibarra parameter, for $m_f = 25 \text{ TeV}$, $m_{\phi_2} = 20
    \text{ TeV}$, $m_{\phi_1} = 4 \text{ TeV}$ and $w_3 = 0.003$. We only
    consider the case $\theta \in \mathbb{R}$ here.}
  \label{fig:xi3plot}
\end{figure}

The decoupling of $f$ and $\phi_2$ from the relevant flavor physics makes
$w_3$ an effectively free parameter that acts as a lepton-flavor-blind scaling
factor on the couplings of the leptoquark to the third generation of quarks,
while $\theta$ governs their relative sizes for a given leptoquark flavor. We
plot the $x_{i3}$ against real $\theta$ values in Fig.~\ref{fig:xi3plot} for the
mass choices $m_f = 25 \text{ TeV}$, $m_{\phi_2} = 20 \text{ TeV}$ and
$m_{\phi_1} = 4 \text{ TeV}$ with fixed $w_3 = 0.003$. Both the normal and
inverted hierarchies are considered.

\begin{figure}[t]
  \centering
  \includegraphics[scale=0.45]{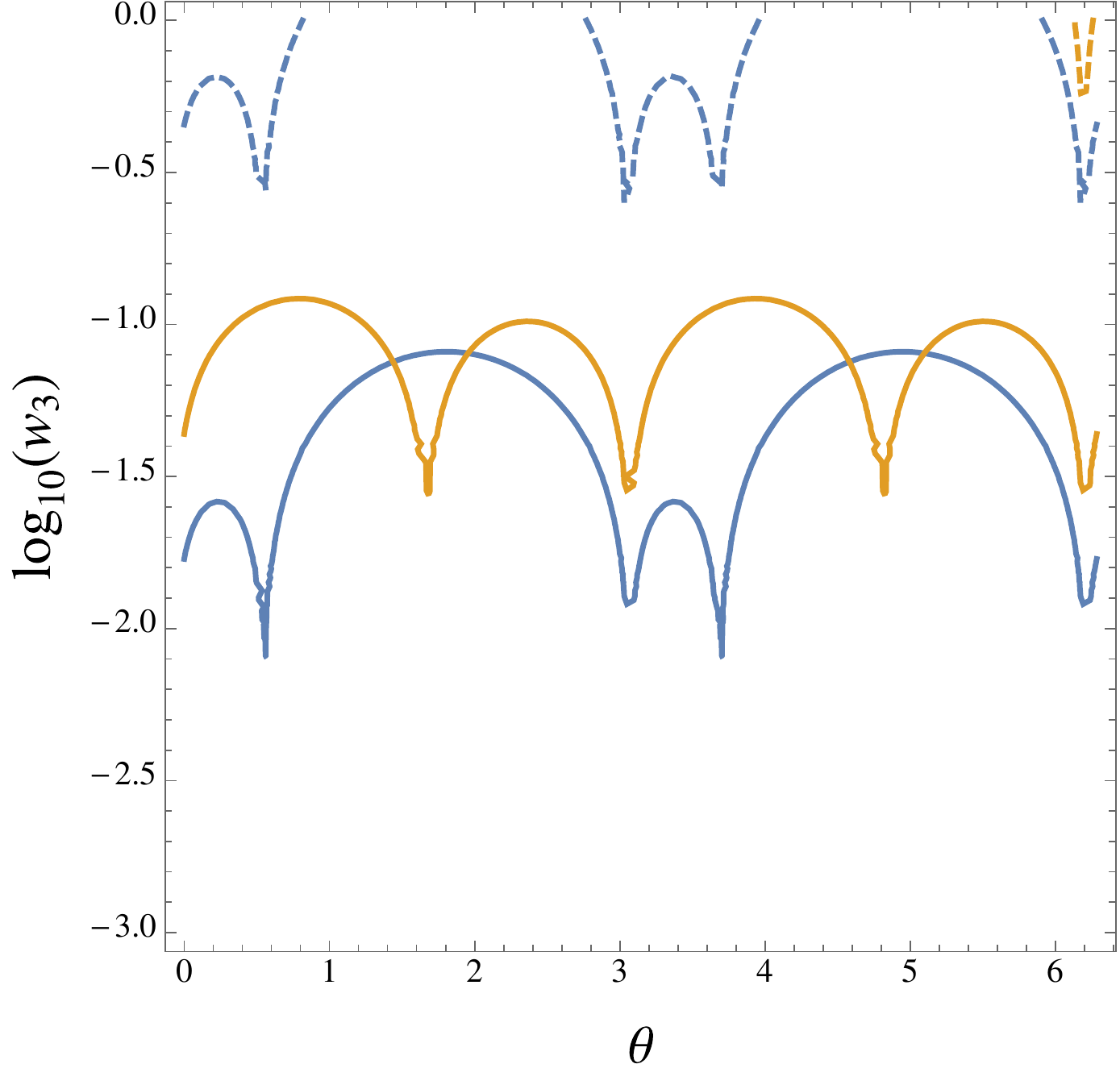}
  \caption{The figure shows the current (solid) and
    expected~\cite{Kurup:2011zza} (dashed) limits from muon--electron conversion
    in nuclei in the $\theta$--$w_3$ plane for normal mass ordering (blue) and
    inverted ordering (orange). The region below each curve is ruled out. The
    dips at $\theta \approx 3.08$ and $\theta \approx 6.22$ stretch to negative
    infinity. Aside from accidental cancellation, the values $\theta \approx
    3.08, 6.22$ ensure that the coupling to the electron vanishes. Only real
    values of $\theta$ are considered.}
  \label{fig:muNeN}
\end{figure}

Both Fig.~\ref{fig:xi3plot} and Eq.~\eqref{eq:ci} indicate that, with the
inclusion of neutrino mass, the couplings to the electron and electron-neutrino
cannot be turned off \textit{ad libitum}. Even a small electron coupling $z_{13}
\neq 0$ can generate dangerous contributions to muon--electron conversion in
nuclei in the presence of $z_{23} \neq 0$, necessary for the model to alleviate
the tensions in the $b \to s$ transition. We plot the current limit from
muon--electron conversion experiments in gold nuclei $\text{Br}(\mu
\ce{^{197}_{79}}\text{Au} \to e \ce{^{197}_{79}}\text{Au}) < 7.0 \cdot
10^{-13}$~\cite{Olive:2016xmw} against $\theta$ and $w_3$ in
Fig.~\ref{fig:muNeN} for both the normal and inverted hierarchies and a range of
masses $m_{\phi_1}$. The prospective limit from the COMET experiment: $\text{Br}
\sim 10^{-16}$~\cite{Kurup:2011zza}, is also shown. A fit to the neutrino
oscillation data while respecting measurements of muon--electron conversion
implies a fine-tuning in $\theta$---or, equivalently, $z_{31}$---to arrange
$|z_{31}| \ll |z_{33}|$, pushing the model into a very specific region of
parameter space. The required $x_{31} \approx 0$ can be arranged with $\theta
\approx 3.08 \pm n\pi$, fixing the ratio $x_{33}/x_{32} = 1.96$ for the normal
neutrino mass hierarchy, and $x_{33}/x_{32} = -0.85$ for the inverted hierarchy.
Comparison with Fig.~\ref{fig:x33x32rat}, however, indicates that neither of the
aforementioned ratios can allow large contributions to $R_{K^{(*)}}$ in the
correct direction, although the inverted hierarchy does slightly better than the
normal mass ordering. This makes a combined explanation of the $b \to s$
anomalies and neutrino mass in this model problematic. If, instead, one required
that this model explain $R_{D^{(*)}}$, $(g-2)_\mu$ and neutrino mass, the values
of $x_{33}$ required are compatible with both the normal and inverted
hierarchies, and the model remains agnostic with respect to its preference.

\section{Conclusions}

We have reconsidered the potential of a scalar leptoquark
$\phi\sim(\mathbf{3},\mathbf{1},-1/3)$ to explain recent $B$-physics
anomalies---the LFU ratios $R_{K^{(*)}}$ and $R_{D^{(*)}}$, anomalies in
branching ratio data and angular observables in the $b \to s$ transition, as
well as the anomalous magnetic moment of the muon.

The leptoquark can reduce the tension in the $R_{D^{(*)}}$ observables to within
$1\sigma$ of their current experimental values at the price of a sizeable
coupling to the right-handed tau and charm quark. The explanation loses
viability for masses above about $10 \text{ TeV}$. The leptoquark can also
reduce the tensions in the $b \to s$ data, particularly the LFU observables
$R_K$ and $R_{K^*}$, albeit at some expense to the explanation of $R_{D^{(*)}}$.
Explicitly, the region of parameter space in which $R_{D^{(*)}}$ is accommodated
to within $1\sigma$ implies $R_{K^{(*)}}$ values differing from SM prediction by
$< 1\%$, and coupling textures explaining $R_{K^{(*)}}$ to within $1\sigma$ keep
$R_{D^{(*)}}$ within theoretical uncertainty from SM prediction. At best, we
find that the model can accommodate the combined tension from both $R_{K^{(*)}}$
and $R_{D^{(*)}}$ to within $2\sigma$ as well as eliminate the tension in
$(g-2)_\mu$, a remarkable feat for a single-particle extension of the SM.

A crucial new ingredient for this model's explanation of $R_{D^{(*)}}$ is the
consideration of the area of parameter space in which the coupling $y_{32}$ is
large. The combination of right- and left-handed couplings induces scalar and
tensor operators, which lift the chirality suppression of the $B$-meson decays
and consequently produce a sizeable new-physics contribution. Moreover the
tensor contribution resolves a possible tension induced by the scalar
contribution to leptonic charmed $B$-meson decays, $B_c\to\tau\nu$. In our
numerical scans we found that the right-handed Yukawa coupling $y_{32}$ need
take $\mathscr{O}(1)$ values, while the left-handed couplings $x_{22}$ and
$x_{32}$ and the right-handed coupling $y_{22}$ are required to be small.
Interestingly, this model predicts a value of $R_{D^*}$ slightly smaller than
that suggested by current data, consistent with the Belle results.

An explanation of $R_{K^{(*)}}$ requires $\mathscr{O}(1)$ couplings of the
leptoquark to the muon, a scenario in conflict with the experimental
measurements of the decays of the $Z$ boson and $D^0$ mesons in the context of
this leptoquark model. Moreover, the tension between $R_{K^{(*)}}$ and the
lepton universality ratio $R_{D^{(*)}}^{\mu/e}$, pointed out in
Ref.~\cite{Becirevic:2016oho}, is naturally relieved for leptoquark masses
$m_\phi \gtrsim 1 \text{ TeV}$. Consequently, the best fit to $R_{K^{(*)}}$
(requiring large, negative values of $C_{LL}^\phi$) is obtained for large
leptoquark masses of $\sim 5 \text{ TeV}$ with a large hierarchy between the
left-handed couplings $|x_{32}|\gg|x_{33}|$ to avoid constraints from $\tau\to
\mu$ LFV transitions.

Apart from the anomalies in lepton flavor universality ratios, the leptoquark
can easily account for the anomalous magnetic moment of the muon by an
appropriate choice of the product of couplings $y_{23} z_{23}$. Moreover, the
leptoquark appears naturally in models of neutrino
mass~\cite{Mahanta:1999xd,AristizabalSierra:2007nf,Angel:2013hla,Cai:2014kra}.
We explicitly demonstrate the possibility to explain $R_{D^{(*)}}$ in the
two-loop neutrino mass model proposed in Ref.~\cite{Angel:2013hla}.

At a future $100 \text{ TeV}$ proton--proton collider the pair-production cross
section of the leptoquark will be substantially enhanced compared to the LHC
with about $1 \text{ fb}$ for a $5 \text{ TeV}$
leptoquark~\cite{Arkani-Hamed:2015vfh} and thus will be able to probe most of
the relevant parameter space for the $B$-physics anomalies studied here.

\section*{Acknowledgments}

This work was supported in part by the Australian Research Council. We thank
Innes Bigaran, Alexander Ermakov and Phillip Urquijo for discussions on the
flavor anomalies. We also thank Damir Be\v{c}irevi\'{c}, Christoph Bobeth, Jorge
Camalich, Andreas Crivellin, Diptimoy Ghosh, Gudrun Hiller, Soumitra Nandi,
Mariano Quiros, Olcyr Sumensari and Javier Virto for helpful correspondence. JG
thanks the organisers of the `Instant workshop on $B$-meson anomalies' held at
CERN for their hospitality and the many valuable discussions that took place
there. All Feynman diagrams were generated using the Ti\textit{k}Z-Feynman
package for \LaTeX~\cite{Ellis:2016jkw}.

\bibliography{mybib}

\end{document}